\documentclass[12pt]{article}

\usepackage{scicite}
\usepackage{times}
\usepackage{graphics}
\usepackage{epsfig}
\usepackage{multirow}
\usepackage{ulem}

\newenvironment{sciabstract}{%
\begin{quote} \bf}
{\end{quote}}

\newcounter{lastnote}

\title{Circumstellar Material in Type Ia Supernovae via Sodium Absorption Features}

\author
{A. Sternberg$^{1\ast}$, A. Gal-Yam$^{1\ast}$, J. D. Simon$^{2}$, D. C. Leonard$^{3}$,
R. M. Quimby$^{4}$, \\ 
M. M. Phillips$^{5}$, N. Morrell$^{5}$, I. B. Thompson$^{2}$, I. Ivans$^{6}$, J. L. Marshall$^{7}$, \\
A. V. Filippenko$^{8}$, G. W. Marcy$^{8}$, J. S. Bloom$^{8}$, F. Patat$^{9}$, R. J. Foley$^{10}$, D. Yong$^{11}$,\\
 B. E. Penprase$^{12}$, D. J. Beeler$^{12}$, C. Allende Prieto$^{13,14}$, G. S. Stringfellow$^{15}$ \\
\small{$^{1}$Benoziyo Center for Astrophysics, Faculty of Physics,
  Weizmann Institute of Science, Rehovot 76100, Israel.} \\
\small{$^{2}$Observatories of the Carnegie Institution of Washington,
  813 Santa Barbara St., Pasadena, CA 91101, USA.}\\
\small{$^{3}$Department of Astronomy, San Diego State University, San
  Diego, CA 92182, USA.}\\
\small{$^{4}$Cahill Center for Astrophysics, California Institute of Technology, Pasadena, CA 91125, USA.} \\
\small{$^{5}$Carnegie Observatories, Las Campanas Observatory, Casilla
  601, La Serena, Chile.} \\
\small{$^{6}$Deparment of Physics \& Astronomy, The University of Utah, Salt Lake City, UT 84112, USA.} \\
\small{$^{7}$Department of Physics, Texas A\&M University, 4242 TAMU,
  College Station, TX 77843, USA.} \\
\small{$^{8}$Department of Astronomy, University of California,
  Berkeley, CA 94720-3411, USA.} \\
\small{$^{9}$European Southern Observatory (ESO), Karl Schwarzschild
  Strasse 2, 85748,}\\
\small{Garching bei M\"unchen, Germany.} \\
\small{$^{10}$Clay Fellow, Harvard-Smithsonian Center for Astrophysics, 60 Garden
  Street, Cambridge, MA 02138, USA.} \\
\small{$^{11}$Research School of Astronomy \& Astrophysics, The
  Australian National University,} \\
\small{Mount Stromlo Observatory, Cotter Rd., Weston ACT 2611,
  Australia.} \\
\small{$^{12}$Department of Physics and Astronomy, Pomona College, 610
  N. College Ave., Claremont, CA 91711, USA.} \\
\small{$^{13}$Instituto de Astrof\'{\i}sica de Canarias, 38205, La
  Laguna, Tenerife, Spain.}\\
\small{$^{14}$Departamento de Astrof\'{\i}sica, Universidad de La
  Laguna, 38206, La Laguna, Tenerife, Spain.}\\
\small{$^{15}$Center for Astrophysics and Space Astronomy, University
  of Colorado, 389-UCB, Boulder, CO 80309, USA.} \\
\\
\footnotesize{$^\ast$To whom correspondence should be addressed;
  E-mail: assaf.sternberg@weizmann.ac.il (A.S.),}\\
\footnotesize{avishay.gal-yam@weizmann.ac.il (A.G).} \\
\footnotesize{The data described in the paper are available for public download from the WIS Experimental} \\
\footnotesize{ Astrophysics Spectroscopy System (http://www.weizmann.ac.il/astrophysics/wiseass/).}
}
\date{}

\begin{document} 

\baselineskip24pt 

\maketitle 

\newpage

\begin{sciabstract}
Type Ia supernovae are key tools for measuring distances on a cosmic scale. They are generally thought to be the thermonuclear explosion of an accreting white dwarf in a close binary system. The nature of the mass donor is still uncertain. In the single-degenerate model it is a main-sequence star or an evolved star, whereas in the double-degenerate model it is another white dwarf. We show that the velocity structure of absorbing material along the line of sight to 35 type Ia supernovae tends to be blueshifted. These structures are likely signatures of gas outflows from the supernova progenitor systems. Thus many type Ia supernovae in nearby spiral galaxies may originate in single-degenerate systems.
\end{sciabstract}

Type Ia supernovae (SNe~Ia) have large and calibratable luminosities, making them good tools for measuring distances on a cosmic scale to gauge the geometry and evolution of the Universe ({\it 1, 2}). Understanding the nature of the progenitor system is important, as progenitor evolution or a changing mix of different progenitors may bias cosmological inferences. The consensus view of SNe Ia is that mass transfer onto a massive carbon-oxygen white-dwarf (WD) star in a close binary leads to a thermonuclear explosion, as the mass of the WD approaches the critical Chandrasekhar mass limit ({\it 3}). In the single-degenerate (SD) model the mass donor is either a main-sequence star or an evolved subgiant or giant star, whereas in the competing double-degenerate (DD) model it is another WD ({\it 4}).

In the SD scenario, nonaccreted material blown away from the system before the explosion should remain as circumstellar matter (CSM) ({\it 5}); thus, detection of CSM in SN~Ia spectra would lend support to the SD model. Patat et al. reported such a detection based on time-variable absorption features of the Na~I doublet (Na~I~D$_1$ and D$_2$; $\lambda_{D_1} = 5896$~\AA\ and $\lambda_{D_2} = 5890$~\AA\ rest-frame wavelengths) in optical spectra of SN 2006X ({\it 6}). These authors suggest that CSM may be common to all SNe~Ia, though variation in its detectability can exist due to viewing angle effects. Although multi-epoch high-resolution spectra of SN 2000cx ({\it 7}) and SN 2007af ({\it 8}) exhibited no absorption features or no time variation in them, additional detections were recently reported for three other events: SNe 2007le, 1999cl, and 2006dd ({\it 9-11}). While it has been suggested that Na~D absorption could not have been caused by CSM due to ionization considerations ({\it 12}), detailed arguments to the contrary have been presented ({\it 9, 13}).

Absorption features from unrelated intervening gas clouds are expected to have random velocity offsets with respect to the SN, whereas absorption due to winds blown by the progenitor system is expected to be always blueshifted with respect to the SN, as the source of light is behind the outflowing absorbing material. We thus searched for this signature in a large sample of high-resolution, single-epoch observations of SNe~Ia. Our sample consisted of spectra of 35 SNe~Ia and 11 core-collapse (CC) SNe obtained using the Keck HIRES and Magellan MIKE spectrometers. We also studied previously published spectra of six SNe~Ia ({\it 6, 14--17}) and seven CC~SNe ({\it 16--22}). A bias may exist if other spectra were obtained but not published. Therefore, we report results separately with and without these historical SNe (hereafter ``the extended sample", for which results are given in square brackets). Detailed analyses of a few events from our sample have been published: SN 2006X ({\it 6}), SN 2007af ({\it 7}), SN 2007le ({\it 9}), and SN 2008D ({\it 23}).

The HIRES observations used setups with spectral resolution ($R=\lambda/\delta\lambda$) of $\sim$ 50,000--52,000, giving full width at half-maximum (FWHM) intensity of $\sim 0.115$--0.12~\AA\, (or $\sim 5.5$--6 km s$^{-1}$) in the vicinity of the Na~I~D lines. The MIKE observations were obtained with $R \approx$ 30,000--36,000, giving FWHM $\approx 0.165$--0.2~\AA\, (or $\sim 8.4$--10 km s$^{-1}$) near Na~I~D. We reduced the HIRES spectra using the MAKEE (MAuna Kea Echelle Extraction) pipeline ({\it 24}). The MIKE data were reduced using the latest version of the MIKE pipeline ({\it 25}).

We obtained the redshift and morphological classification of the host galaxies of our SN sample from the NASA/IPAC Extragalactic Database (NED; {\it 26}), the SIMBAD astronomical database ({\it 27}), or using images from the Digital Sky Survey (DSS; {\it 28}). Our sample host-galaxy redshifts $z$ are between 0.0019 [0.0015] and 0.06. Tables S1 and S2 list our complete SN sample with host-galaxy properties.

The redshifts of the host galaxies are sufficiently large to allow us to differentiate between absorption resulting from material in the Milky Way (MW; i.e., at $z \approx 0$; used as a control sample) and from material in the host galaxies of the SNe. Out of the 35 [41] events in our SN~Ia sample, 22 [28] events exhibit absorption features consistent with the object's host-galaxy redshift and 13 events do not [supporting online  material (SOM) text S1]. In the CC~SN sample, 9 [16] events exhibit absorption compatible with the SN host-galaxy redshifts and 2 events do not (SOM text S1). Of a total of 46 [59] events (SNe~Ia and CC~SNe), 42 [51] exhibit Na~I~D absorption due to material in the Milky Way whereas 4 [8] do not.

We normalized the spectra across the continuum and translated wavelengths into velocities relative to the wavelength of the minimum of the most prominent absorption feature of both of the D lines (SOM text S2), both for the features in the host galaxies and in the Milky Way when detected, using the Doppler shift,
\begin{equation}
v/c \approx (\lambda-\lambda_0)/\lambda_0\,,
\end{equation}
where $c$ is the speed of light in vacuum, $\lambda_0$ is the wavelength of the zero-velocity component, and $v$ is the velocity of the component that was Doppler shifted from $\lambda_0$ to $\lambda$. We categorized the absorption features exhibited in our spectra into three classes of structures as follows (see Fig. 1 for graphic examples):
\begin{itemize}
\item[(i)] {\it Blueshifted}: One prominent absorption feature with weaker features at shorter wavelengths with respect to it.
\item[(ii)] {\it Redshifted}: One prominent absorption feature with weaker features at longer wavelengths with respect to it.
\item[(iii)] {\it Single/Symmetric}: A single absorption feature, or several features with both blue and redshifted structures of similar magnitude.
\end{itemize}
Classification results are presented in Table 1, Figure 2, and Figure S1. \\
 
\begin{figure}
\vspace{0 cm}
\includegraphics[width=16.0cm]{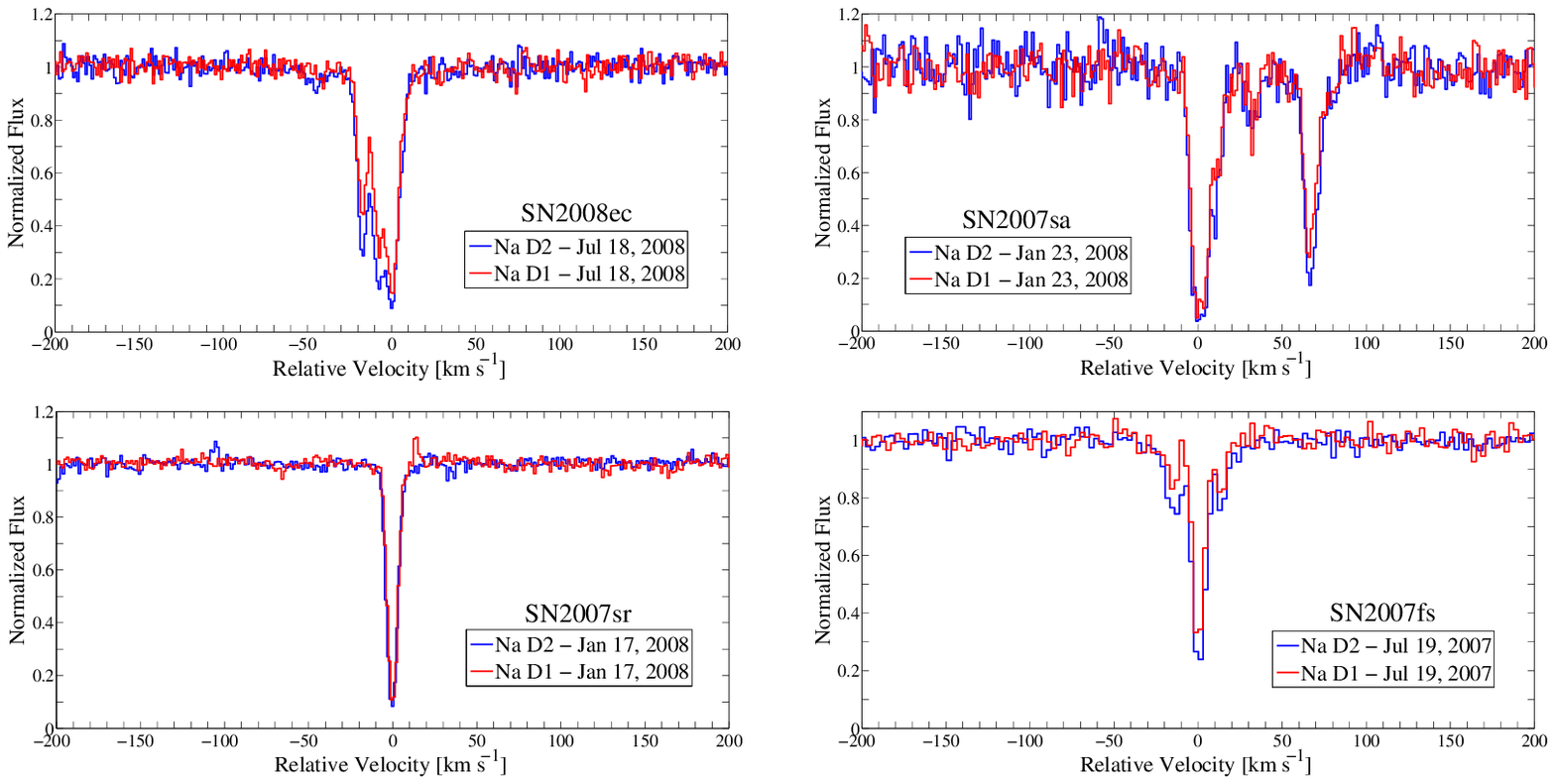} \\
\noindent {\bf Fig. 1:} Examples of spectra classified as {\it blueshifted} (top left), {\it redshifted} (top right), and {\it single} (bottom left) or {\it symmetric} (bottom right). Flux is normalized and velocities are given in km s$^{-1}$ relative to the prominent absorption-feature minimum (relative to D$_2$ in blue and to D$_1$ in red).
\end{figure}

\vskip 0.5 cm
\begin{figure}
\vspace{0 cm}
\includegraphics[width=16.0cm]{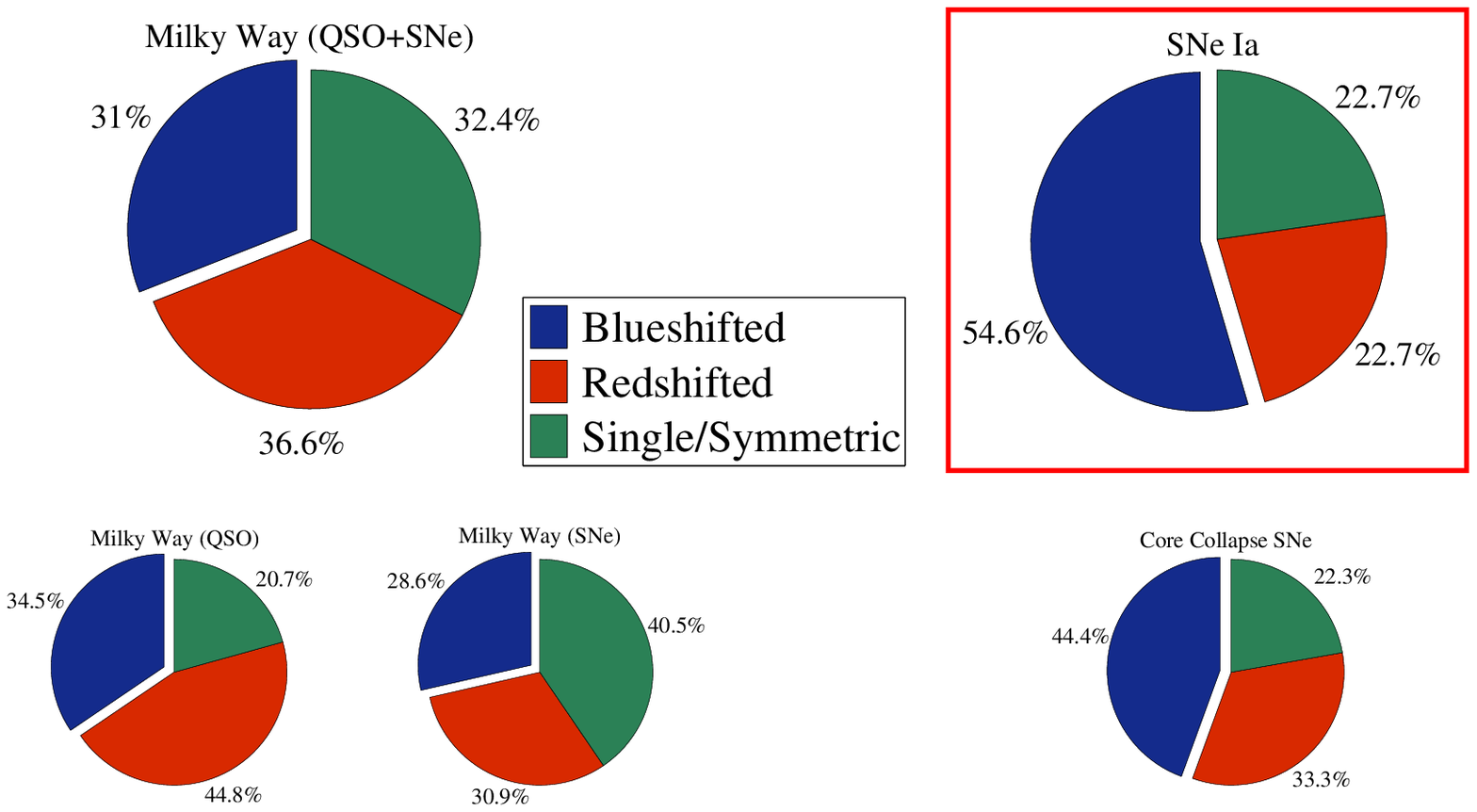} \\
\noindent {\bf Fig. 2:} Pie-chart visualization of the results given in Table 1. The results for the extended samples are shown in Fig. S1.
\end{figure}

We used the Galactic Na~I~D absorption along the line of sight to our SNe as a control sample, extending it with Galactic absorption along the line of sight toward quasi-stellar objects (QSOs) published by Ben Bekhti et al. ({\it 29}). To imitate being observed from outside of the Milky Way we inverted the velocities of the Galactic features. Figures S2--S5 show the spectra of the SN~Ia sample and Figure S6 shows the spectra of the CC~SN sample. 

The absorption features we observe are narrow, with typical separations and velocity dispersions of a few tens of km s$^{-1}$, and are distinct from features due to material ejected in the explosion itself that are wide ($\sim 10^4$ km s$^{-1}$), reflecting the expansion velocity of the SN ejecta. It is apparent from Table 1, Figure 2, and Figure S1 that the SN~Ia sample displays a strong preference for blueshifted structures, whereas the Milky Way sample shows no significant preference.

To test whether our SN~Ia sample could be a random draw out of a uniform distribution, we calculated the probability of observing a set of 22 [28] spectra exhibiting the preference we detected in our sample, or a set even more extreme (i.e., 12 [16] or more blueshifted spectra and 5 [6] or fewer redshifted spectra); the result was low 2.23\% [0.54\%]. Moreover, a K-S test rejects the null hypothesis that the SN~Ia and MW samples are from the same distribution at a $1\%$ significance level.

Out of the 17 SNe~Ia that were classified as blueshifted or redshifted, 2 events (11.8\%; SNe 2006X and 2008fp) exhibit saturated features. Out of the 7 CC~SNe classified as blueshifted or redshifted, 4 events (57.1\%; SNe 2006bp, 2008cg,x 2008D, and 2008J) exhibit saturated features. Three SNe~Ia in our sample occurred in elliptical galaxies (SNe 2006ct, 2006eu, and 2007on; see Table S1), and their spectra do not exhibit Na~I~D absorption features. Eight of our SNe~Ia occurred in lenticular galaxies, of which two exhibit absorption features (one blueshifted and one single) and 6 do not. Therefore, our sample provides a good representation of SNe~Ia in nearby ($z \leq 0.06$) spiral galaxies.

We calculated the velocities and column density ($N$) of the absorbing material by fitting them with Voigt profiles useing VPFIT [({\it 30}); see Figure 3]. SN~Ia features have a log($N$) similar to the that of the MW features, lower on average than those in CC~SNe. The relative velocities of the SN~Ia features range is between $-150.8$ to +139.4 km s$^{-1}$ (with an additional single feature at +193.7 km s$^{-1}$), whereas the CC~SN relative velocities range is between $-87.2$ to +116.6 km s$^{-1}$, and the MW velocity range is between $-79.5$ to +116.9 km s$^{-1}$. The cumulative distribution of the relative number of features as a function of the velocity for each sample is given in Fig. 4.

\vskip 0.5 cm
\begin{figure}
\vspace{0 cm}
\includegraphics[width=16.0cm]{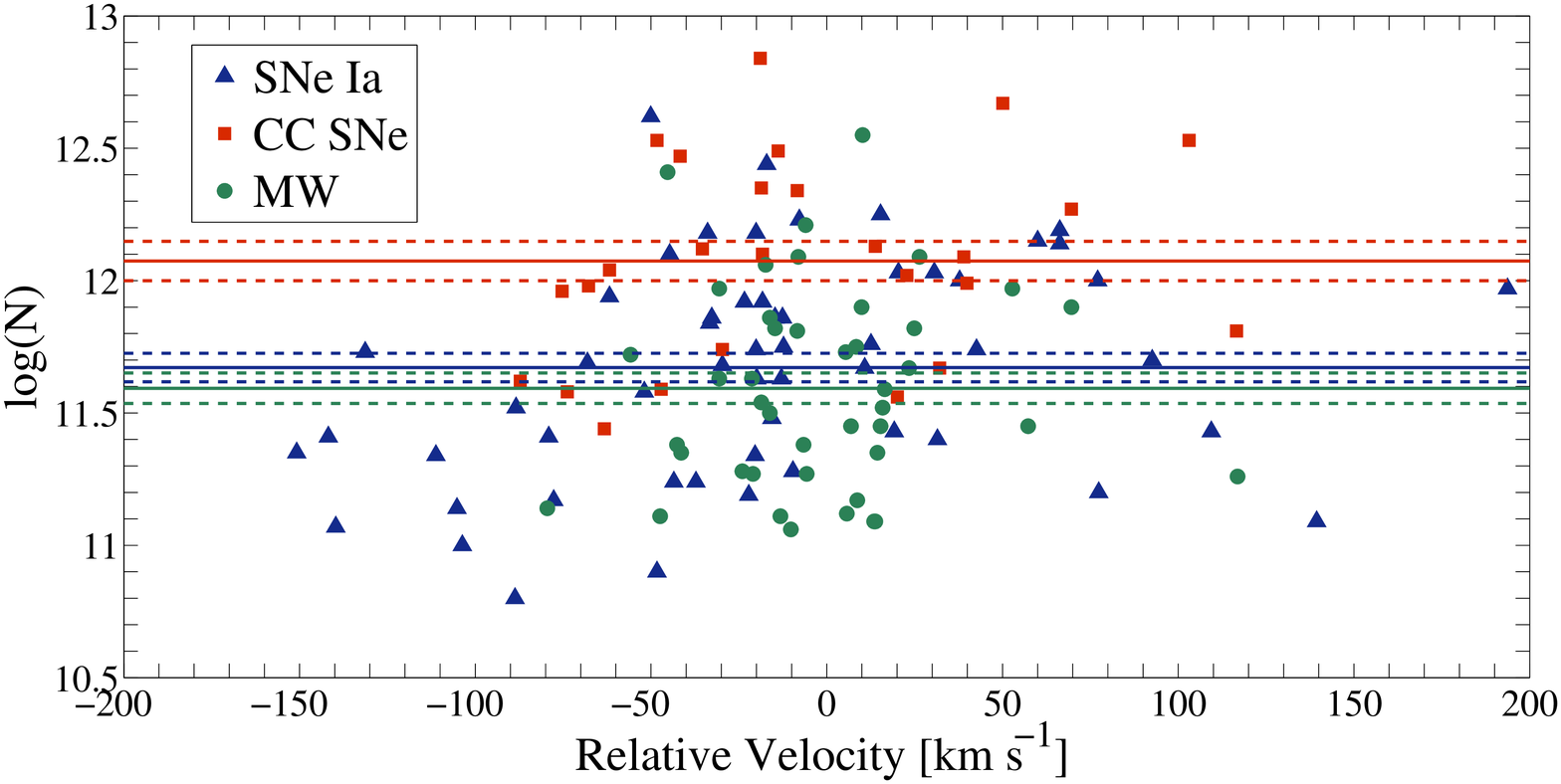} \\
\noindent {\bf Fig. 3:} A plot of log($N$) of the absorbing material calculated using VPFIT as a function of velocity relative to the zero-velocity component, for the blueshifted and redshifted components observed in our sample. Zero-velocity components are omitted. The solid lines are the log($N$) averages for the different samples: SNe~Ia (blue), log($N$)$_{\rm avg} = 11.67 \pm 0.054$; CC~SNe (red), log($N$)$_{\rm avg} = 12.07 \pm 0.075$; and MW (green), log($N$)$_{\rm avg} = 11.59 \pm 0.058$. Column densities are given in cm$^{-2}$. Dashed lines mark the uncertainty in the mean of the corresponding solid lines.
\end{figure}

\vskip 0.5 cm
\begin{figure}
\vspace{0 cm}
\includegraphics[width=16.0cm]{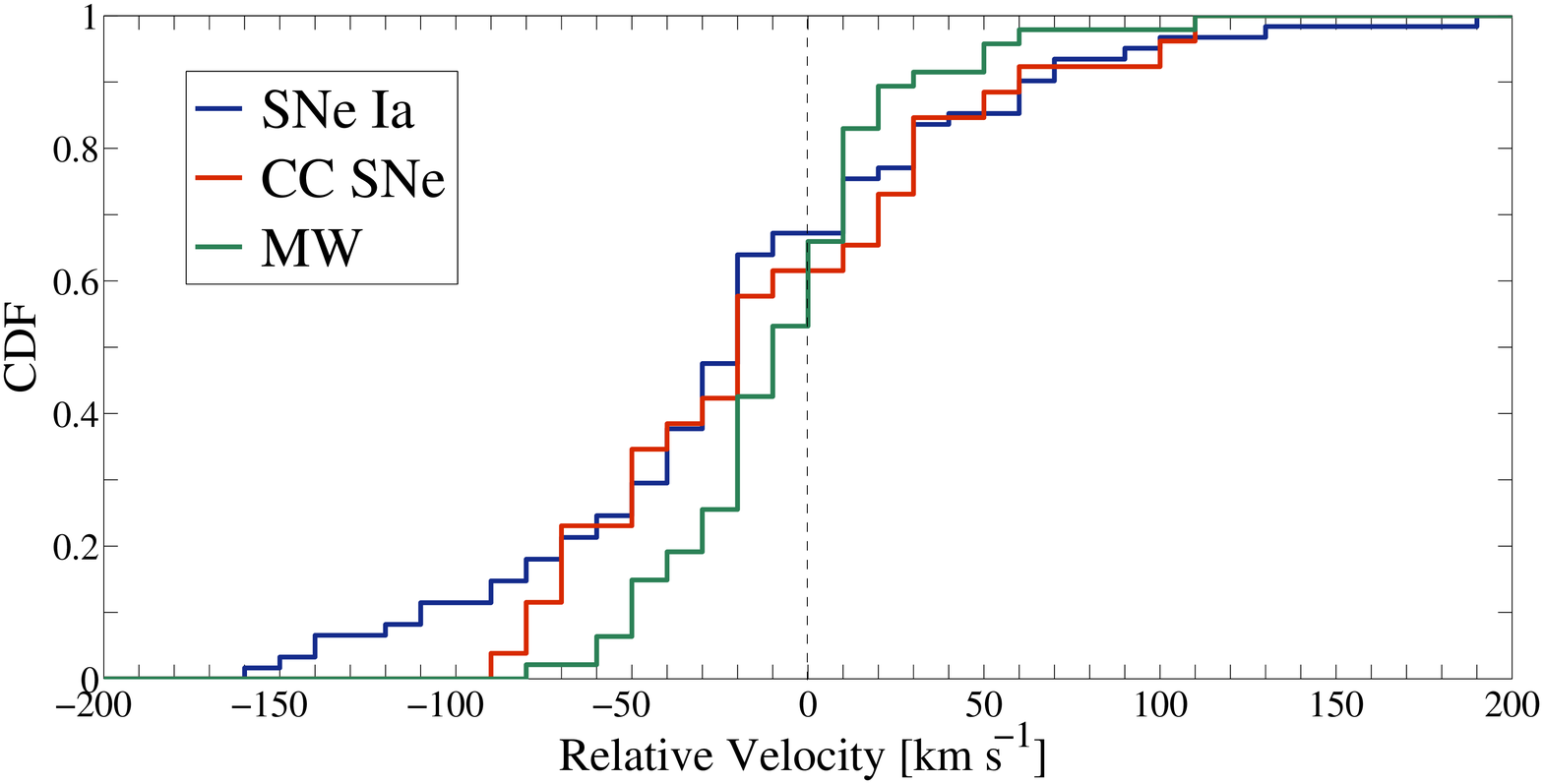} \\
\noindent {\bf Fig. 4:} A plot of the cumulative density function of our three samples (SNe~Ia, CC~SNe, and MW). The zero-velocity components are omitted in order to minimize the effect of the different sample sizes. The color scheme is the same as that used in Figure 3. The overabundance of SN~Ia high-velocity bluesifted absorption components ($-150$ km s$^{-1} \leq v_{\rm rel} \leq -100$ km s$^{-1}$) in comparison with the CC~SN sample is questionable, as a K-S test cannot reject (at a $5\%$ significance level) the null hypothesis that both samples are from the same parent distribution.
\end{figure}

\begin{table}                                                          
\caption{Classification of absorption features. }
\begin{center}
\begin{tabular}{| c || c | c | c || c |}
\hline
Sample & Blueshifted & Redshifted & Single/Symmetric & Total \\
\hline
\hline
SNe~Ia & 12 [16] & 5 [6] & 5 [6] & 22 [28] \\
\hline
CC~SNe & 4 [8] & 3 [5] & 2 [3] & 9 [16] \\
\hline
MW~(SNe) & 12 [13] & 13 [16] & 17 [22] & 42 [51] \\
\hline
MW~(QSO) & 10 & 13 & 6 & 29\\
\hline
MW (QSO+SNe) & 22 [23] & 26 [29] & 23 [28] & 71 [80] \\
\hline
\end{tabular}
\end{center}
\end{table}

The narrow absorption features we measured could possibly be produced by,
\begin{itemize}
\item[(i)] intervening clouds of gas in the host galaxy;
\item[(ii)] a systemic wind blown by the host galaxy;
\item[(iii)] nonrandomly moving interstellar matter (ISM) - a relic from a star-forming event; and 
\item[(iv)] material ejected from the progenitor system before its explosion.
\end{itemize}
In the first case, intervening clouds of gas in galaxies may have a small random velocity relative to the local rotation velocity; therefore, the resulting absorption features are equally likely to be either blueshifted or redshifted relative to the SN, in contrast to the preference exhibited in our SN~Ia sample (shown to be inconsistent with a random draw from a uniform parent distribution). The structure of the absorption features observed in the Milky Way is due to the movement of ISM gas clouds, and our results are in good agreement with a uniform distribution from randomly moving clouds (see Table 1, Fig. 2, and Fig. S1). The CC~SN sample is unfortunately quite small, and is consistent with either a uniform distribution or one similar to that of the SN~Ia sample.

If the second explanation was viable, we would expect to see a signature of this wind in the Milky Way sample, as a prevalence of blueshifted absorption (i.e., outflowing). However, the MW could be a special case. We would also expect to see a similar, or even greater, effect in our CC~SN sample, but this is not seen. Nevertheless, the CC~SN sample is too small to reach statistical significance. Furthermore, if this model was correct, blueshift due to galactic outflows would be common, and imply a large velocity span, at small host inclination (i.e., hosts closer to being face on). Figure S7 shows that the absorption features of SN~Ia events that occurred in spiral hosts (see Figure S9--S11 for host images) show no preference to low-inclination hosts, nor is there a correlation between the velocity span and the inclination.

In addition, galactic winds are observed mainly in galaxies with a star-formation-rate (SFR) surface density threshold above $\sim0.1$M$_\odot$ yr$^{-1}$ kpc$^{-2}$, and are rare in local ordinary disk galaxies [({\it 31}) and references therein]. Moreover, galactic winds are thought to be expelled into the halo and then cool and fall back onto the disk [the ``galactic fountain" process; ({\it 32--33})]. Because of its low first ionization potential (5.1 eV) sodium is a good tracer for neutral gas; thus, Na~I~D absorption is expected to be observed in the later stage of the galactic fountain process (i.e., the inflow) when the gas is mostly neutral, and less during the earlier outflow stage when the gas is expected to be mostly ionized ({\it 34--35}), resulting in a redshifted preference. Absorption from gas in galactic halos has larger equivalent widths and column densities of Ca II H \& K, ${\cal N}$(Ca II) relative to Na I D, while ISM and CSM have ${\cal N}$(Na~I)/${\cal N}$(Ca~II) of order unity [({\it 36} and references therein]. We have spectral coverage of Ca II H \& K for 12 out of the 22 SN~Ia events that exhibit Na I D absorption (seven of which are blueshifted). All these events exhibit ${\cal N}$(Na~I)/${\cal N}$(Ca~II)$\leq 1$. This implies that the absorbing material is not in the galactic halo.

It is plausible that most of the SNe in our SN~Ia sample are ``prompt" events, arising from progenitor systems with short ($<500$ Myr) delay times from star formation to explosion ({\it 37--38}). If this is the case, the massive stars that were formed at the same time as the progenitor system could have driven a wind and blown a cavity in the ISM, such that the remaining gas clouds along the line of sight would be blueshifted from our perspective. However, even in spiral galaxies, SNe~Ia do not follow massive stars as traced by H~II regions ({\it 39}), and so are unlikely to reside close to recent star-formation regions. This process might be evident in our observations of CC~SNe which show an excess of strong (and even saturated) absorption compared to the SNe~Ia. 

Hence, we conclude that the blueshifted preference is due to an inherent feature of the SNe themselves. The low velocities and narrowness of the features suggest that these absorption features are due to CSM that was ejected by the progenitor system prior to the SN explosion. Features with relatively low velocities (i.e., of order a few 10 km s$^{-1}$) may be attributed to a wind blown by a red giant companion, but those with higher velocities are incompatible with such winds.

Patat et al. ({\it 6}) proposed that the observed features in spectra of SN 2006X arise from shells of successive recurrent nova eruptions that swept up the stellar wind of the companion [SOM text S6 of ({\it 6})]. This scenario has gained support from a study of the Na~I~D, Ca~II H\&K, and K~I blueshifted absorption lines observed in spectra of RS Ophiuchi, a symbiotic recurrent nova ({\it 40}). These features are similar to the features exhibited in our SN~Ia sample (SOM text S3), i.e., at least one type of SD system is shown to be able to produce such features.

Our findings are consistent with the SD model ({\it 41-42}) as the progenitor system of a certain fraction of SNe~Ia in nearby spiral galaxies ({\it 43}). But how large is this fraction? The detection of the absorption features might be angle dependent; thus, we can only estimate a lower limit. If we assume that we should observe the same amount of blueshifted and redshifted components due to ISM, we can attribute any blueshifted excess to CSM, implying at lower limit of $25\%$ [$26\%$] of SNe~Ia events that occur in spiral galaxies. Alternatively, assuming the MW blueshifted fraction represents galactic ISM, we get an estimate of at least $20\%$ [$24\%$] of SNe~Ia in spiral hosts.

\subsection*{References and Notes}
\begin{itemize}
\item[1.]
A. G. Riess et al., {\it Astron. J.} {\bf116}, 1009 (1998).
\item[2.]
S. Perlmutter et al., {\it Astrophys. J.} {\bf 517}, 565 (1999).
\item[3.]
J. Whelan, I. Iben, {\it Astrophys. J.} {\bf186}, 1007 (1973).
\item[4.]
I. Iben, A. V. Tutukov, {\it Astrophys. J. Supp.} {\bf54}, 335 (1984).
\item[5.]
D. Branch, M. Livio, L. R. Yungelson, F. R. Boffi, E. Baron, {\it Publ. Astron. Soc. Pac.} {\bf 107}, 1019 (1995).
\item[6.]
F. Patat et al., {\it Science} {\bf 317}, 924 (2007).
\item[7.]
F. Patat et al., {\it Astron. Astrophys.} {\bf474}, 931 (2007).
\item[8.]
J. D. Simon et al., {\it Astrophys. J.} {\bf671}, 25 (2007).
\item[9.]
J. D. Simon et al., {\it Astrophys. J.} {\bf702}, 1157 (2009).
\item[10.]
S. Blondin et al., {\it Astrophys. J.} {\bf693}, 207 (2009).
\item[11.]
M. Stritzinger et al., {\it Astron. J.}, {\bf 140}, 2036 (2010).
\item[12.]
N. N. Chugai, {\it Astron. Lett.} {\bf 34}, 389 (2008).
\item[13.]
F. Patat, N. L. J. Cox, J. Parrent, D. Branch, {\it Astron. Astrophys.} {\bf 514}, 78 (2010).
\item[14.]
C. C. Steidel, R. M. Rich, J. K. McCarthy, {\it Aston. J.} {\bf 99}, 5 (1990).
\item[15.]
S. D'Odorico et al., {\it Astron. Astrophys.} {\bf 215}, 21 (1989).
\item[16.]
L. C. Ho, A. V. Filippenko, {\it Astrophys. J.} {\bf 444}, 165 (1995).
\item[17.]
J. Sollerman et al., {\it Astron. Astrophys.} {\bf 429}, 559 (2005).
\item[18.]
S. D'Odorico, M. Pettini, D. Ponz, {\it Astrophys. J.} {\bf 299}, 852 (1985).
\item[19.]
A. Vidal-Madjar et al., {\it Astron. Astrophys.} {\bf 177}, L17 (1987).
\item[20.]
A. Fassia et al., {\it Mon. Not. R. Astron. Soc.} {\bf 318}, 1093 (2000).
\item[21.]
D. C. Leonard, A. V. Filippenko, R. Chornock, R. J. Foley, {\it Publ. Astron. Soc. Pac.} {\bf 114}, 35 (2002).
\item[22.]
M. Takada-Hidai, W. Aoki, G. Zhao, {\it Publ. Astron. Soc. Japan} {\bf 54}, 899 (2002).
\item[23.]
A. Soderberg et al., {\it Nature} {\bf 453}, 469 (2008).
\item[24.]
MAKEE was written by T. Barlow and can be downloaded for free at \\
http://spider.ipac.caltech.edu/staff/tab/makee/
\item[25.]
D. Kelson, {\it Publ. Astron. Soc. Pac.} {\bf 115}, 688 (2003).
\item[26.] 
http://nedwww.ipac.caltech.edu/ .
\item[27.] 
http://simbad.u-strasbg.fr/simbad/ .
\item[28.] 
http://archive.stsci.edu/dss .
\item[29.]
N. Ben Bekhti, P. Richter, T. Westmeier, M. T. Murphy, {\it Astron. Astrophys.} {\bf 487}, 583 (2008).
\item[30.] 
VPFIT was developed by R. F. Carswell and can be downloaded for free at \\
http://www.ast.cam.ac.uk/~rfc/vpfit.html .
\item[31.]
B. Weiner et al., {\it Astrophys. J.} {\bf 692} ,187 (2009).
\item[32.]
P. R. Shapiro, G. B. Field, {\it Astrophys. J.} {\bf 205}, 762 (1976).
\item[33.]
J. N. Bregman, {\it Astrophys. J.} {\bf 237}, 280 (1980).
\item[34.]
F. Fraternali, T. Oosterloo, R. Sancisi, {\it Astron. Astrophys.} {\bf 424}, 485 (2004).
\item[35.]
H. B. Krug, D. S. N. Rupke, S. Veilleux, {\it Astrophys. J.} {\bf 708}, 1145 (2010).
\item[36.] 
J. A. Baldwin, M. M. Phillips, R. F. Carswell, {\it Mon. Not. R. Astron. Soc.} {\bf 216}, 41 (1985).
\item[37.]
F. Mannucci et al., {\it Astron. Astrophys.} {\bf 433}, 807 (2005).
\item[38.] 
I. Hachisu, M. Kato, K. Nomoto, {\it Astrophys. J.} {\bf 683}, L127 (2008).
\item[39.] 
P. A. James, J. P. Anderson, {\it Astron. Astrophys.} {\bf 453}, 57 (2006).
\item[40.] 
F. Patat et al, {\it Astron. Astrophys.} {\bf 530}, 63 (2011).
\item[41.]
I. Hachisu, M. Kato, K. Nomoto, {\it Astrophys. J.} {\bf 522}, 487 (1999).
\item[42.]
I. Hachisu, M. Kato, K. Nomoto, {\it Astrophys. J.} {\bf 679}, 1390 (2008). 
\item[43.] 
As only three of our SNe~Ia occurred in elliptical galaxies (no Na~I~D absorption detection in all) we do not know if our result also applies to SNe~Ia in elliptical galaxies. This would have to be tested with a larger sample of events in such hosts. Including these three events in our analysis (i.e., conducting our analysis for all SNe~Ia regardless of host) would not significantly change the reported results. Nondetection of absorption in spectra of SNe~Ia in elliptical galaxies favors, though does not prove, a DD progenitor system for these events. As the sample of events with no Na~I~D detection is composed of events in elliptical, lenticular, and the outskirts of spiral galaxies, this may apply to SNe~Ia arising in old environments ({\it 44}), indicating that SNe~Ia in early- and late-type galaxies might have different, or a different mixture of, progenitor systems.
\item[44.]
M. Sullivan et al., {\it Mon. Not. R. Aatron. Soc.} {\bf 406}, 782 (2010).
\item[45.]
D. C. Leonard, A. V. Filippenko, {\it Publ. Astron. Soc. Pac.} {\bf 113}, 920 (2001).
\item[46.]
Three extreme cases are SNe 2007fb, 2007kk, and 2008dt. The prominent feature of SN 2007fb is redshifted $\sim 230$ km s$^{-1}$ compared to the host-galaxy redshift obtained from NED, that of SN 2007kk is redshifted $\sim 130$ km s$^{-1}$, and that of SN 2008dt is redshifted $\sim 120$ km s$^{-1}$. These velocities, although high, are still comparable to those of disk or bulge objects with respect to the core of their galaxy.
\item[47.]
E. R. Seaquist, A. R. Taylor, {\it Astrophys. J.} {\bf 349}, 313 (1990).

\end{itemize} 
\subsection*{Acknowledgments.} We thank the anonymous reviewers for their comments. Some of the data presented herein were obtained at the W.~M. Keck Observatory, which is operated as a scientific partnership among the California Institute of Technology, the University of California, and NASA; the observatory was made possible by the generous financial support of the W. M. Keck Foundation. We thank the staffs of the various observatories at which we obtained data, as well as J. G. Cohen, G. D. Becker, A. L. Kraus, W. L. W. Sargent, E. Norris, G. J. Herczeg, G. Preston, and I. Toro-Martinez for their data contributions. We thank T. Barlow for his work on developing the MAKEE reduction pipeline, and to J. X. Prochaska and B. Weiner for advice. A.G. acknowledges support by The French-Israeli Astrophysics Network program, The Israeli Science Foundation, an EU FP7 Marie Curie IRG Fellowship, The Weizmann-UK program, and a research grant from the Peter and Patricia Gruber Awards. A.V.F.'s group at UC Berkeley has been supported by NSF grant AST-0908886 and the TABASGO Foundation. R.J.F. is supported by a Clay Fellowship. G.S.S. thanks S. Hawley for allocating Director's Discretionary Time at Apache Point Observatory, and the APO observing staff. The data described in the paper are presented in the Supporting Online Material and are available for public download from the WIS Experimental Astrophysics Spectroscopy System (http://www.weizmann.ac.il/astrophysics/wiseass/).

\newpage

\subsection*{Supporting Online Material}
\vskip 1.0 cm
{\bf S1. 3$\sigma$ Upper Limits for Detection of Absorption Features when Features are Not Apparent} 
\vskip 0.5 cm
Thirteen of the SNe~Ia and two CC~SNe in our sample do not exhibit Na~I~D absorption features. Using Equation 4 of Leonard \& Filippenko ({\it 45}), we calculated $W_\lambda(3\sigma)$, the upper detection limit of Na~I~D absorption for these spectra:
\begin{equation}
W_\lambda(3\sigma) = 3\Delta\lambda\Delta I\sqrt{\frac{W_{\rm line}}
{\Delta\lambda}}\sqrt{\frac{1}{B}},
\end{equation}
where $\Delta\lambda$ is the spectral resolution, $\Delta I$ is the root-mean square (rms) fluctuation of the flux about the normalized continuum level, $W_{\rm line}$ is the spread of the D$_2$ or D$_1$ lines (in our sample, we see that the Na I D features are usually within a spread of $\sim 200$ km s$^{-1}$ or $\sim 4$~\AA), and $B$ is the number of bins per resolution element. We used $\Delta\lambda = 0.115$~\AA, $\Delta I = ({\rm S/N})^{-1}$, $W_{\rm line} = 4$~\AA, and $B=5$; we calculated the signal-to-noise ratio (S/N) over a range of 2~\AA\ of continuum between the D$_2$  and D$_1$ lines using the IRAF splot tool. Table S3 lists the values of S/N and $W_\lambda(3\sigma)$ for the nondetection events.

Among the spectra that exhibit Na~I~D absorption features, those of SNe 2006E, 2007af, and 2007rw have the weakest ones (see Fig. S2, S3, and S6). Fig. S12 shows the Gaussian fits to the D$_2$ lines of these events using the parameters given in Table S4. It is apparent from Table S3 that in the events whose S/N is above 10, features like those exhibited by SNe 2006E or 2007af should have been apparent; hence, we conclude that any Na~I~D absorption must be weaker than any of the detected events. The line in the SN 2007rw spectrum has an equivalent width that is less than $W_\lambda(3\sigma)$ (=54.9 m\AA).

\vskip 1.0 cm
{\bf S2. The Zero-Velocity Component}
\vskip 0.5 cm
Ideally, we would like to know the progenitor system's velocity in order to use it to define our zero-velocity frame. However, this velocity is not easily determined. In some of the cases (SNe 2006cm, 2006X, 2007af, 2007le, 2007sr, 2008C, 2009ds, and SNF 20080612-003), a distinct H$\alpha$ emission line (from superposed H~II regions) can
be seen in the spectra, enabling us to determine the local frame of reference (LFR) of the galaxy along the line of sight. These features  cannot be attributed to the SNe~Ia explosions themselves as they are narrow emission lines. In most of the cases the progenitor system should have a velocity close to that of the LFR. When present, the H$\alpha$ emission, and thus most probably the LFR, coincides (within a few km s$^{-1}$ to a few 10 km s$^{-1}$, in all but one case) with the velocity of the most prominent Na~I~D absorption feature, and blueshifted components extend out to 150 km s$^{-1}$. 

In cases when H$\alpha$ is not detected, we calculate the expected location of the Na~I~D lines using the cataloged galaxy redshifts from NED ({\it 26}), and find these to coincide with the most prominent Na~I~D absorption to within a few tens of km s$^{-1}$; weaker features extend up to 200 km s$^{-1}$ ({\it 46}). Though we cannot determine the absolute velocity of the blueshifted absorption with respect to the progenitor system in each individual case, we can say with high confidence that these velocities span a few tens of km s$^{-1}$ to 200 km s$^{-1}$. Note that the strong preference we see in SNe~Ia for dominant strong features with bluer weaker features does not depend on the location of the velocity zero point. This preference is naturally explained by progenitor outflows and inconsistent with random ISM clouds.

\vskip 1.0 cm
{\bf S3. Velocity of Recurrent Novae Shells}
\vskip 0.5 cm
As a first-order approximation, Patat et al. ({\it 40}) show that a shell formed by nova ejecta sweeping up the stellar wind of the companion star expands at a velocity of roughly
\begin{equation}
u=\frac{2\mu v_0+u_w}{1+\mu},
\end{equation}
where $v_0$ is the scale velocity of the nova ejecta ($\rho\propto\exp(-v/v_0)$), $u_w$ is the velocity of the companion's wind, and $\mu=M_e/M_w$ is the mass ratio between the nova ejecta mass and the swept-up mass. They show that for typical values of these parameters the velocity of such a shell will be $\sim 50$ km s$^{-1}$ relative to the systemic velocity of the recurrent nova system.

When the zero velocity is taken to be the minimum of the deepest absorption feature, as is done in our analysis, the bluest feature of RS Oph has a velocity of $\sim 70$ km s$^{-1}$ and the features show resemblance to features in our SN~Ia sample [see fig. 10 of ({\it 40})]. 

Nevertheless, some of our SN~Ia events exhibit features with higher velocities. The consistency of these features with this first-order approximation needs to be checked. Using the same parameters used for the slower feature ($M_e=10^{-7}\,M_\odot$, $v_0=9.1\times10^7$ cm s$^{-1}$, $u_w=20$ km s$^{-1}$), and assuming $u=150$ km s$^{-1}$ we get $\mu=0.0778$, or $M_w\approx1.3\times10^{-6}\,M_\odot$. This value implies a red giant mass loss of $\dot{M}_{RG}\approx1.3\times10^{-6}/\Delta t\approx6.5\times10^{-8}\,M_\odot$ yr$^{-1}$ (the latter assuming $\Delta t=20$ yr) which is a reasonable value ({\it 47}).

\begin{table}[h]
\begin{center}
Table S1: SN~Ia sample host-galaxy morphology classification
\begin{tabular}{|c|c|c|c|}
\hline
Absorption classification & Object name & Host galaxy & Type \\
\hline
\hline
\multirow{12}{*}{Blueshifted} & SN 2006X & NGC 4321 & Sbc \\
\cline{2-4}
& SN 2007fb & UGC 12859 & Sbc \\
\cline{2-4}
& SN 2007kk & UGC 2828 & Sbc \\
\cline{2-4}
& SN 2007le & NGC 7721 & Sc \\ 
\cline{2-4}
& SN 2008C & UGC 3611 & S0/a \\
\cline{2-4}
& SN 2008dt & NGC 6261 & S0/a \\
\cline{2-4}
& SN 2008ec & NGC 7469 & Sa \\
\cline{2-4}
& SN 2008fp & ESO 428-G14 & S0 \\
\cline{2-4}
& SNF20080612-003 & 2MASX J16152860+1913344  & Sc$^\ast$ \\
\cline{2-4}
& SN 2009ds & NGC 3905 & Sc \\
\cline{2-4}
& SN 2009ig & NGC 1015 & Sa \\
\cline{2-4}
& SN 2009iw & IC 2160 & Sbc \\
\hline
\hline
\multirow{5}{*}{Redshifted} & SN 2007af & NGC 5584 & Scd \\
\cline{2-4}
& SN 2007sa & NGC 3499 & S0/a \\ 
\cline{2-4}
& SN 2009le & ESO 478-G006 & Sbc \\
\cline{2-4}
& SN 2009mz & NGC 5426 & Sc pec \\
\cline{2-4}
& SN 2010A & UGC 2019 & Sab \\ 
\hline
\hline
\multirow{5}{*}{Single/symmetric} & SN 2006cm & UGC 11723 & Sb edge-on \\
\cline{2-4}
& SN 2006E & NGC 5338 & S0 \\
\cline{2-4}
& SN 2007fs & ESO 601-G5 & Sb \\
\cline{2-4}
& SN 2007sr & NGC 4038 & Sm pec \\
\cline{2-4}
& SN 2010ev & NGC 3244 & Scd \\
\hline
\hline
\multirow{12}{*}{None} & SN 2006ct & 2MASX J12095669+4705461 & E$^\ast$ \\
\cline{2-4}
& SN 2006eu & MCG +08-36-16 & E \\
\cline{2-4}
& SN 2007gj & ESO 298-28 & Sbc \\
\cline{2-4}
& SN 2007hj & NGC 7461 & S0 \\
\cline{2-4}
& SN 2007on & NGC 1404 & E1 \\
\cline{2-4}
& SNF20080514-002 & UGC 8472 & S0 \\
\cline{2-4}
& SN 2008dh & 2MASX J00351050+2315184 & Sb$^\ast$ \\
\cline{2-4}
& SN 2008ee & NGC 307 & S0 \\
\cline{2-4}
& SN 2008ge & NGC 1527 & S0 \\
\cline{2-4}

& SN 2008hv & NGC 2765 & S0 \\
\cline{2-4}
& SN 2008ia & ESO 125-G006 & S0 \\
\cline{2-4}
& SN 2009ev & NGC 5026 & S0/a \\
\cline{2-4}
& SN 2009nr & UGC 8255 & Scd \\
\hline
\end{tabular}
\end{center}
$^\ast$Morphology classified using images from DSS (Digital Sky Survey) ({\it 28}).
\end{table}

\begin{table}[h]
\begin{center}
Table S2: CC~SN sample host-galaxy morphology classification
\begin{tabular}{|c|c|c|c|}
\hline
Absorption classification & Object name & Host galaxy & Type \\
\hline
\hline
\multirow{4}{*}{Blueshifted} & SN 2006bp & NGC 3953 & Sbc \\ 
\cline{2-4}
& SN 2006cu & UGC 9530 & S \\ 
\cline{2-4}
& SN 2008ax & NGC 4490 & Sd pec \\
\cline{2-4}
& SN 2008cg & FGC 1965 & Scd \\
\hline
\hline
\multirow{3}{*}{Redshifted} & SN 2006be & IC 4582 & S \\
\cline{2-4}
& SN 2008D & NGC 2770 & Sc \\ 
\cline{2-4}
& SN 2008J & MCG -02-7-33 & Sbc \\
\hline
\hline
\multirow{2}{*}{Single/Symmetric} & SN 2003gd & NGC 628 & Sc \\
\cline{2-4}
& SN 2007rw & UGC 7798 & Im \\
\hline
\hline
\multirow{2}{*}{None} & SN 2006ca & UGC 11214 & Scd \\
\cline{2-4}
& SN 2007pk & NGC 579 & Scd \\
\hline
\end{tabular}
\end{center}
\end{table}

\begin{table}[h]
\begin{center}
Table S3: S/N and $W_\lambda(3\sigma)$ for spectra with no apparent Na~I~D absorption features
\begin{tabular}{|c|c|c|c|}
\hline
 Object name & Type & S/N$^\ast$ & $W_\lambda(3\sigma)$ (m\AA)$^{\ast\ast}$ \\
\hline
\hline
SN 2006ct &  & 6.47 & 144 \\ 
\cline{1-1} \cline{3-4}
SN 2006eu &  & 11.11 & 84 \\
\cline{1-1} \cline{3-4}
SN 2007gj &  & 12.52 & 74 \\
\cline{1-1} \cline{3-4}
SN 2007hj &  & 7.22 & 129 \\
\cline{1-1} \cline{3-4}
SN 2007on &  & 39.95 & 23 \\
\cline{1-1} \cline{3-4}
SNF20080514-002 &  & 26.6 & 35 \\
\cline{1-1} \cline{3-4}
SN 2008dh & Ia & 7.52 & 124 \\
\cline{1-1} \cline{3-4}
SN 2008ee &  & 23.41 & 40 \\
\cline{1-1} \cline{3-4}
SN 2008ge &  & 111.27 & 15.5 \\
\cline{1-1} \cline{3-4}
SN 2008hv &  & 85.94 & 20 \\
\cline{1-1} \cline{3-4}
SN 2008ia &  & 55.03 & 31.4  \\
\cline{1-1} \cline{3-4}
SN 2009ev &  & 76 & 16  \\
\cline{1-1} \cline{3-4}
SN 2009nr &  & 108.53 & 15.9 \\
\hline
\hline
SN 2006ca & II & 22.30 & 42 \\
\hline
SN 2007pk & IIn-pec & 6.47 & 144 \\
\hline
\end{tabular}
\end{center}
$^\ast$Calculated over a range of 2~\AA\ of continuum between the D$_2$ and D$_1$ lines using the IRAF splot tool. \\
$^{\ast\ast}$For each one of the Na~I~D lines.
\end{table}

\begin{table}[h]
\vspace{-2.5 cm}
\begin{center}
Table S4: Gaussian fit parameters for SN 2006E and SN 2007af D$_2$ features
\begin{tabular}{|c|c|c|c|c|}
\hline
Object name & $\lambda_0$ (\AA) & FWHM (m\AA) & $W$ (m\AA) & S/N \\
\hline
\hline
SN 2006E & 5906.7 & 192 & 92 & 12.21$^\ast$ \\
\hline
\multirow{2}{*}{SN 2007af} & 5921.7 & 286 & 127.5 & \\
                          & 5922.1 & 168 & 39.3 &
\multirow{-2}*{47.72$^{\ast\ast}$} \\
\hline
SN 2007rw & 5941.54 & 115 & 21 & 16.57$^{\ast\ast\ast}$ \\
\hline
\end{tabular} \\
\end{center}
$^\ast$Calculated over the range 5908--5910~\AA. \\ 
$^{\ast\ast}$Calculated over the range 5924--5926~\AA. \\
$^{\ast\ast\ast}$Calculated over the range 5943--5945~\AA.
\end{table}

\begin{figure}
    \begin{tabular}{c c c}
           \hskip -1 cm
           \includegraphics[width=6.0cm]{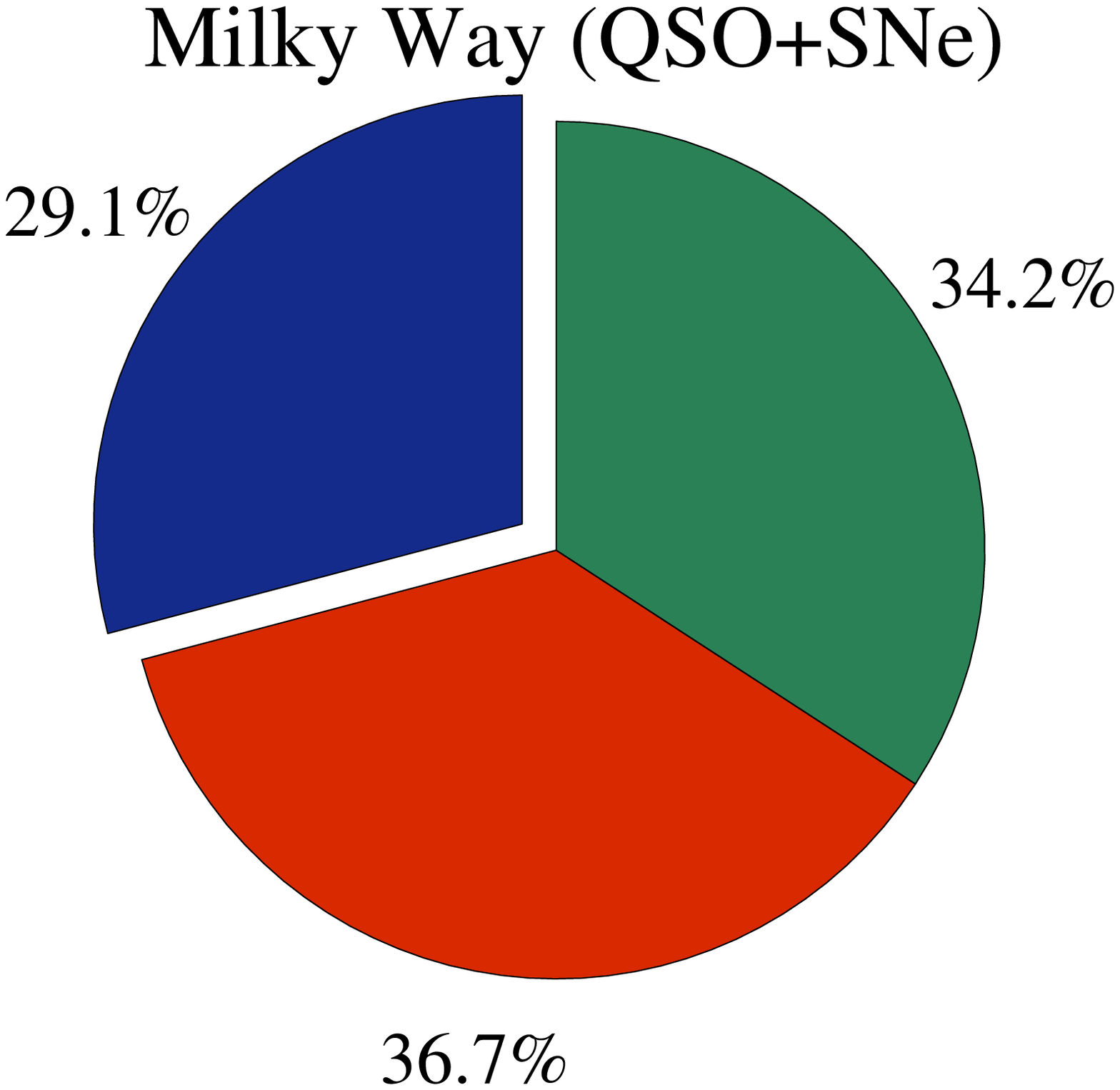}  &
           \hspace{-3cm}
           \includegraphics[width=5.0cm]{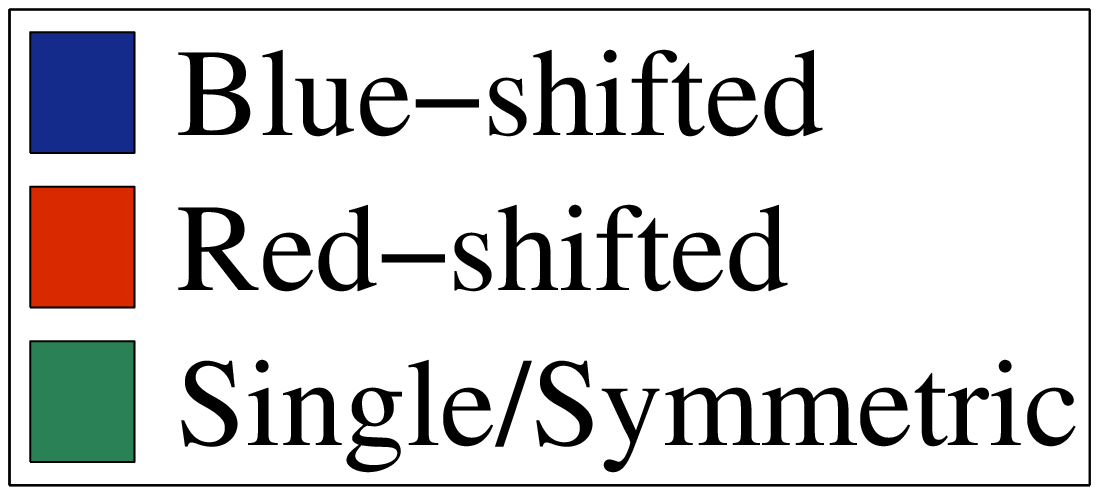}  &
           \hspace{-0.5cm}
           \includegraphics[width=6.0cm]{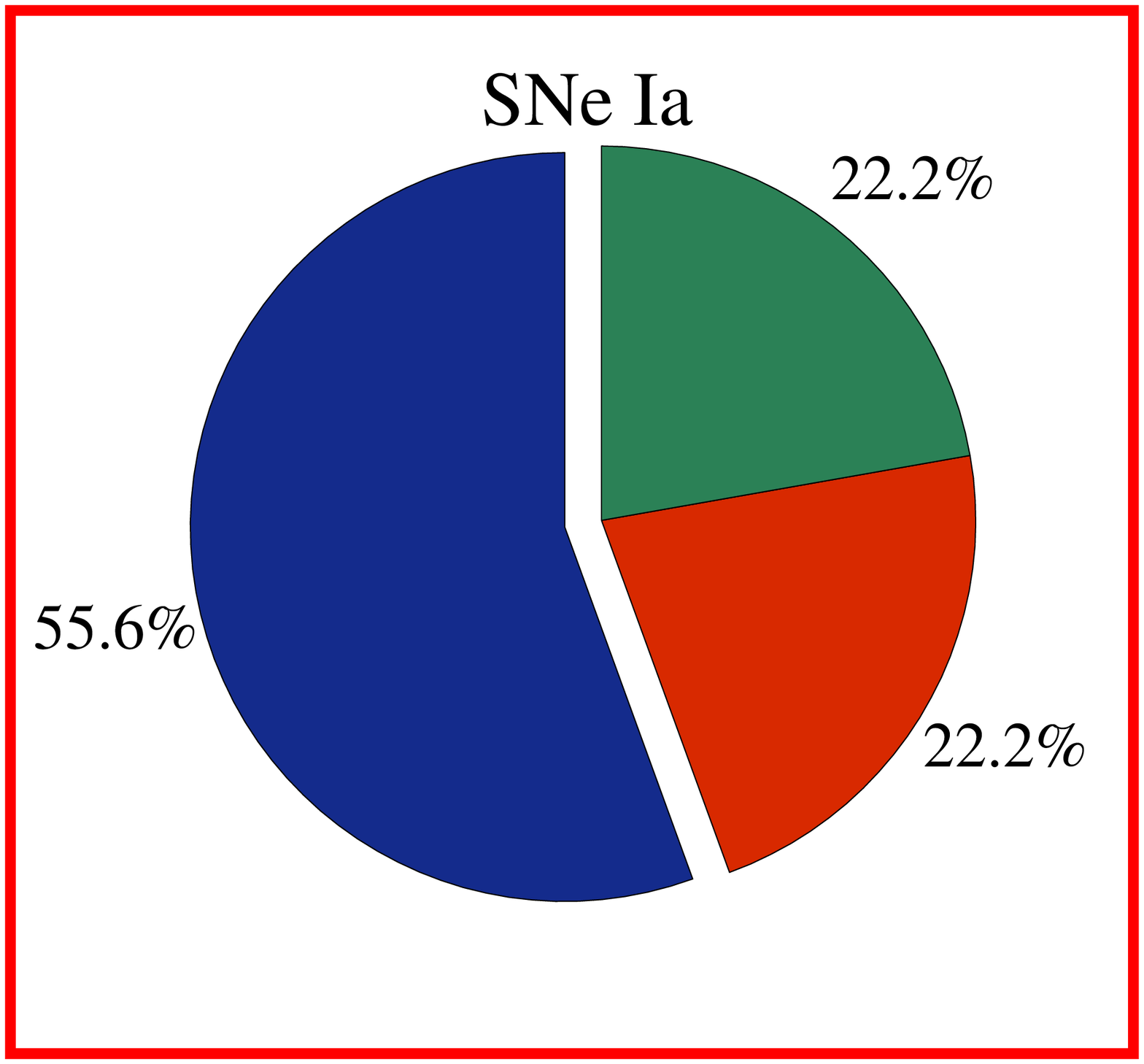}  \\
           \vspace{0cm}
          \includegraphics[width=4.0cm]{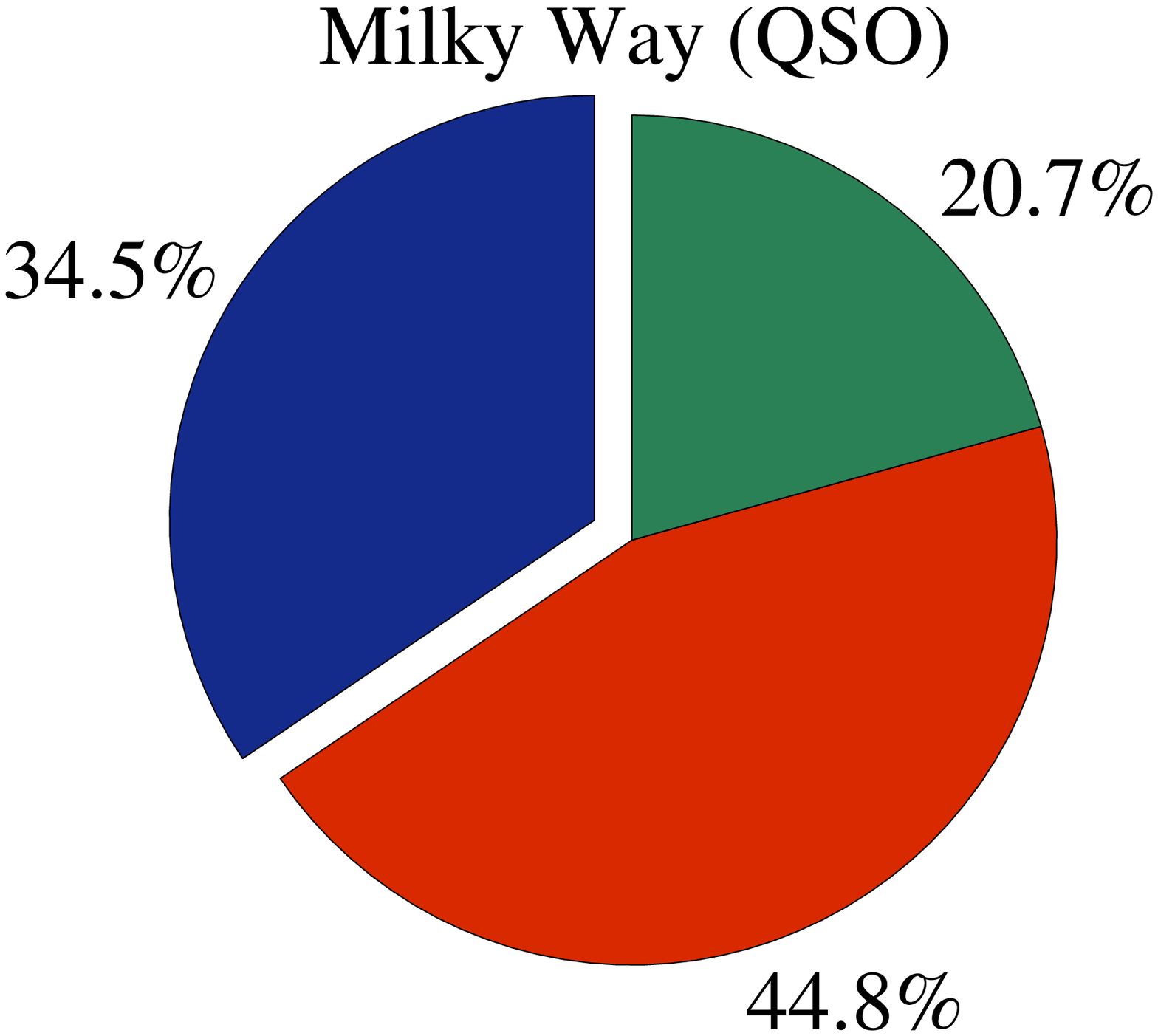} 
           \hspace{0cm}
           \includegraphics[width=4.0cm]{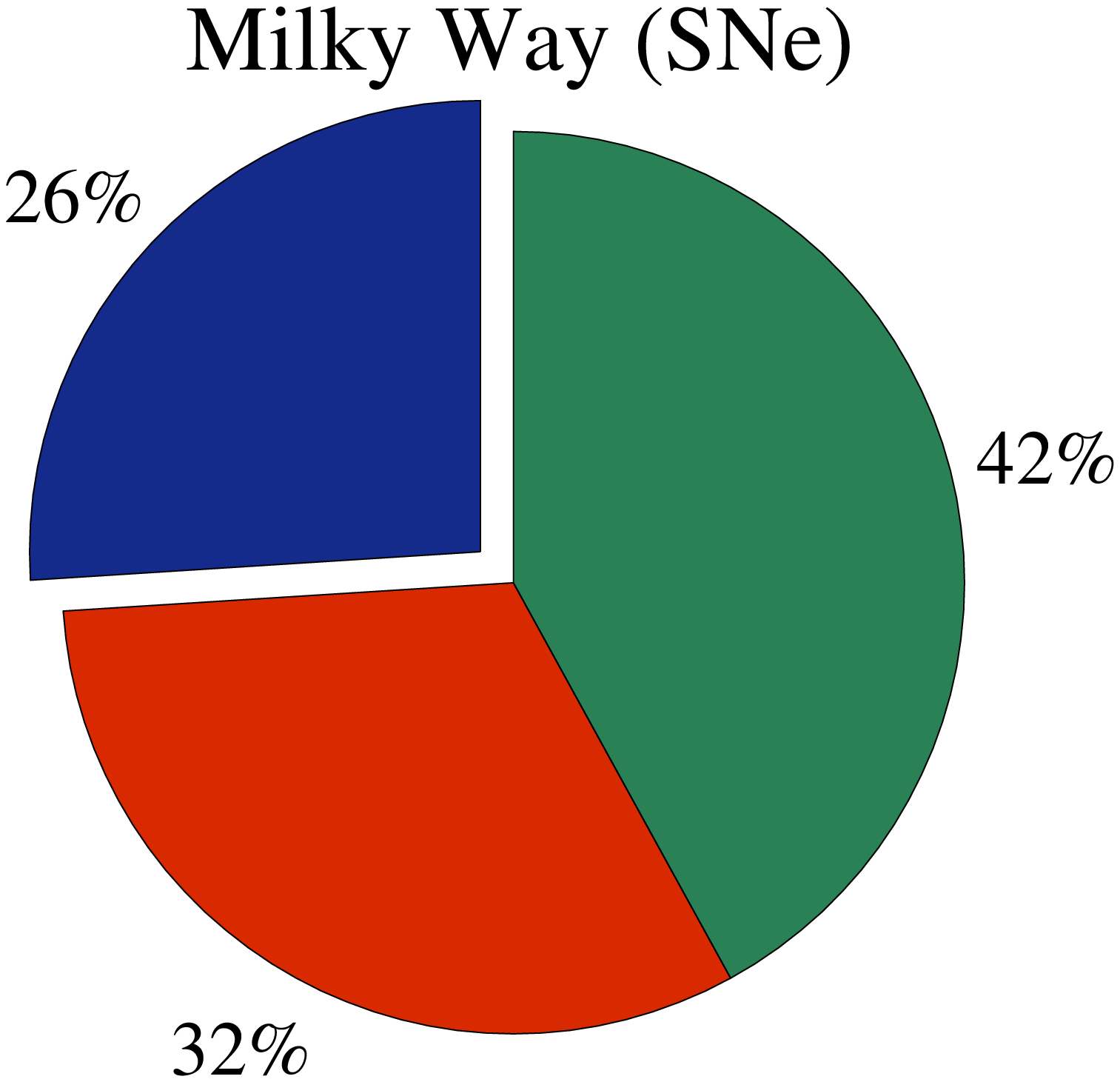}  &   &
%           \vspace{-1cm}
           \includegraphics[width=4.0cm]{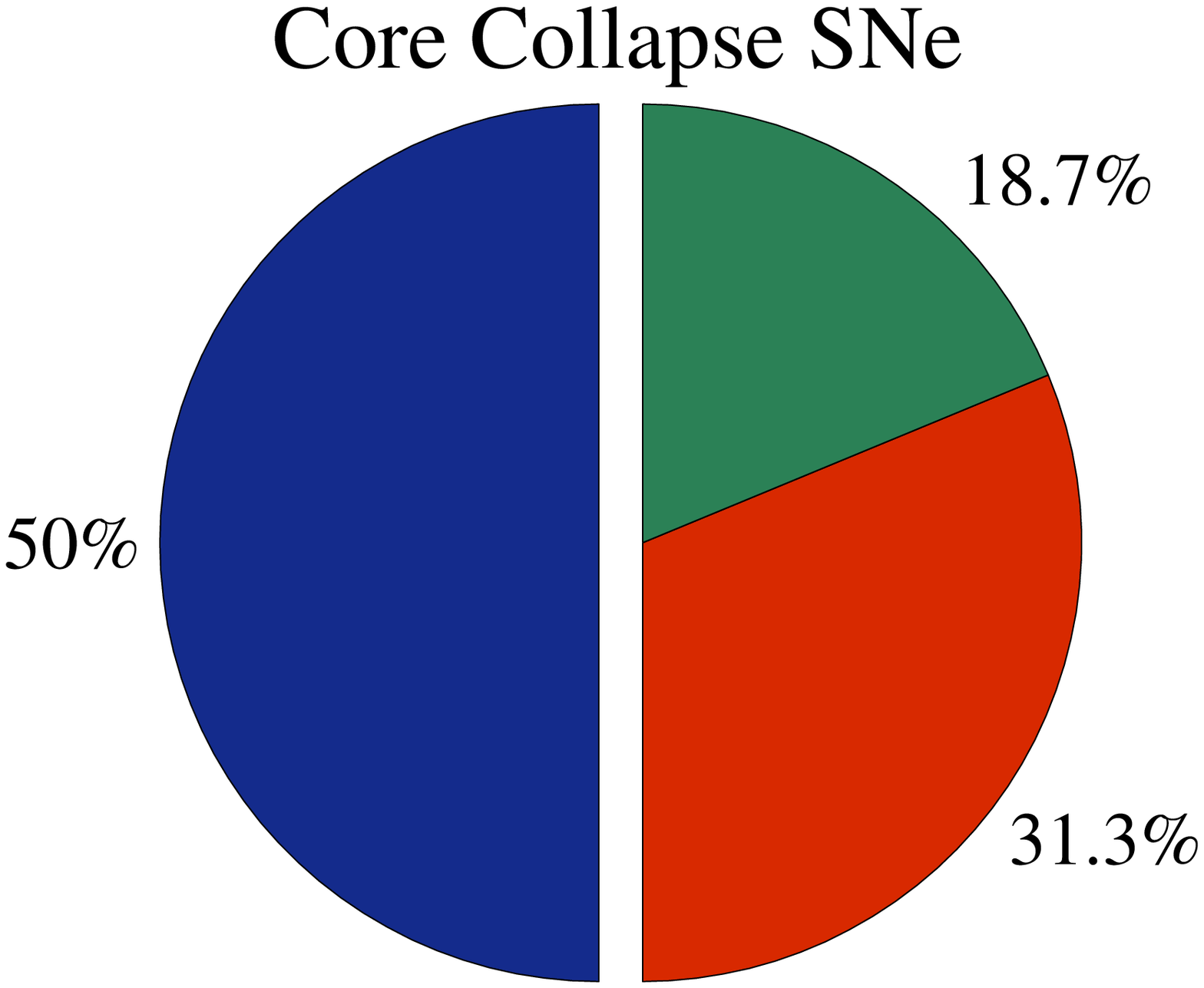}  \\
    \end{tabular}
Figure S1: Pie-chart visualization of the results given in Table 1 for
the extended samples.
\end{figure}

\begin{figure}
\vspace{-2.5 cm}
\begin{tabular}{c c}
\includegraphics[width=7.0cm]{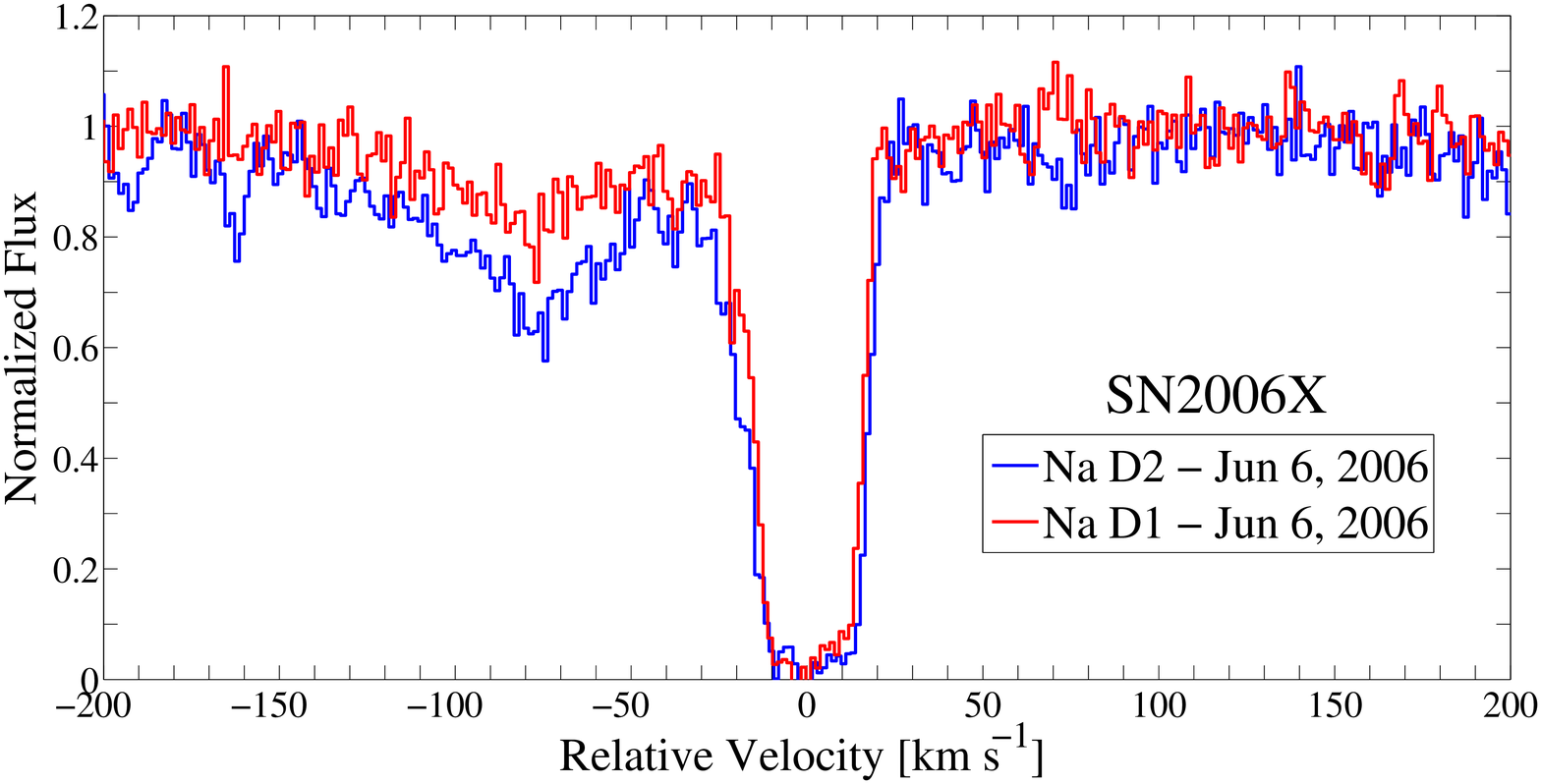} & 
\hspace{0.cm}
\includegraphics[width=7.0cm]{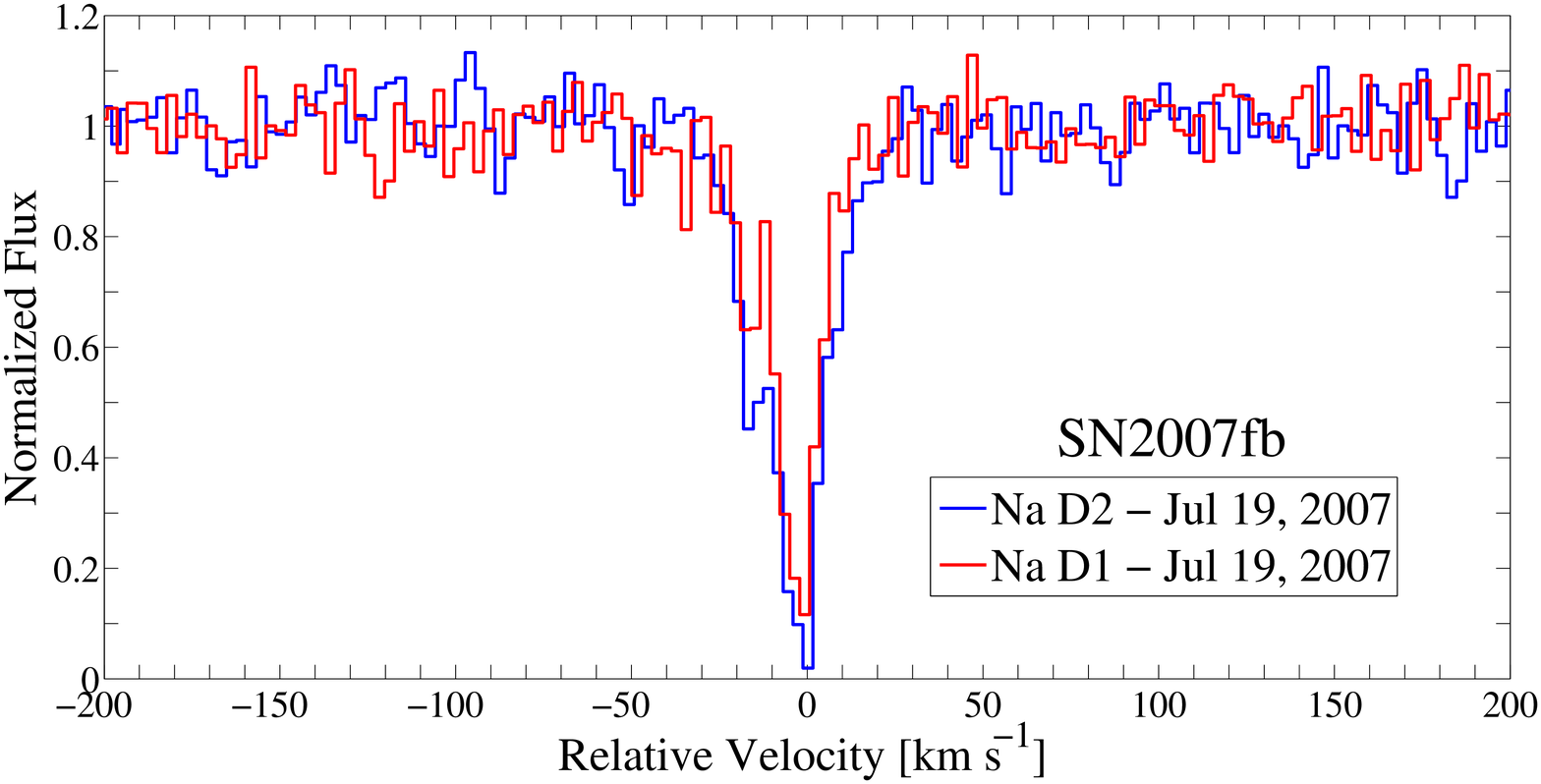}  \\
\includegraphics[width=7.0cm]{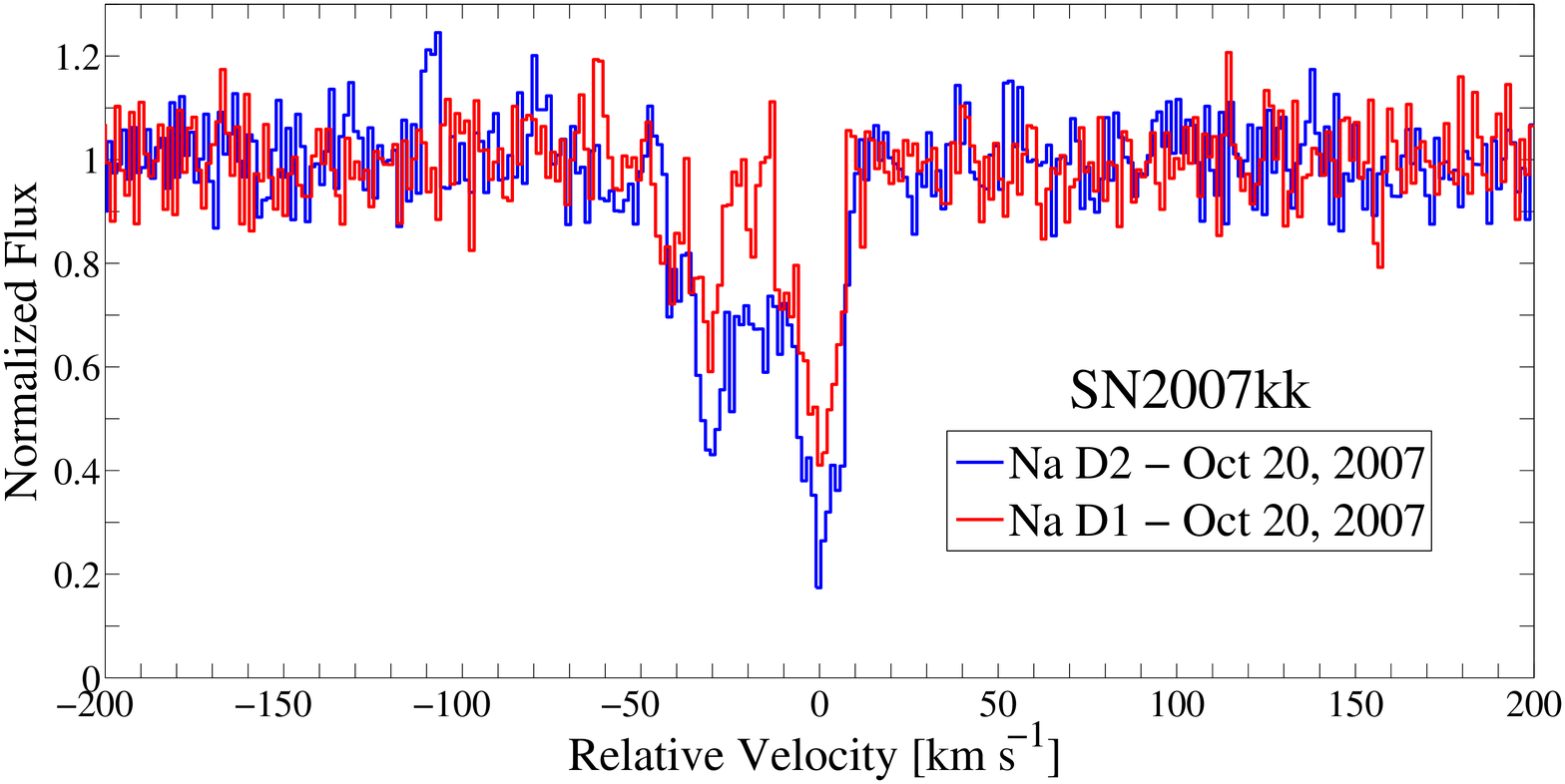} &
\hspace{0.cm}
\includegraphics[width=7.0cm]{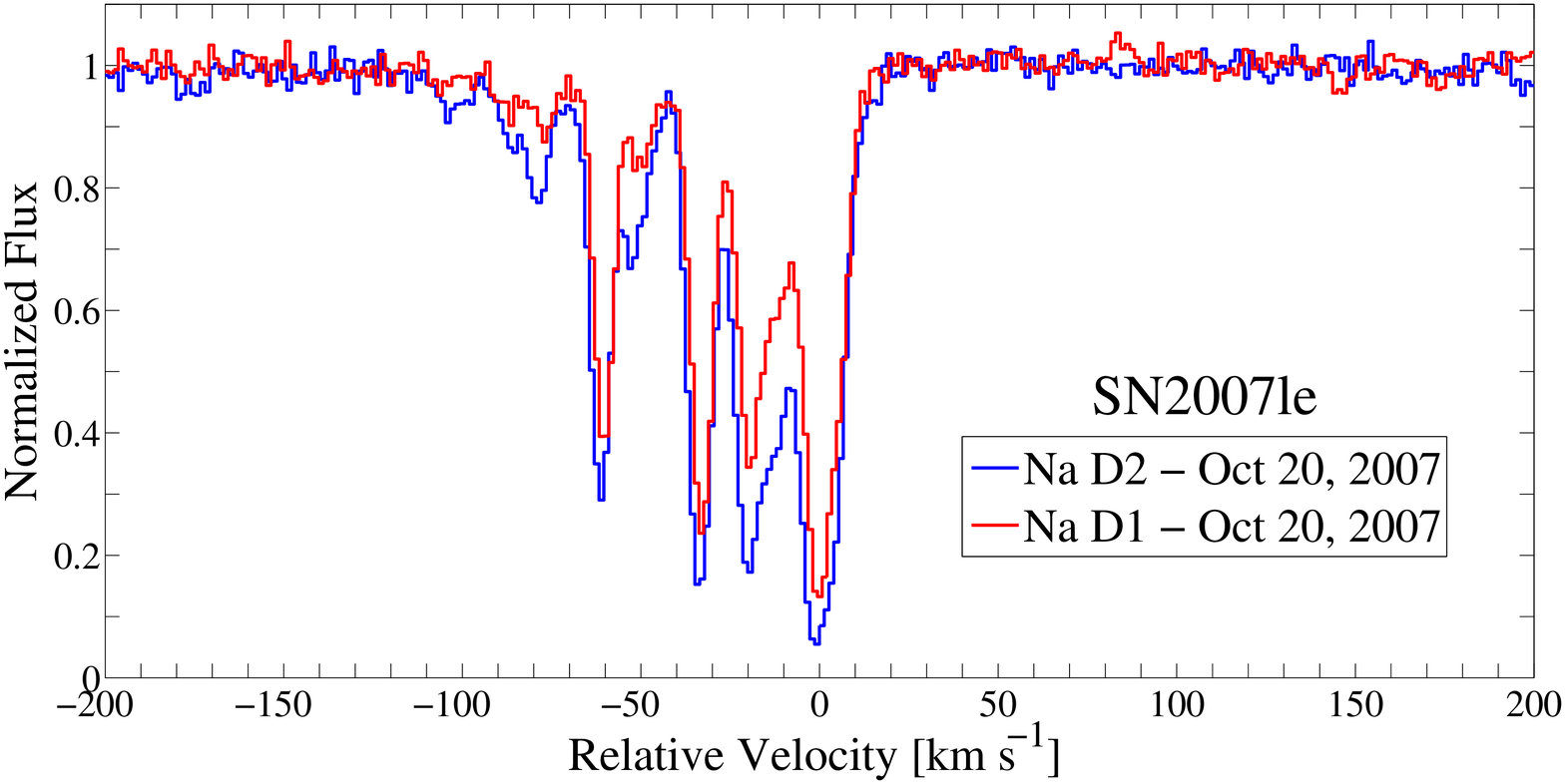}  \\
\includegraphics[width=7.0cm]{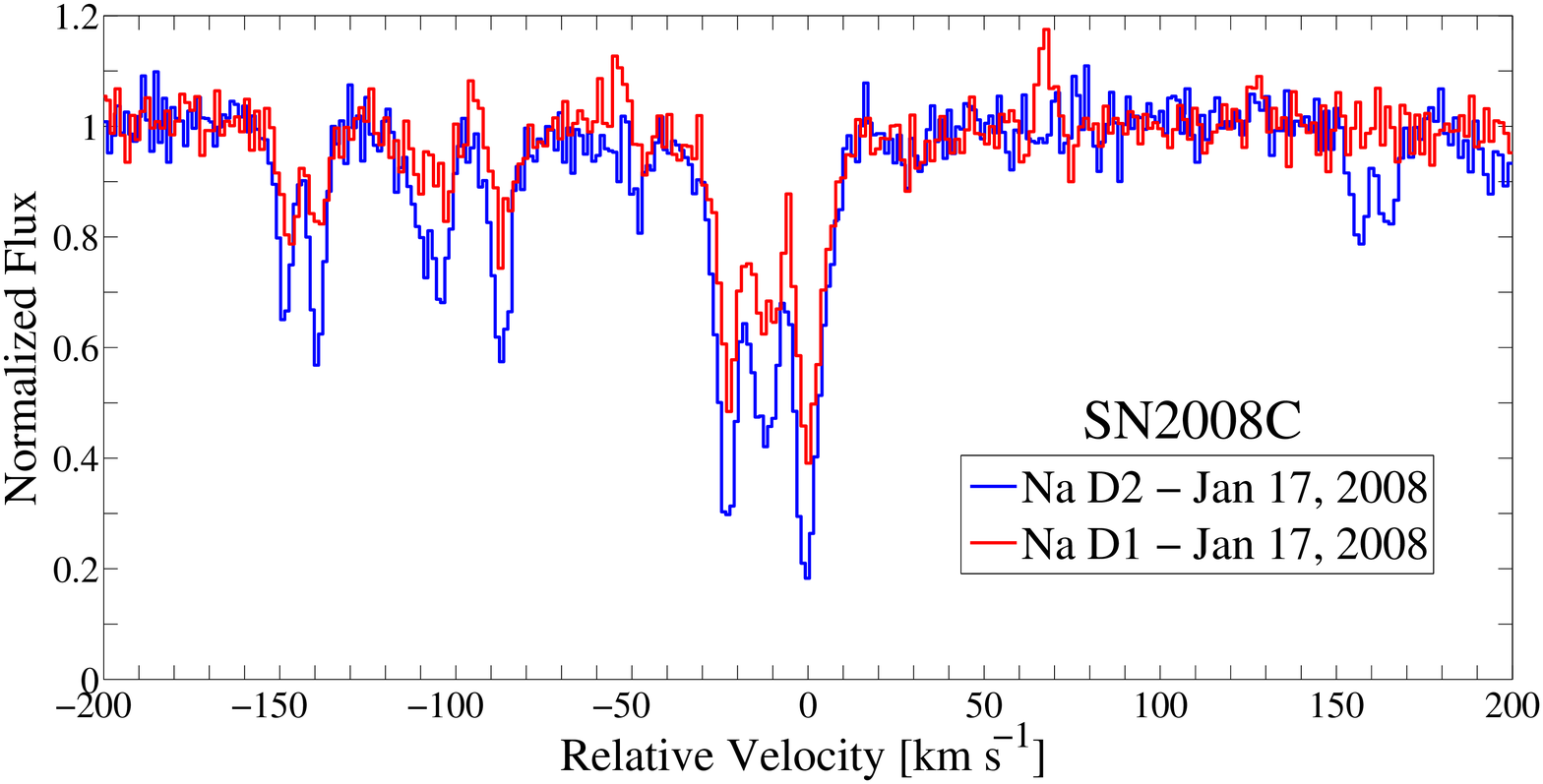} &
\hspace{0.cm}
\includegraphics[width=7.0cm]{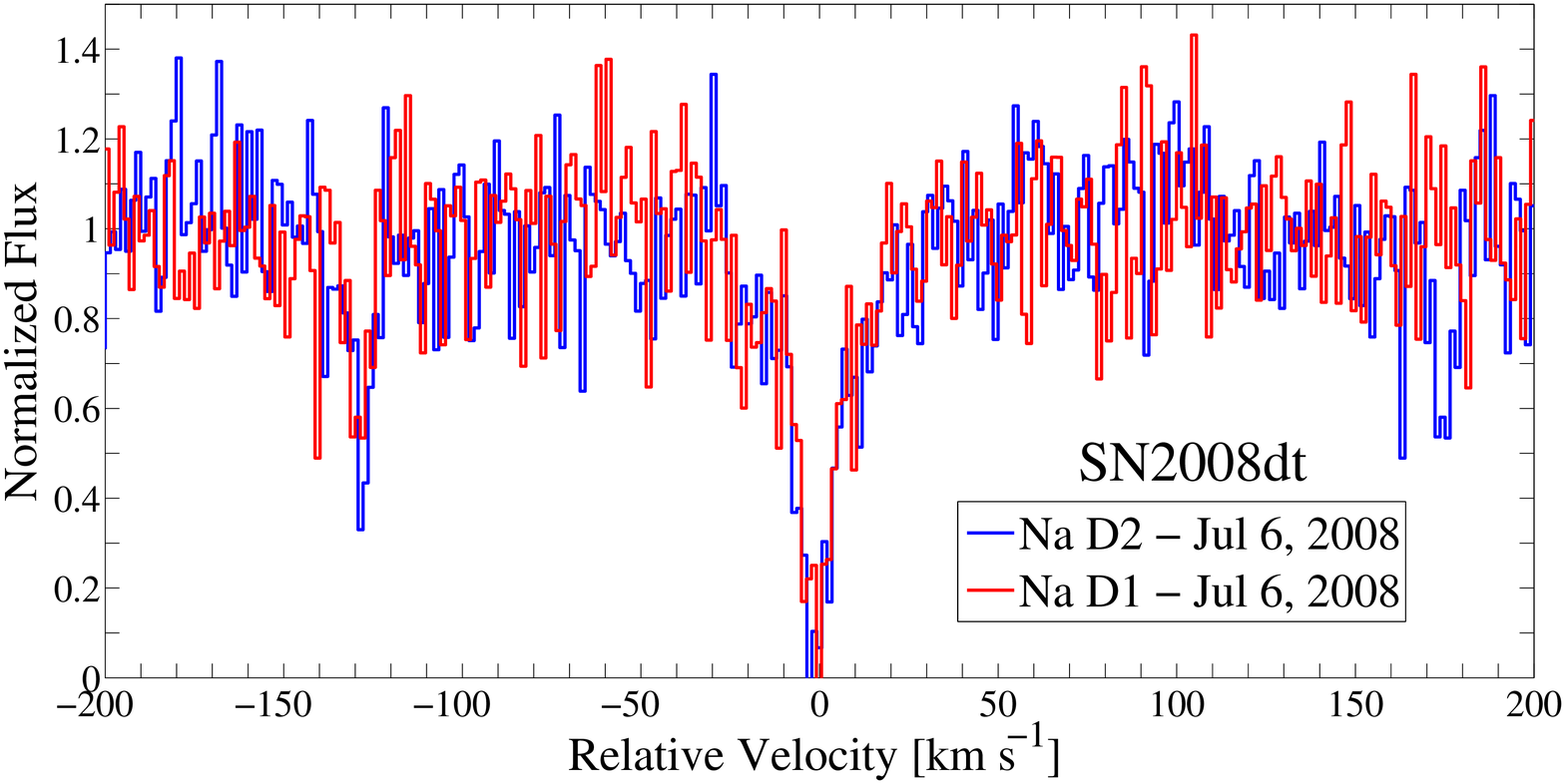}  \\
\includegraphics[width=7.0cm]{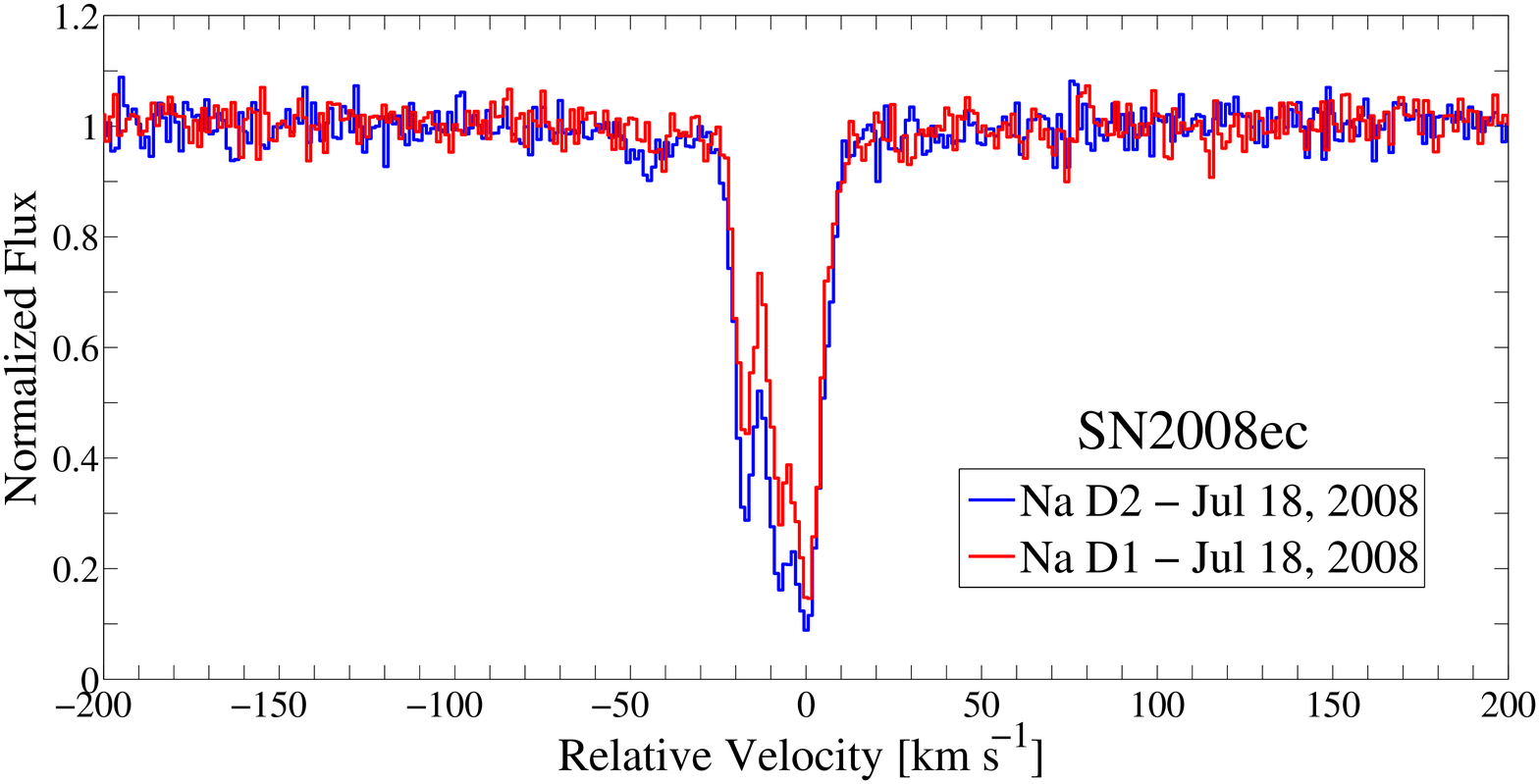} &
\hspace{0.cm}
\includegraphics[width=7.0cm]{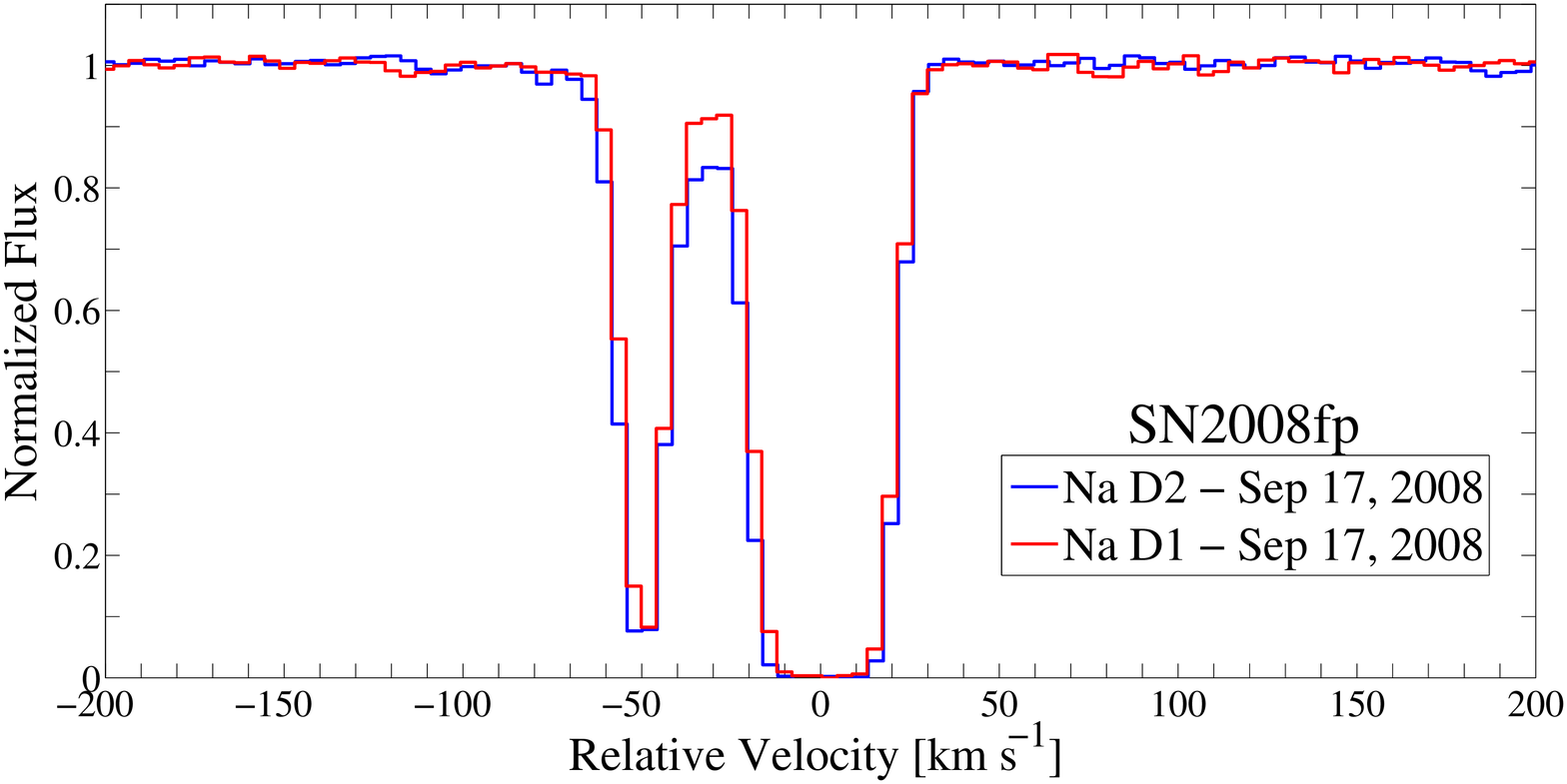}  \\
\includegraphics[width=7.0cm]{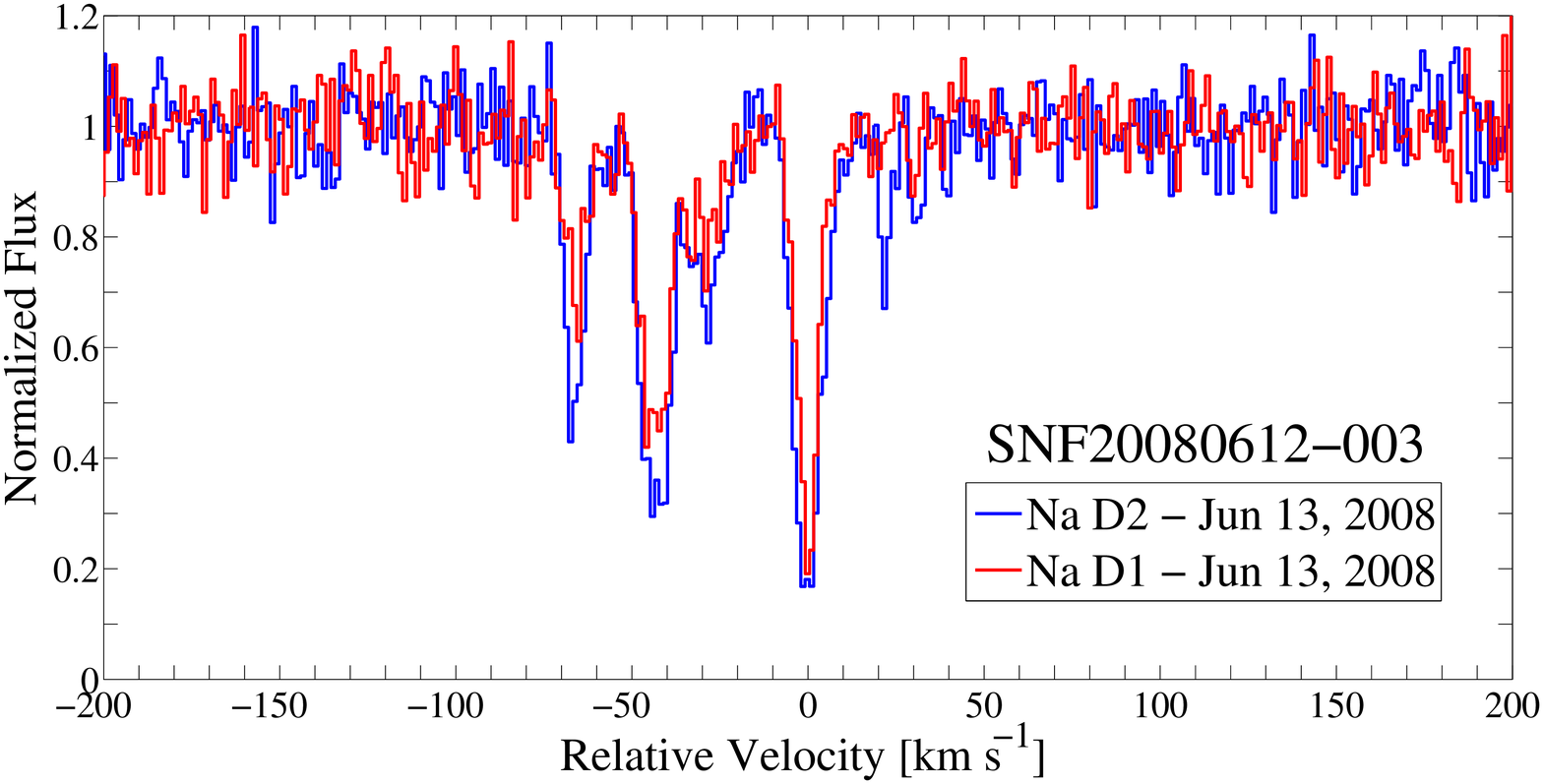} &
\hspace{0.cm}
\includegraphics[width=7.0cm]{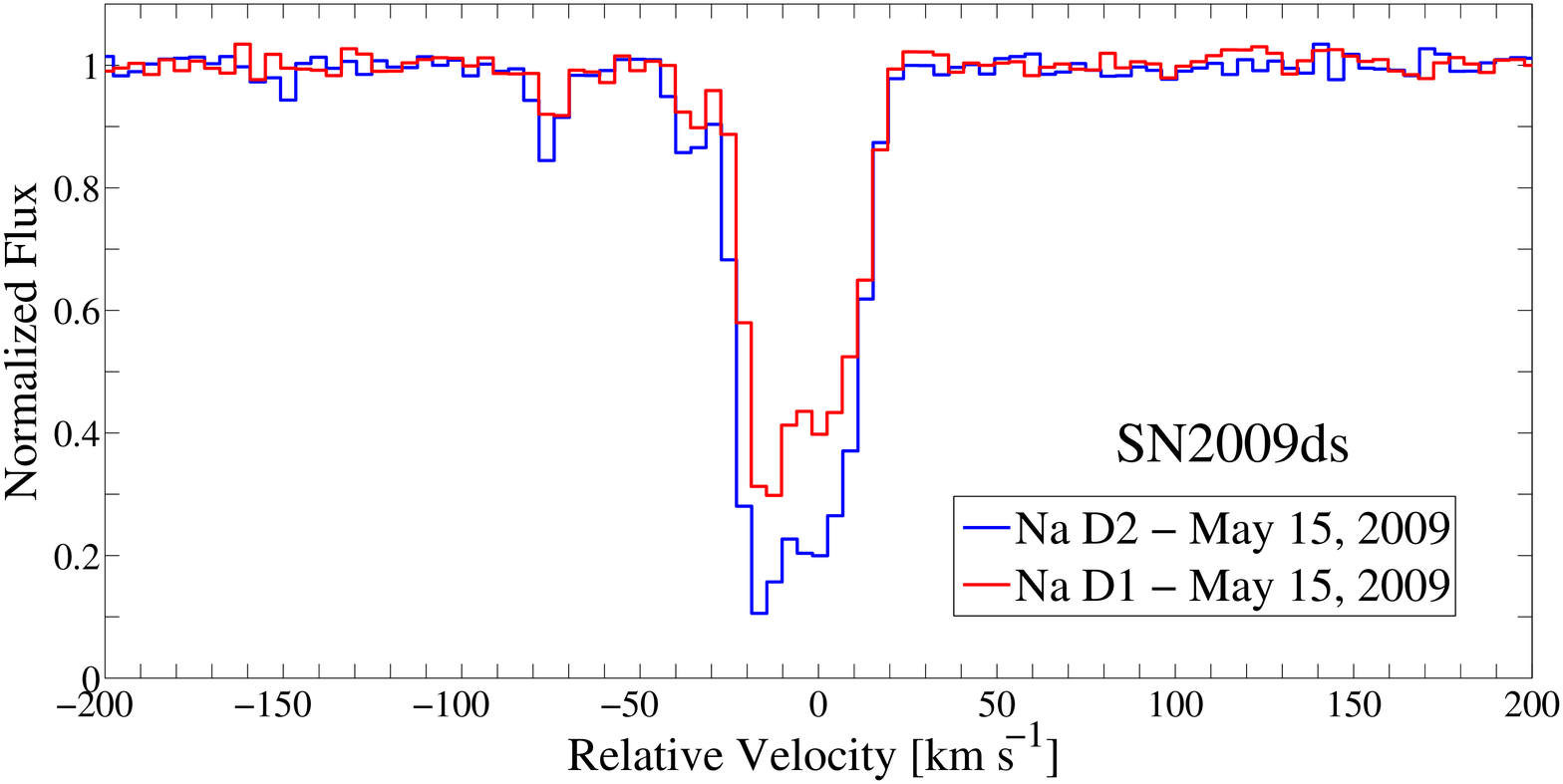}  \\
\includegraphics[width=7.0cm]{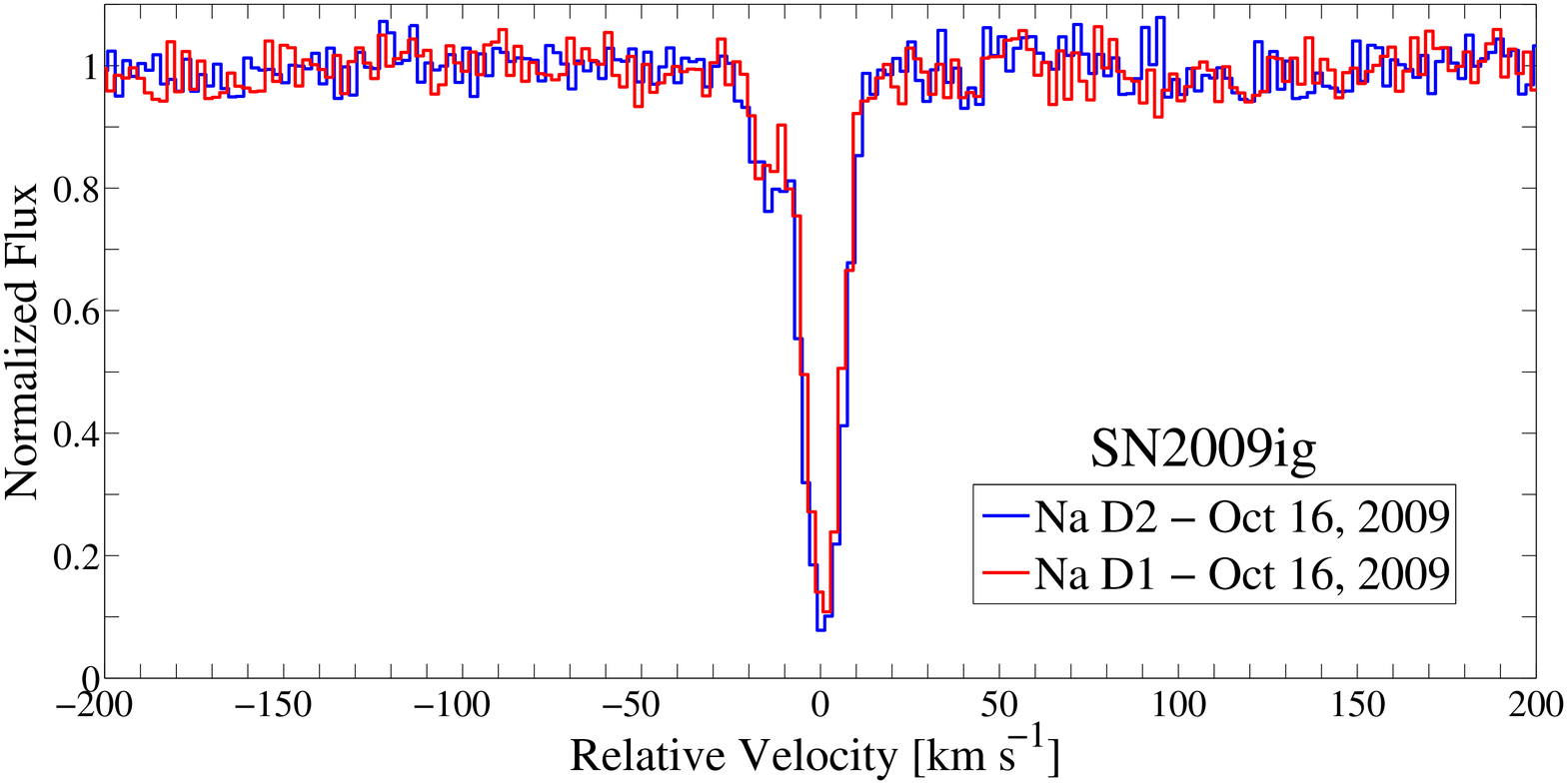} &
\hspace{0.cm}
\includegraphics[width=7.0cm]{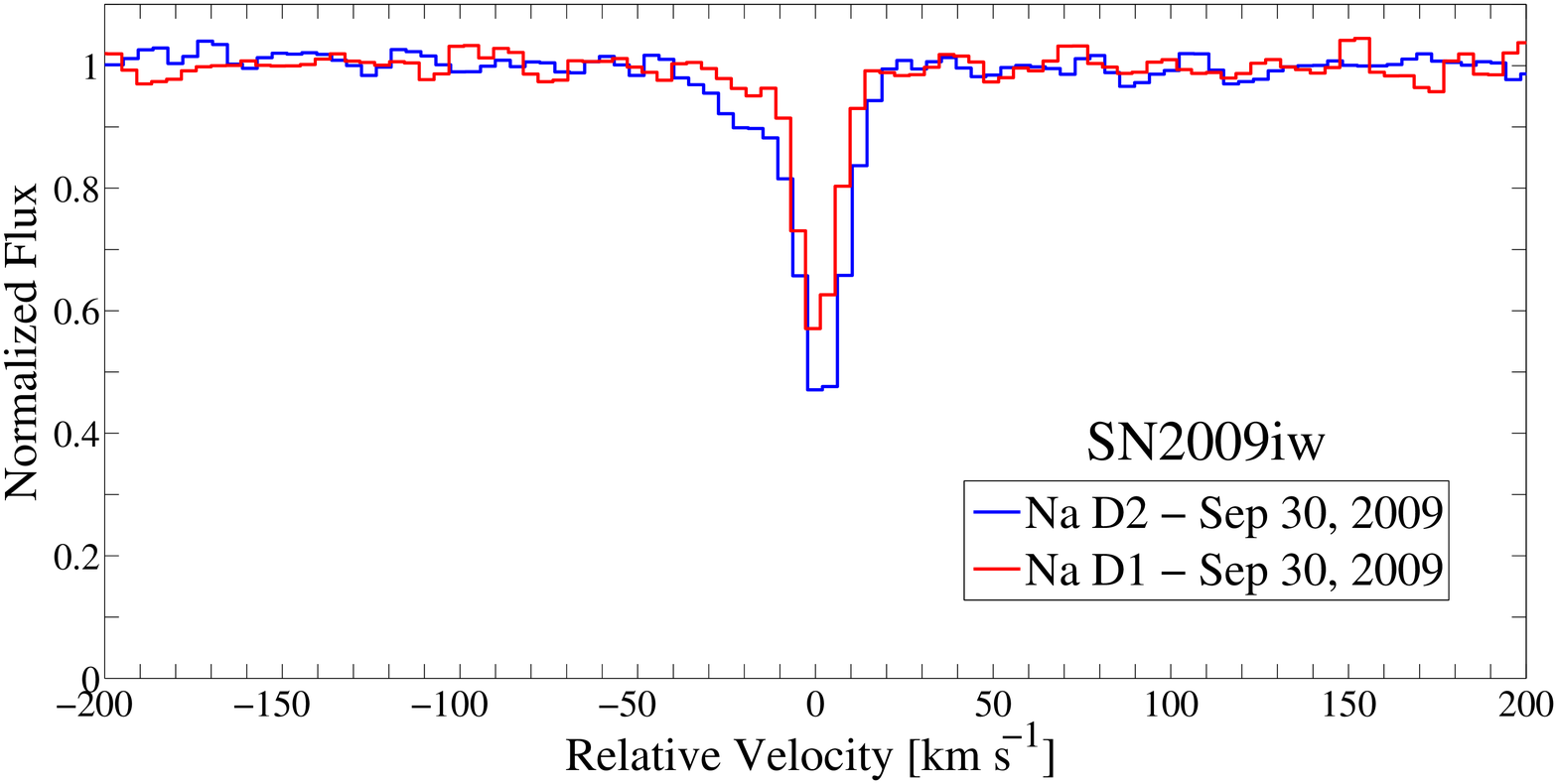}
\end{tabular} \\
Figure S2: SN~Ia spectra that exhibit blueshifted structures. The
blue/red color scheme is as described in the caption of Fig. 1. Two
additional spectra of SN 1991T and SN 1998es are given in Fig. S5 of
({\it 6}). Spectra of SN 2006X and SN 2007le from additional epochs can
be found in Patat et al. (2007) ({\it 6}; structure of blueshifted
components seen better) and Simon et al. (2009) ({\it 9}), respectively.
\end{figure}

\begin{figure}
\vspace{0. cm}
\hspace{0. cm}
\begin{tabular}{c c}
\includegraphics[width=7.0cm]{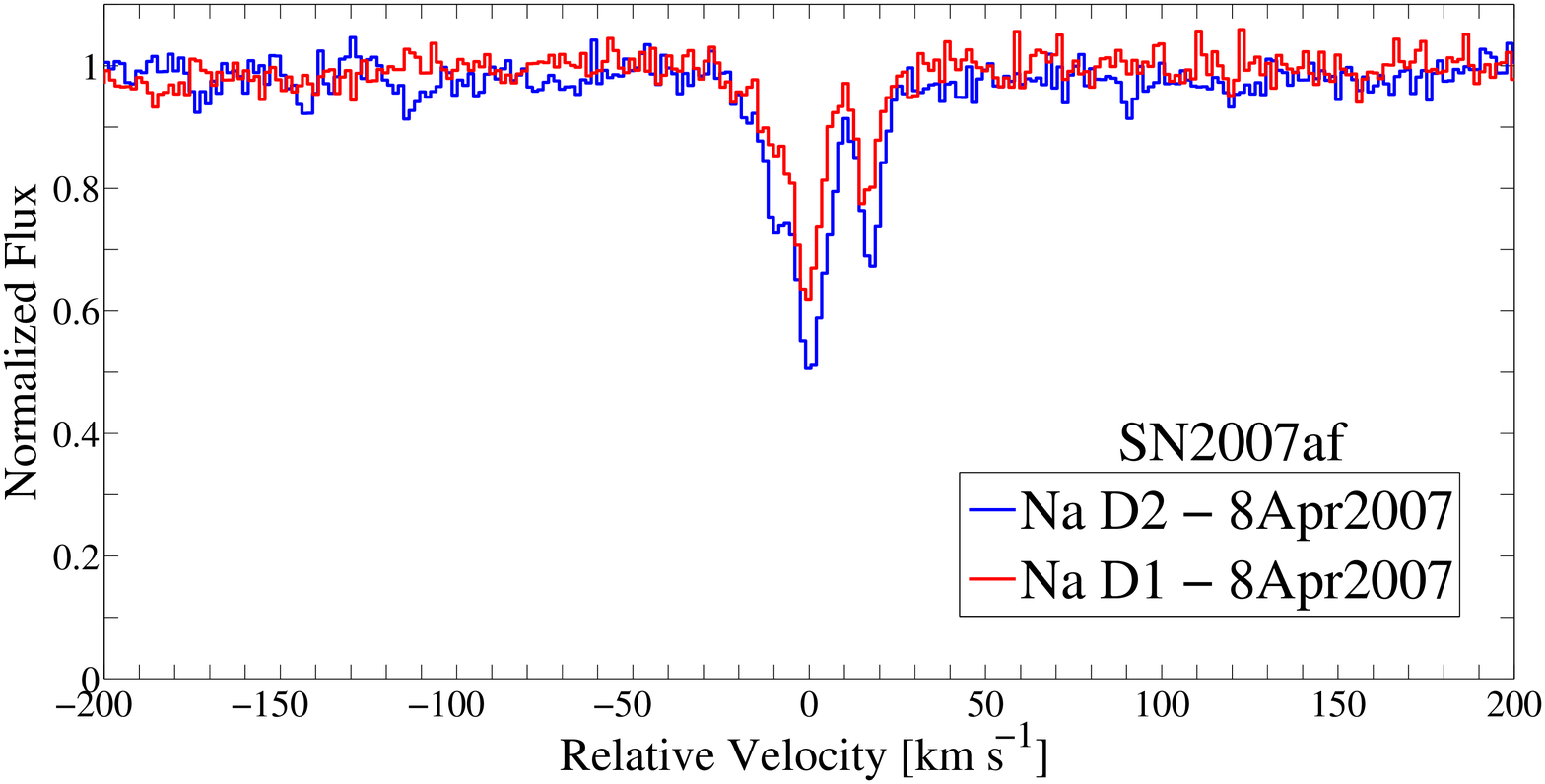} &
\includegraphics[width=7.0cm]{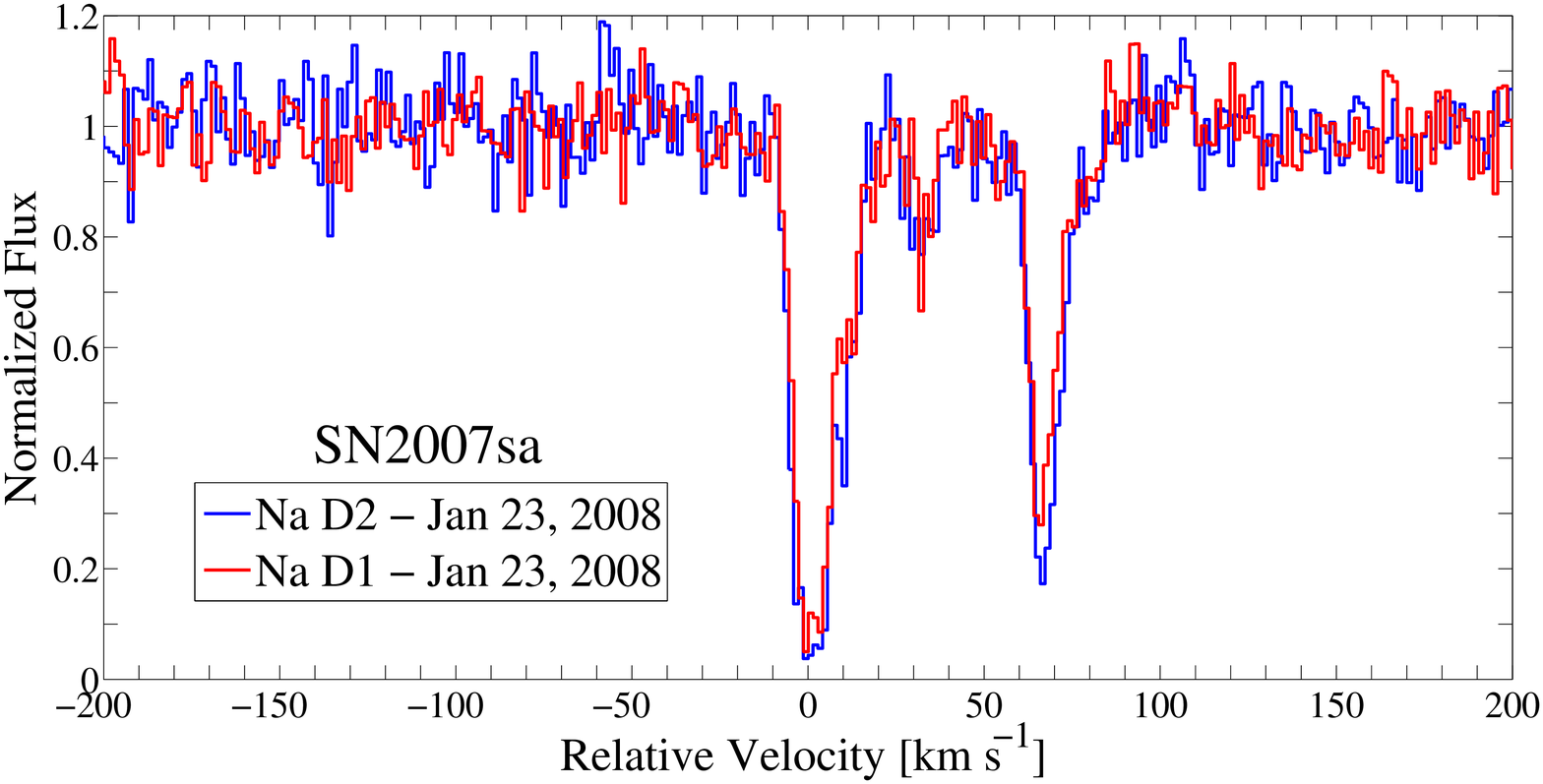} \\
\includegraphics[width=7.0cm]{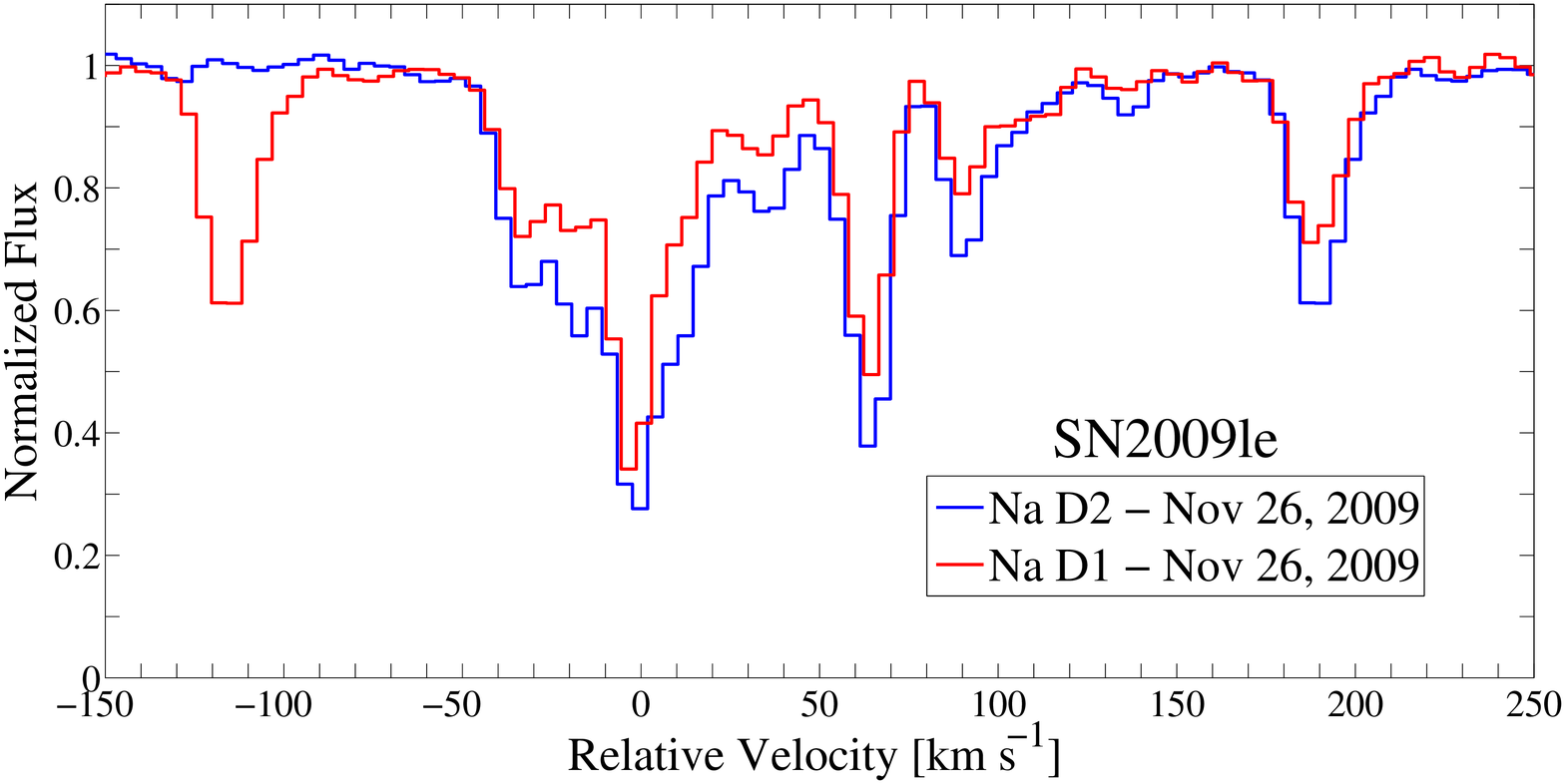} &
\includegraphics[width=7.0cm]{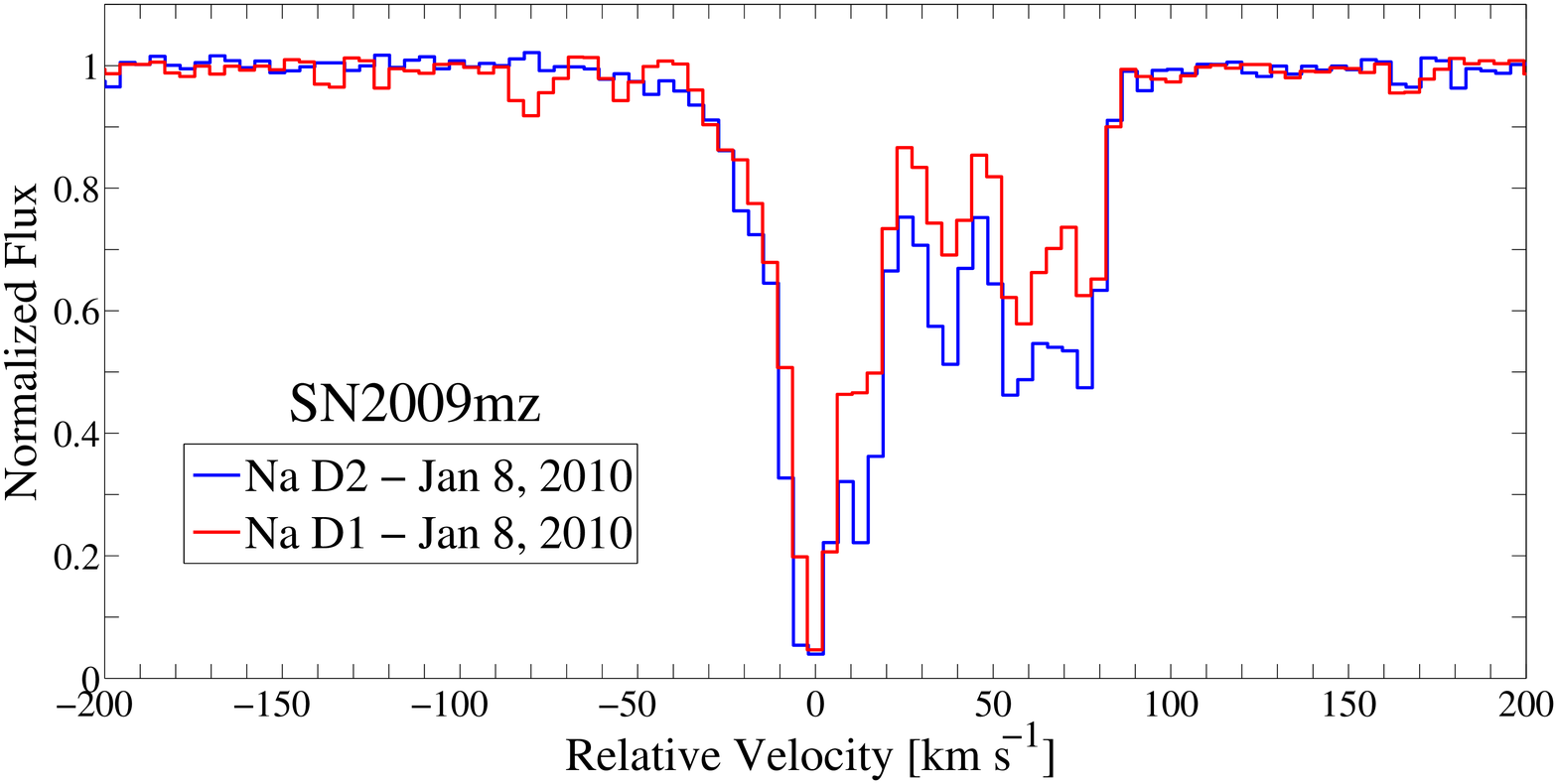} \\
\end{tabular} \\
\vspace{-0.4cm}
\begin{center}
\begin{tabular}{c}
\includegraphics[width=7.0cm]{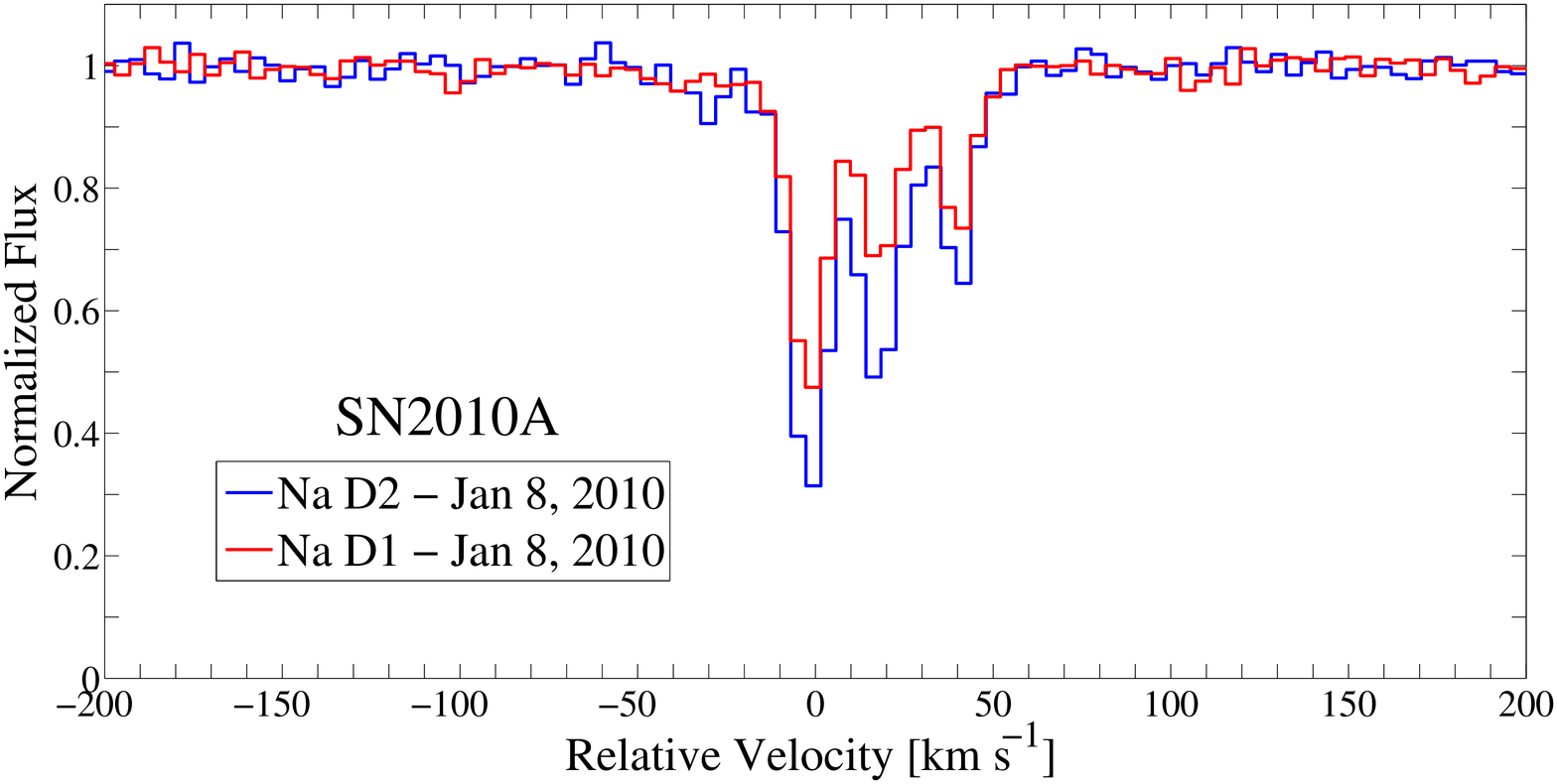} 
\end{tabular}
\end{center}
Figure S3: SN~Ia redshifted structure spectra. Color scheme identical
to that described in the caption of Fig. 1. Spectra of SN 2007af from
additional epochs can be found in Simon et al. (2007) ({\it 8}).
\end{figure}

\begin{figure}
\vspace{0.0 cm}
\hspace{0. cm}
\begin{tabular}{c c}
\includegraphics[width=7.0cm]{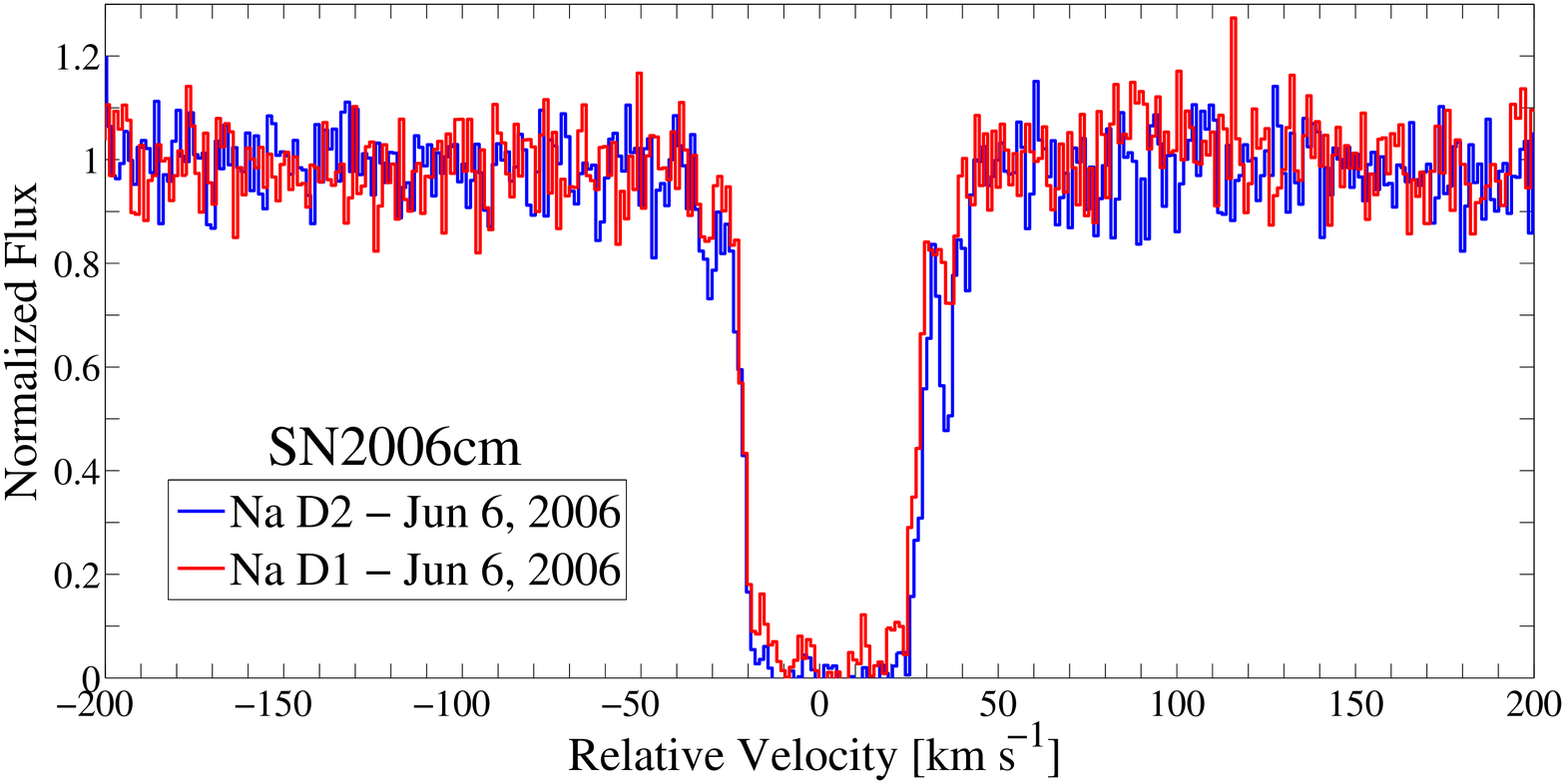} & 
\hspace{0.cm}
\includegraphics[width=7.0cm]{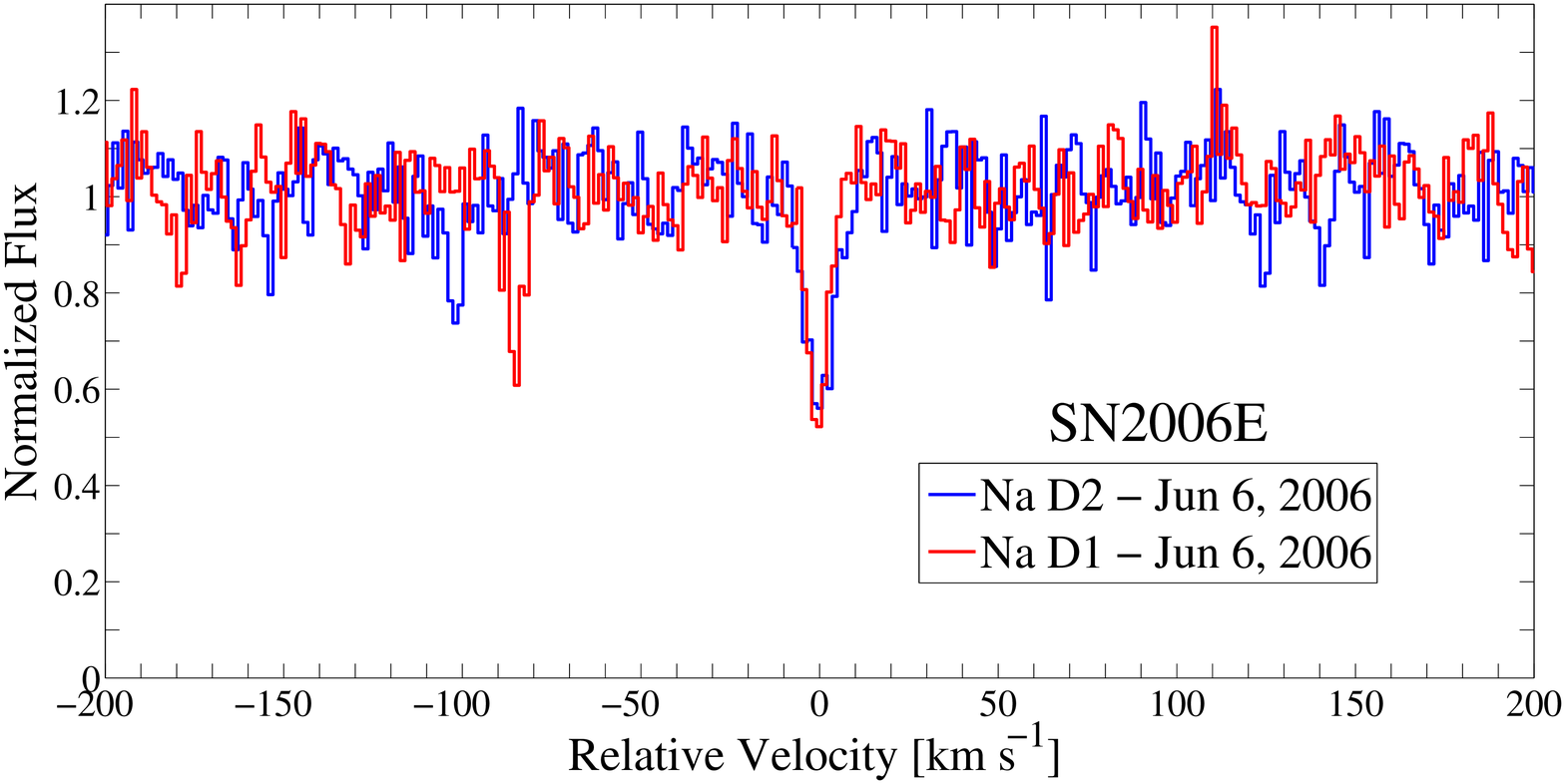}  \\
\includegraphics[width=7.0cm]{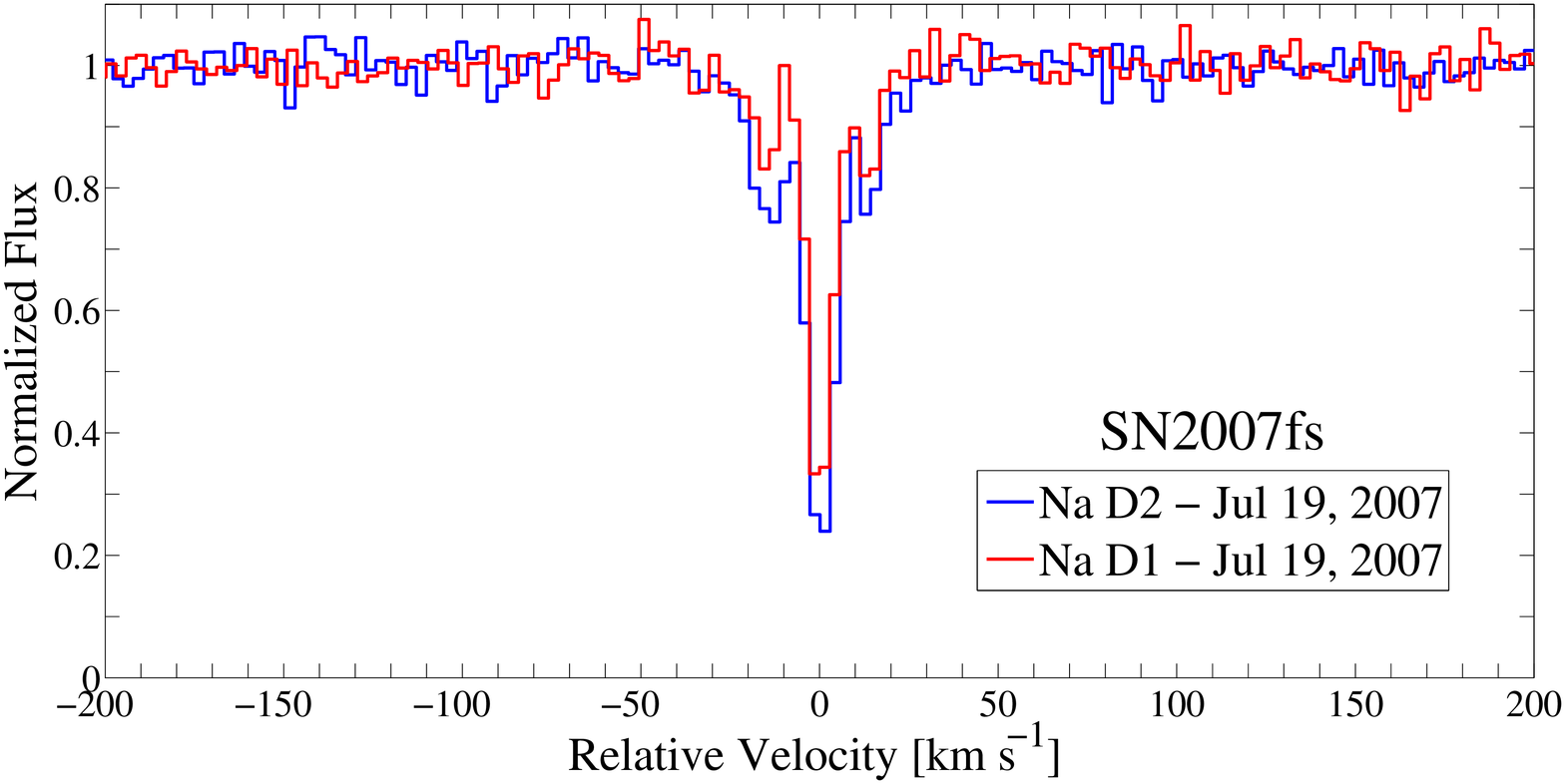} &
\hspace{0.cm}
\includegraphics[width=7.0cm]{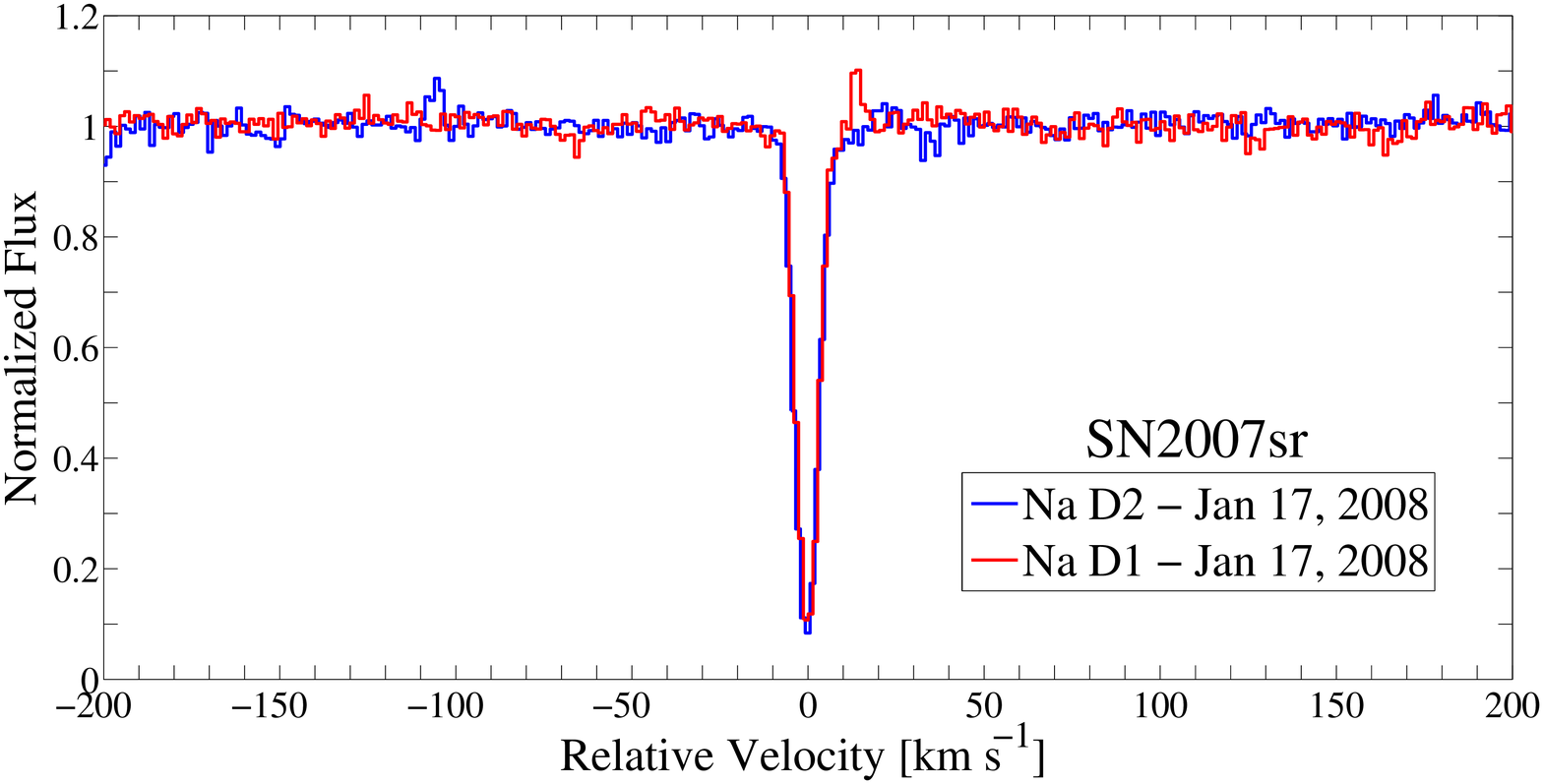} 
\end{tabular} \\
\vspace{0.cm}
\begin{center}
\begin{tabular}{c}
\includegraphics[width=7.0cm]{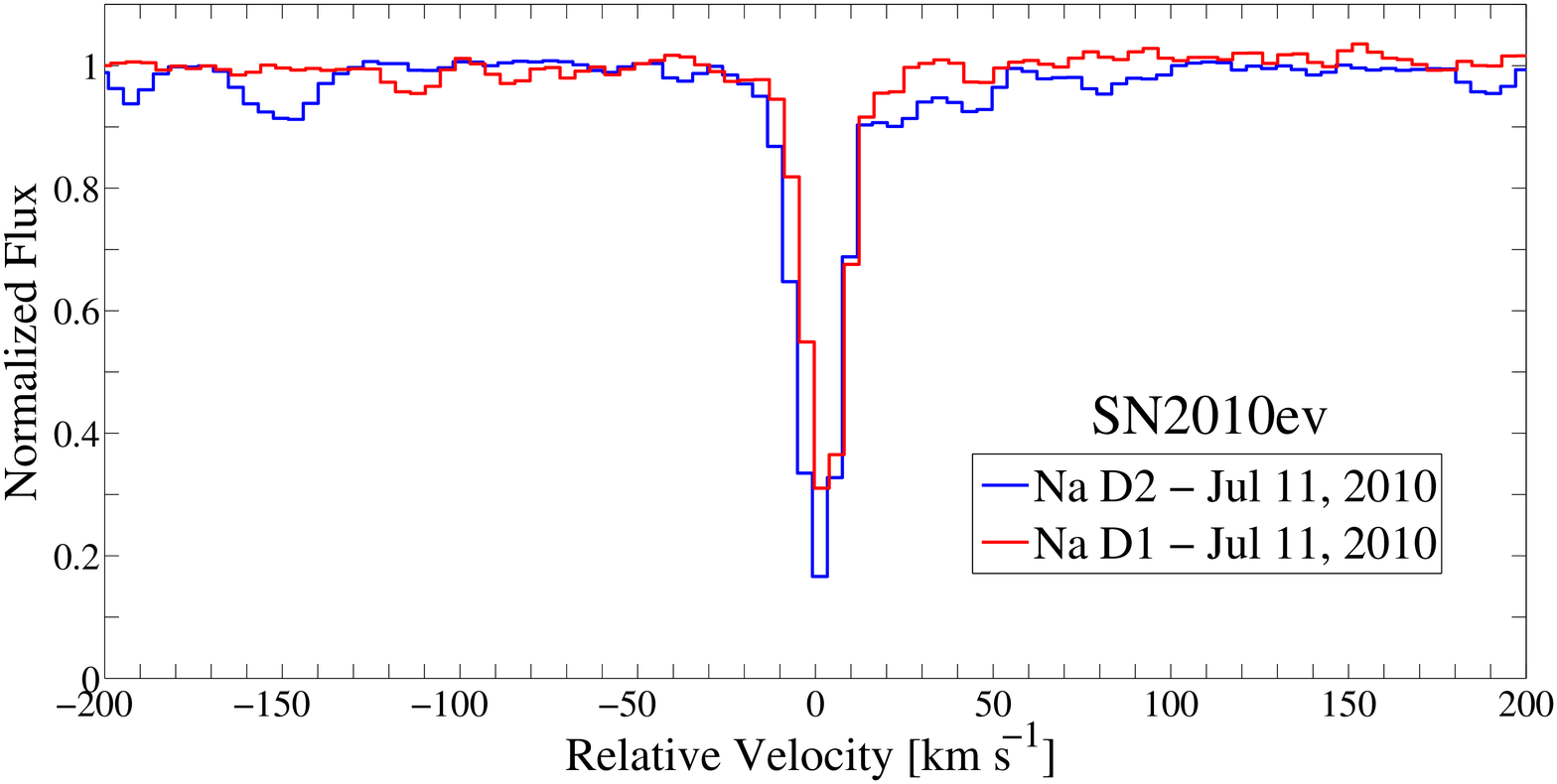} 
\end{tabular}
\end{center}
Figure S4: SN~Ia spectra that exhibit single/symmetric absorption
structures. The blue/red color scheme is as in Fig. 1.
\end{figure}

\begin{figure}
\vspace{-3.0 cm}
\hspace{0. cm}
\begin{tabular}{c c}
\includegraphics[width=7.0cm]{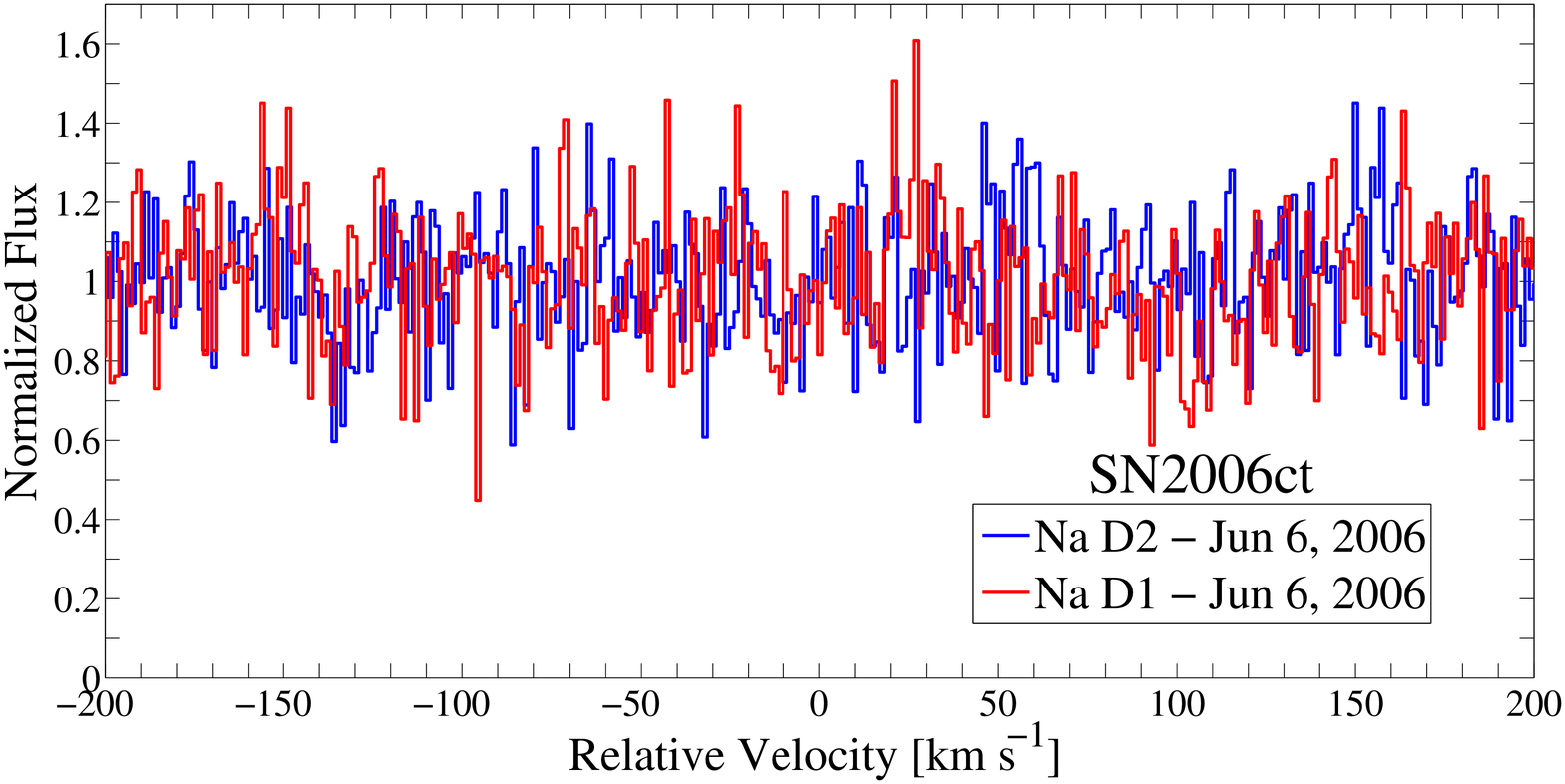} & 
\hspace{0.cm}
\includegraphics[width=7.0cm]{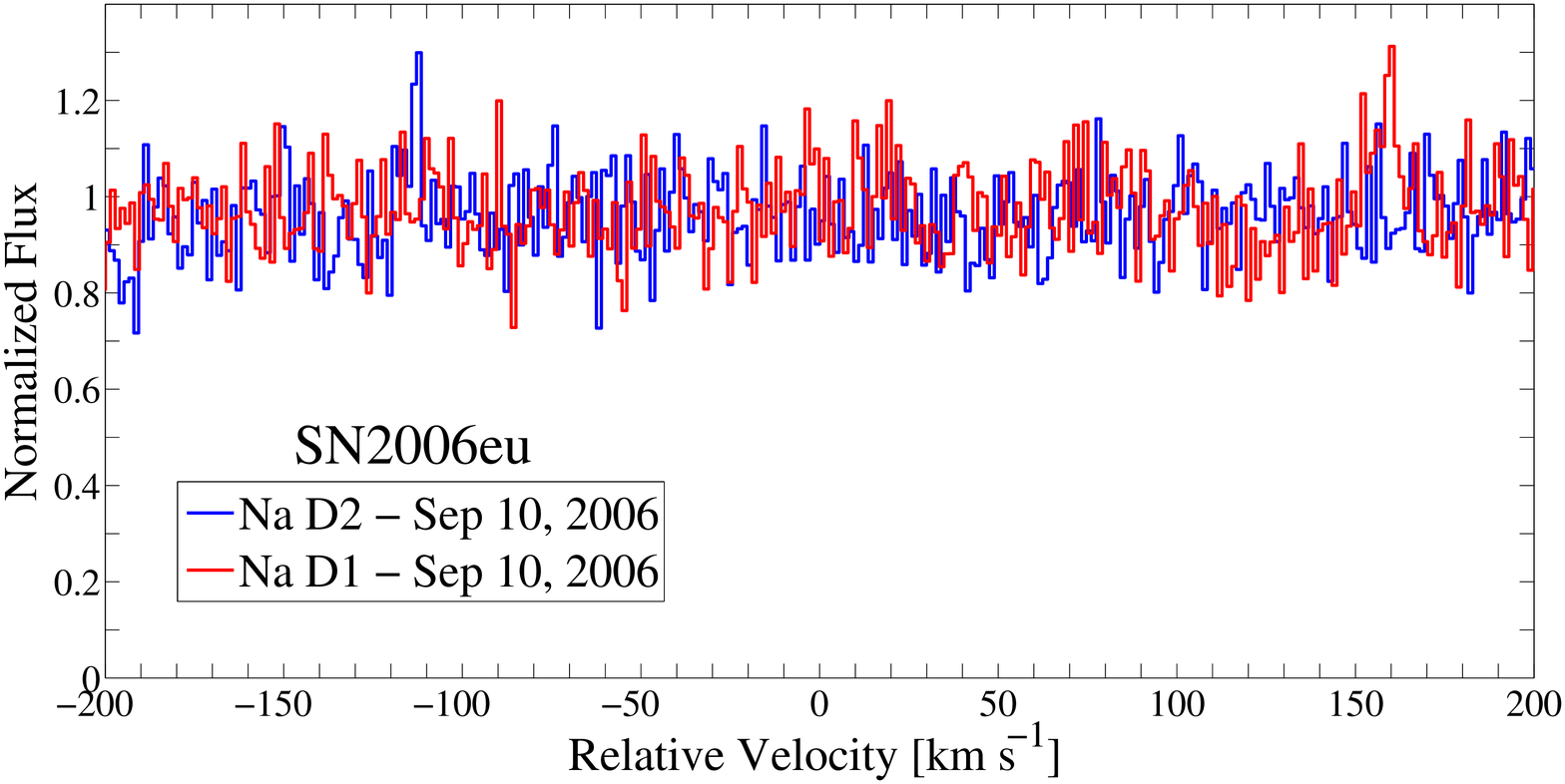}  \\
\includegraphics[width=7.0cm]{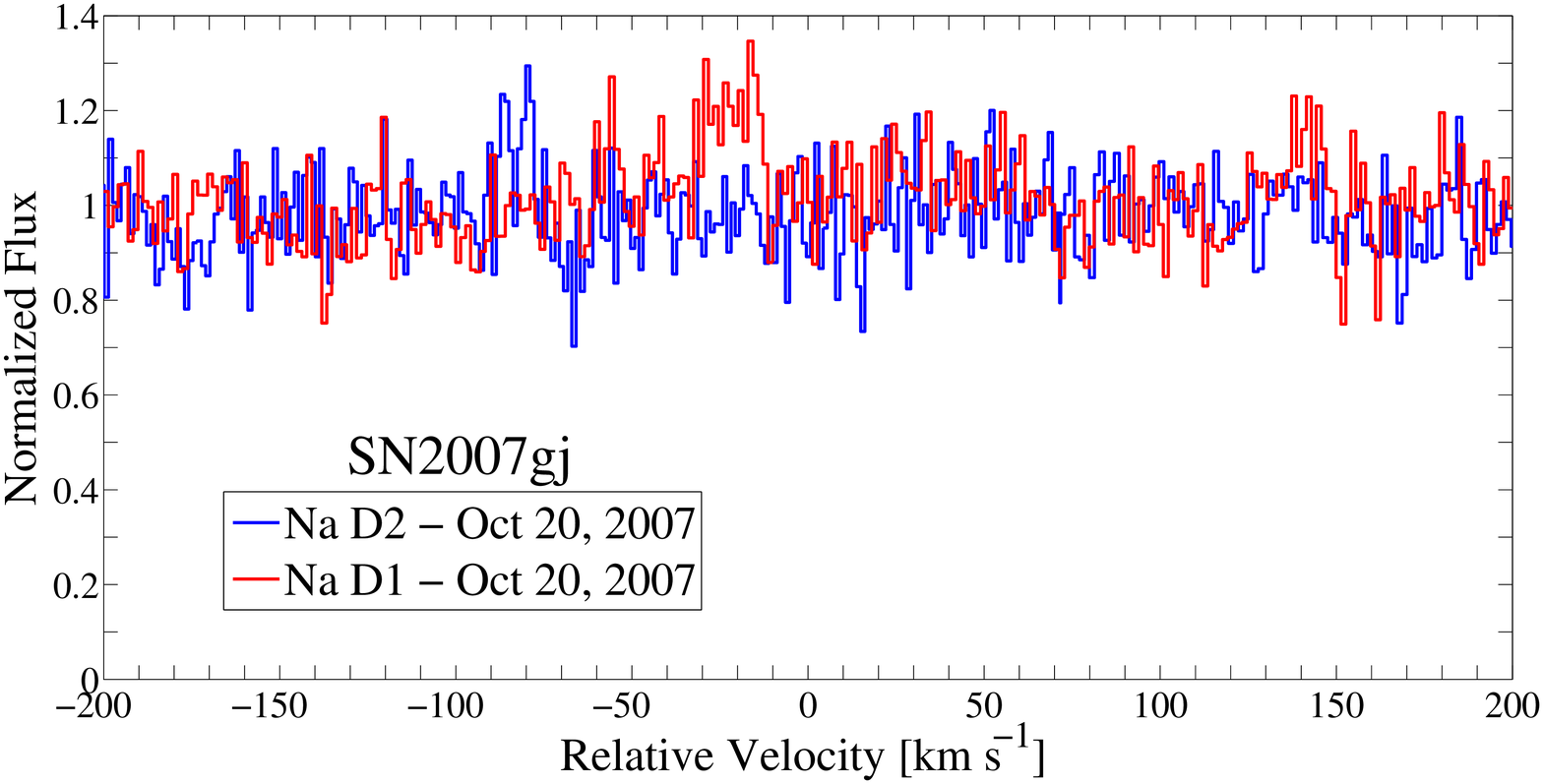} &
\hspace{0.cm}
\includegraphics[width=7.0cm]{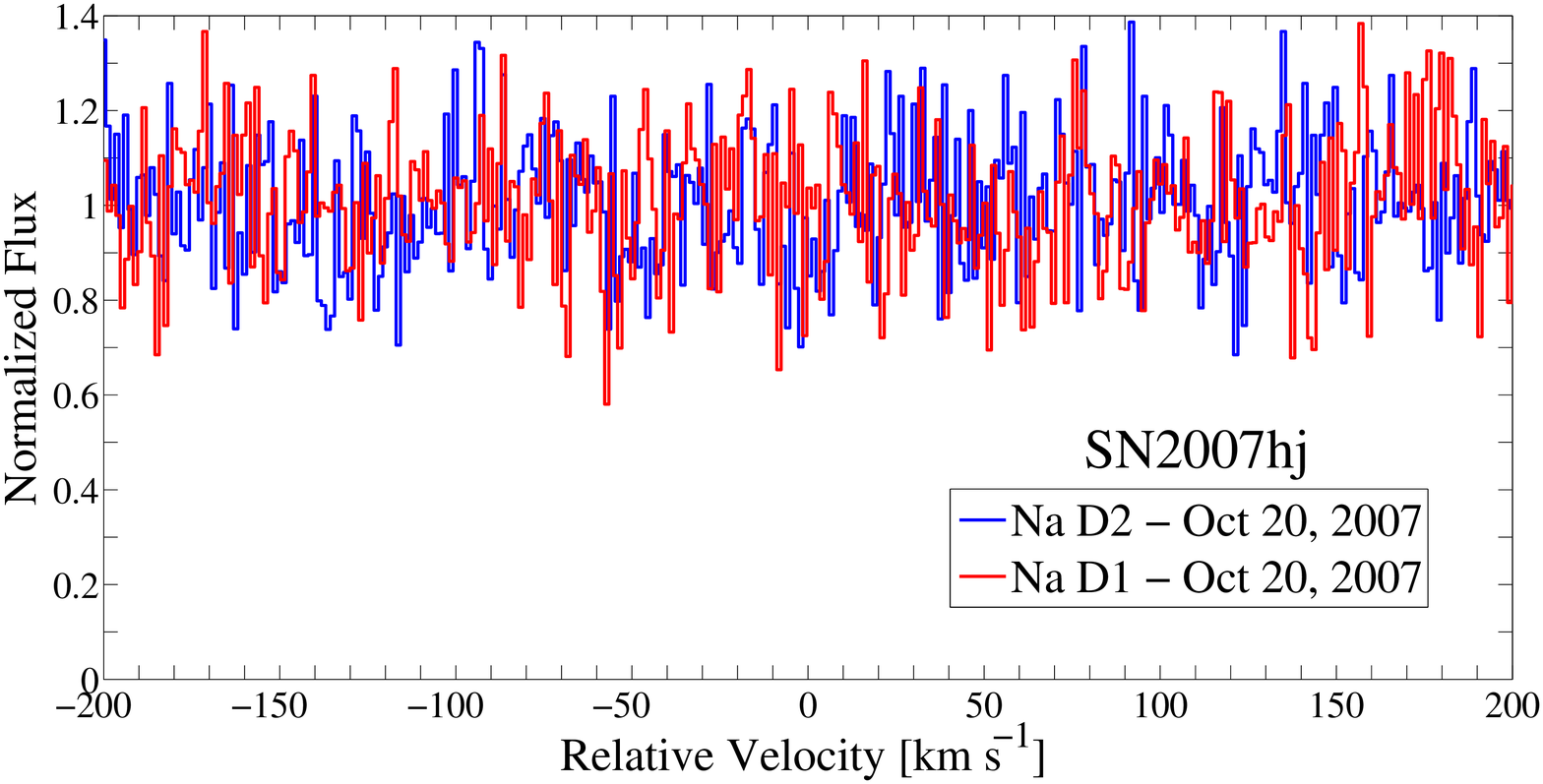}  \\
\includegraphics[width=7.0cm]{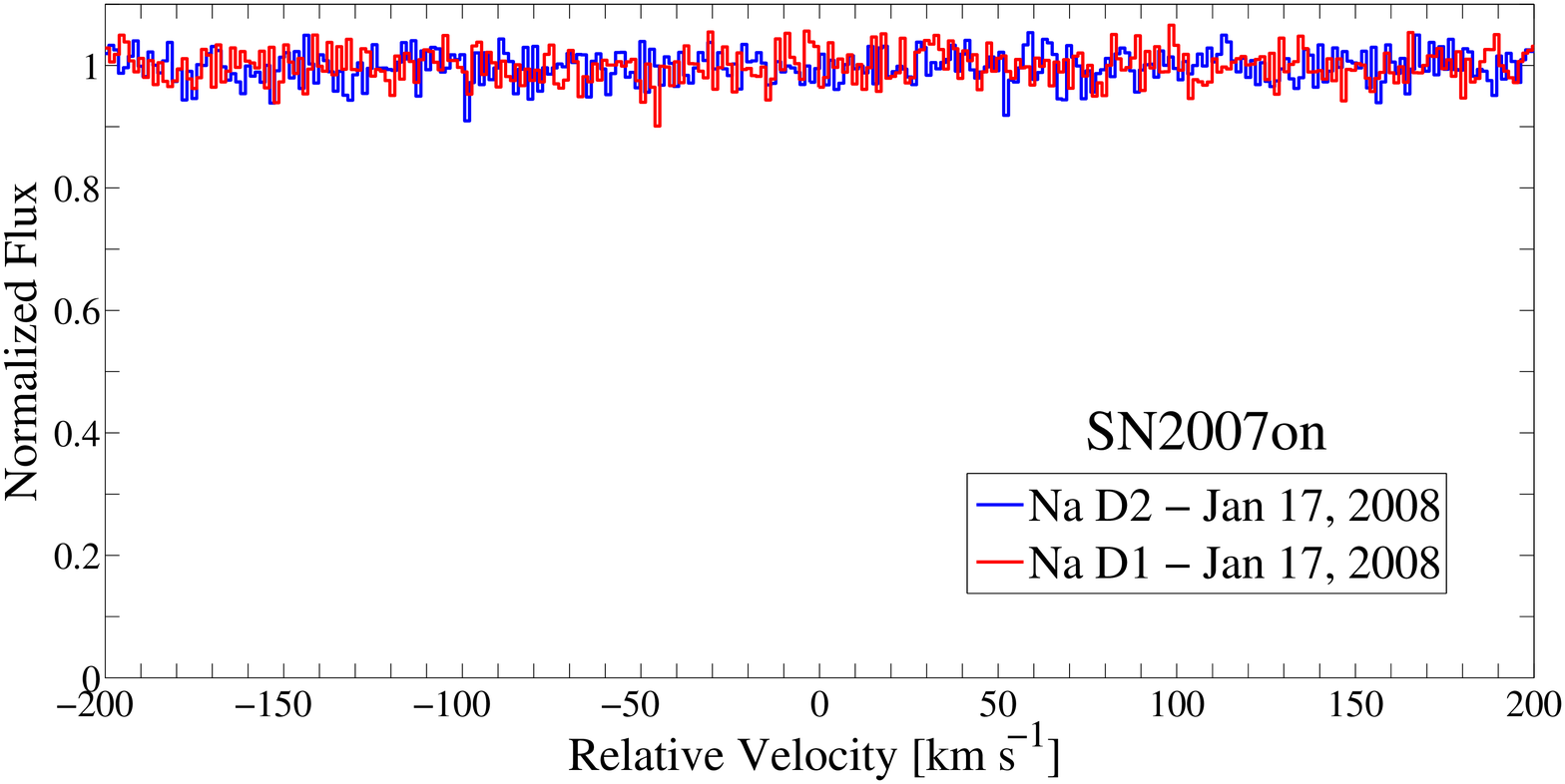} &
\hspace{0.cm}
\includegraphics[width=7.0cm]{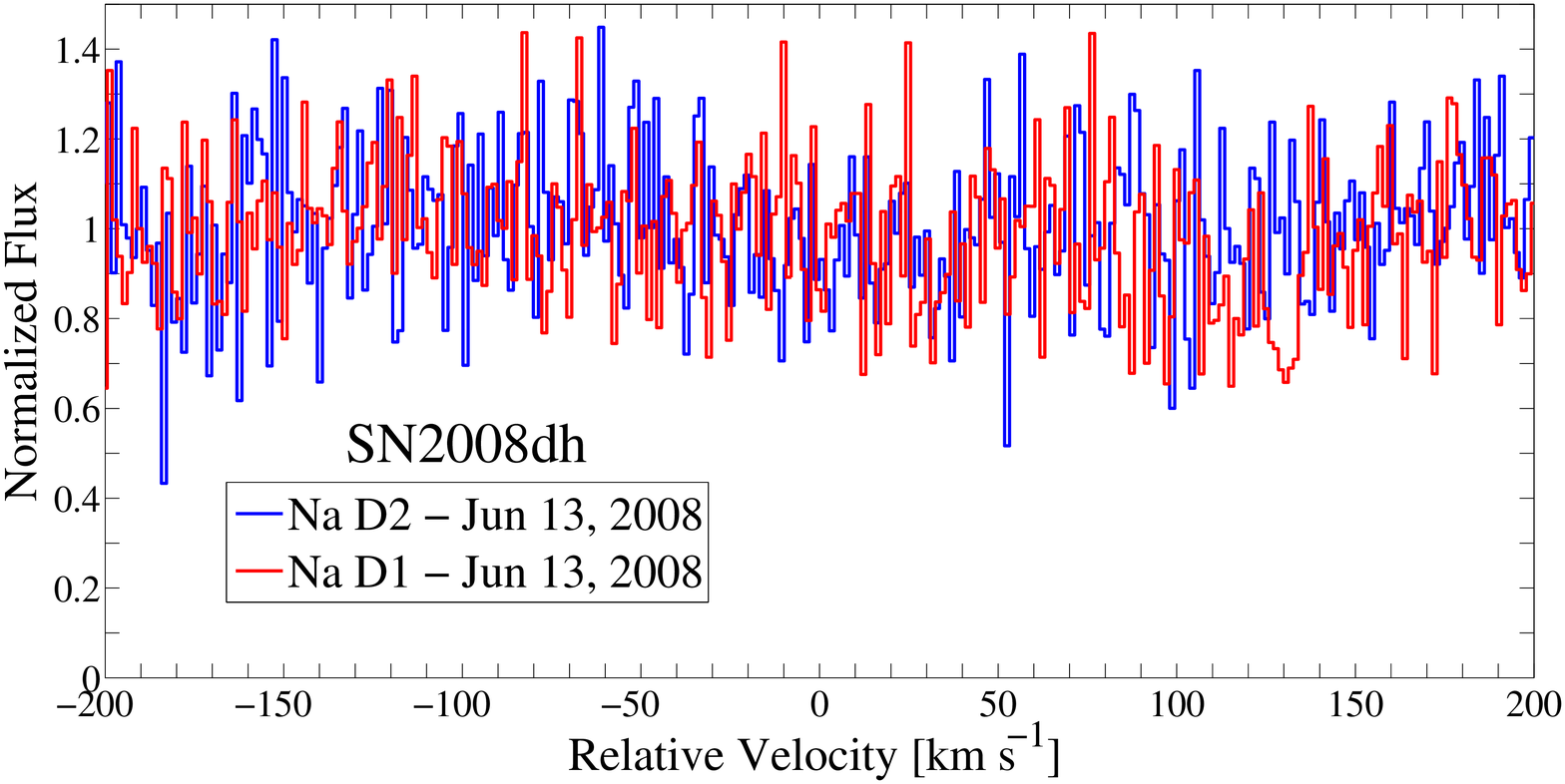}  \\
\includegraphics[width=7.0cm]{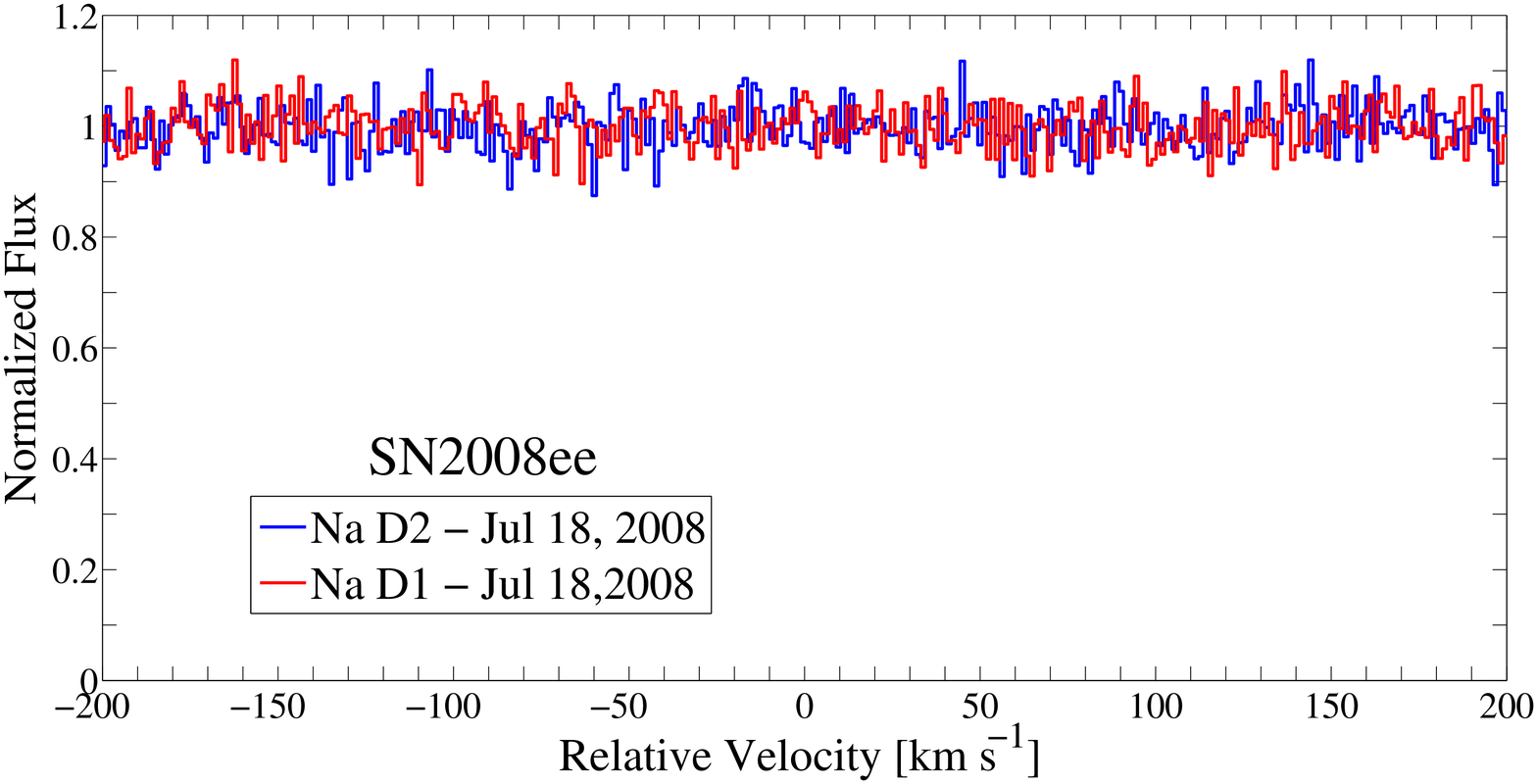} &
\hspace{0.cm}
\includegraphics[width=7.0cm]{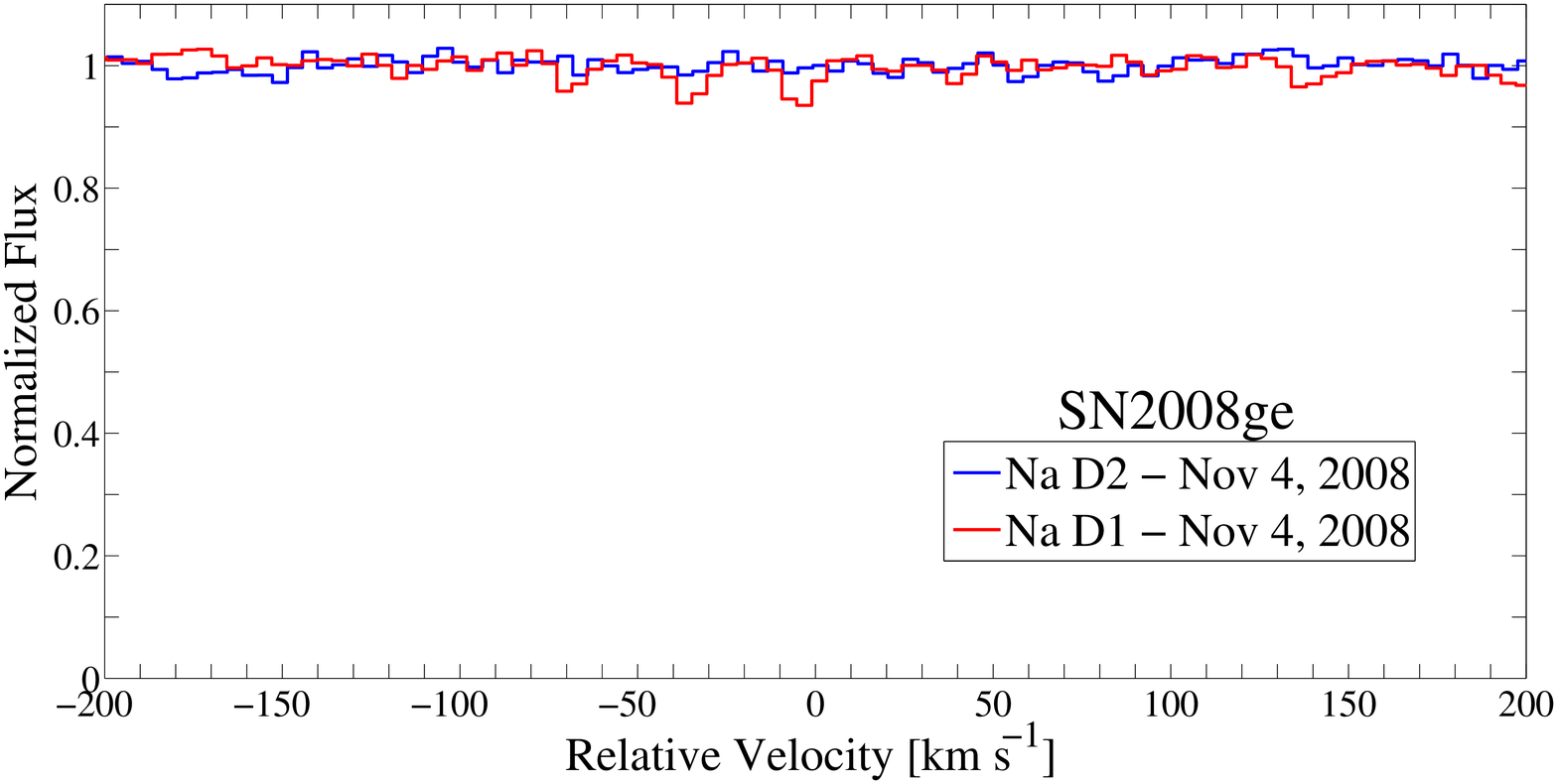} \\
\includegraphics[width=7.0cm]{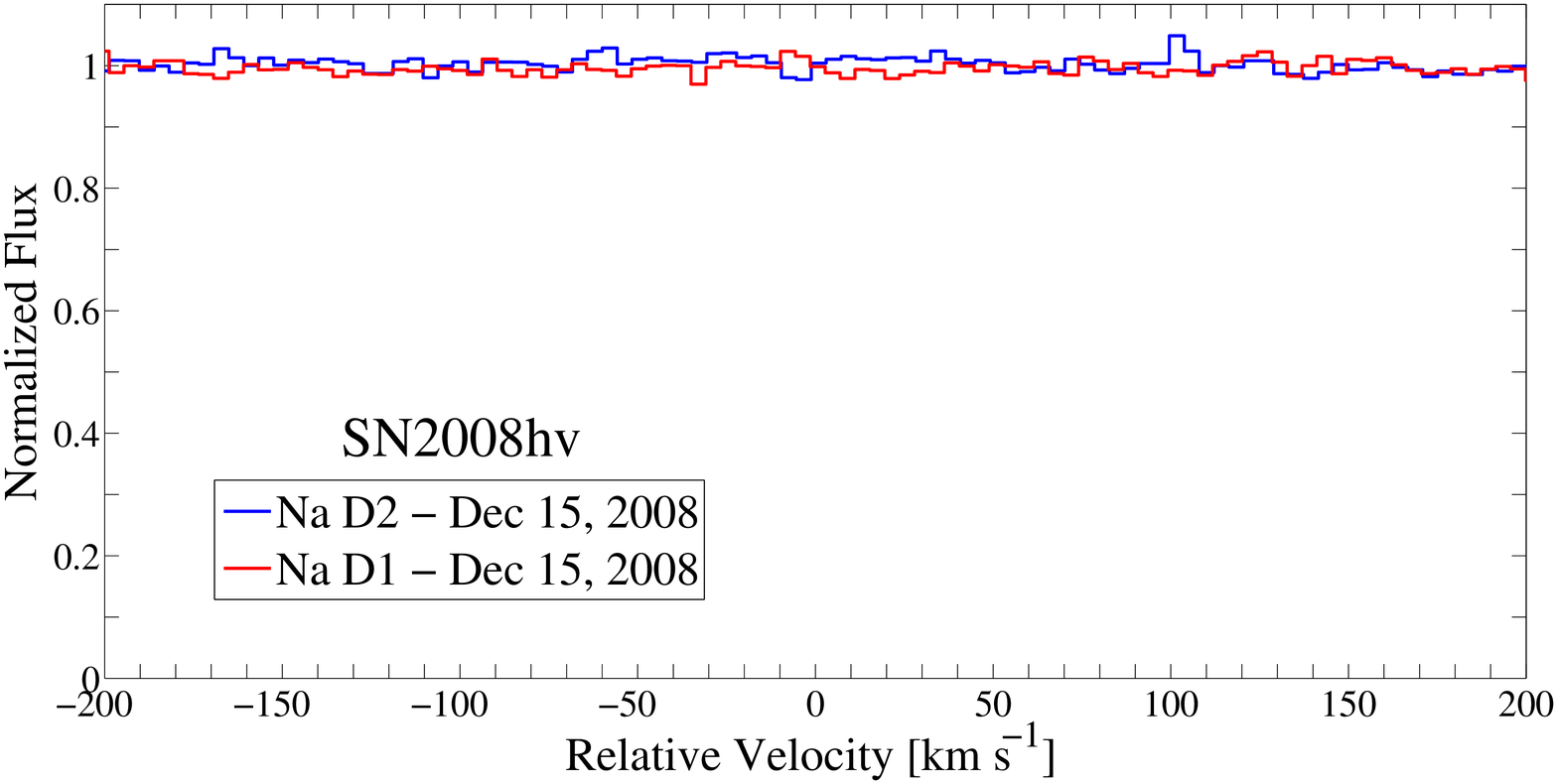} &
\hspace{0.cm}
\includegraphics[width=7.0cm]{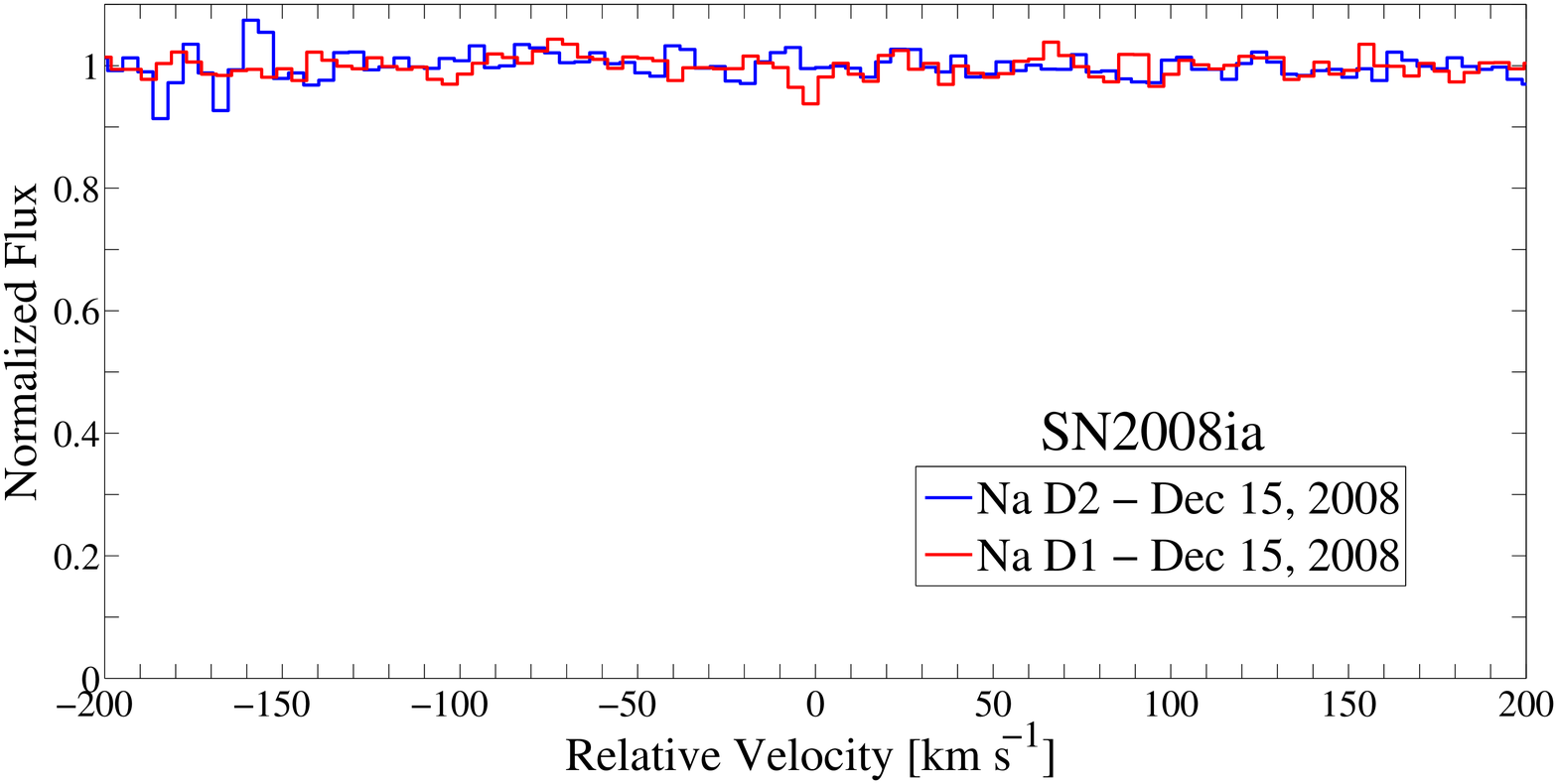} \\
\includegraphics[width=7.0cm]{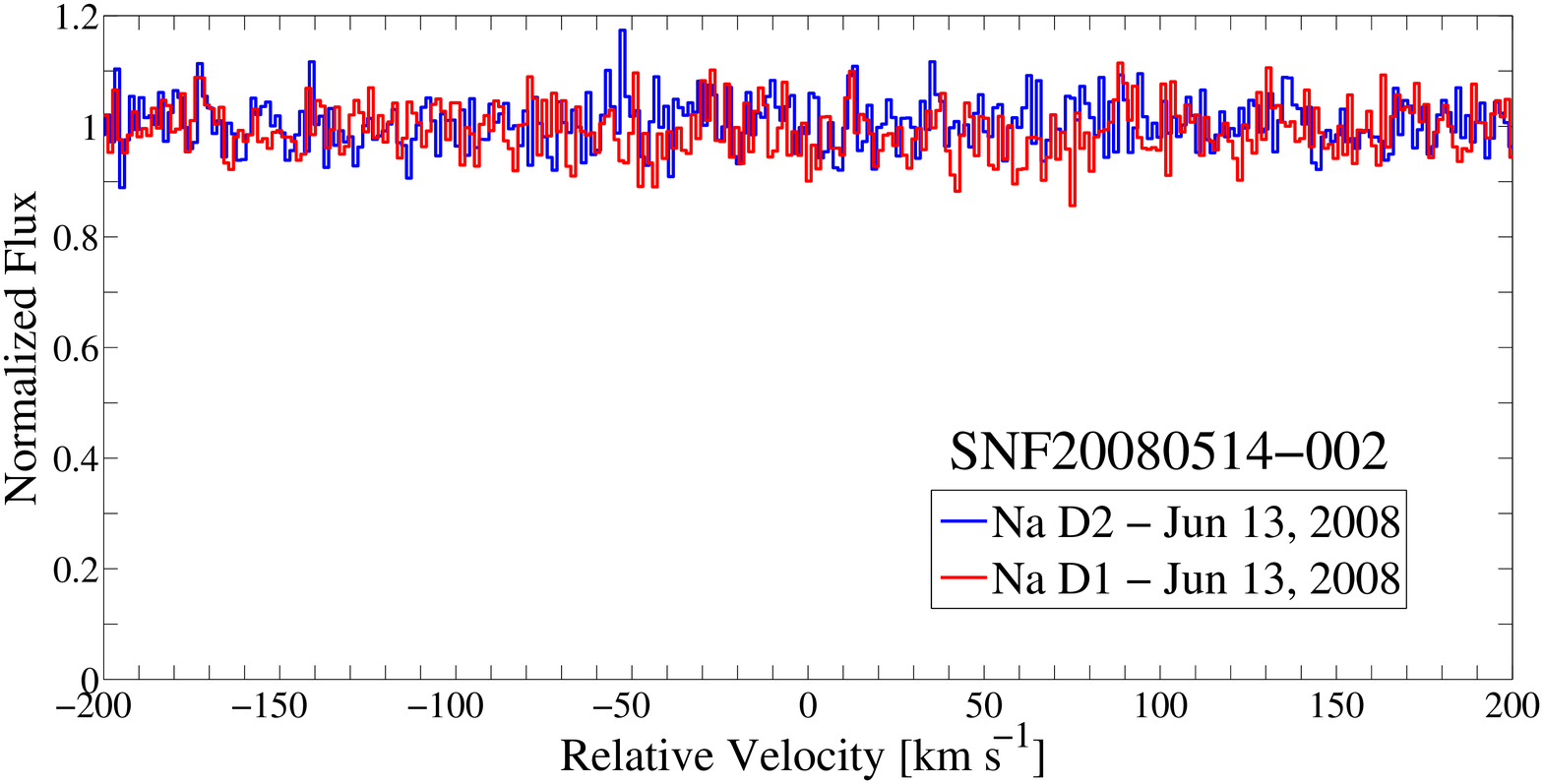} &
\hspace{0.cm}
\includegraphics[width=7.0cm]{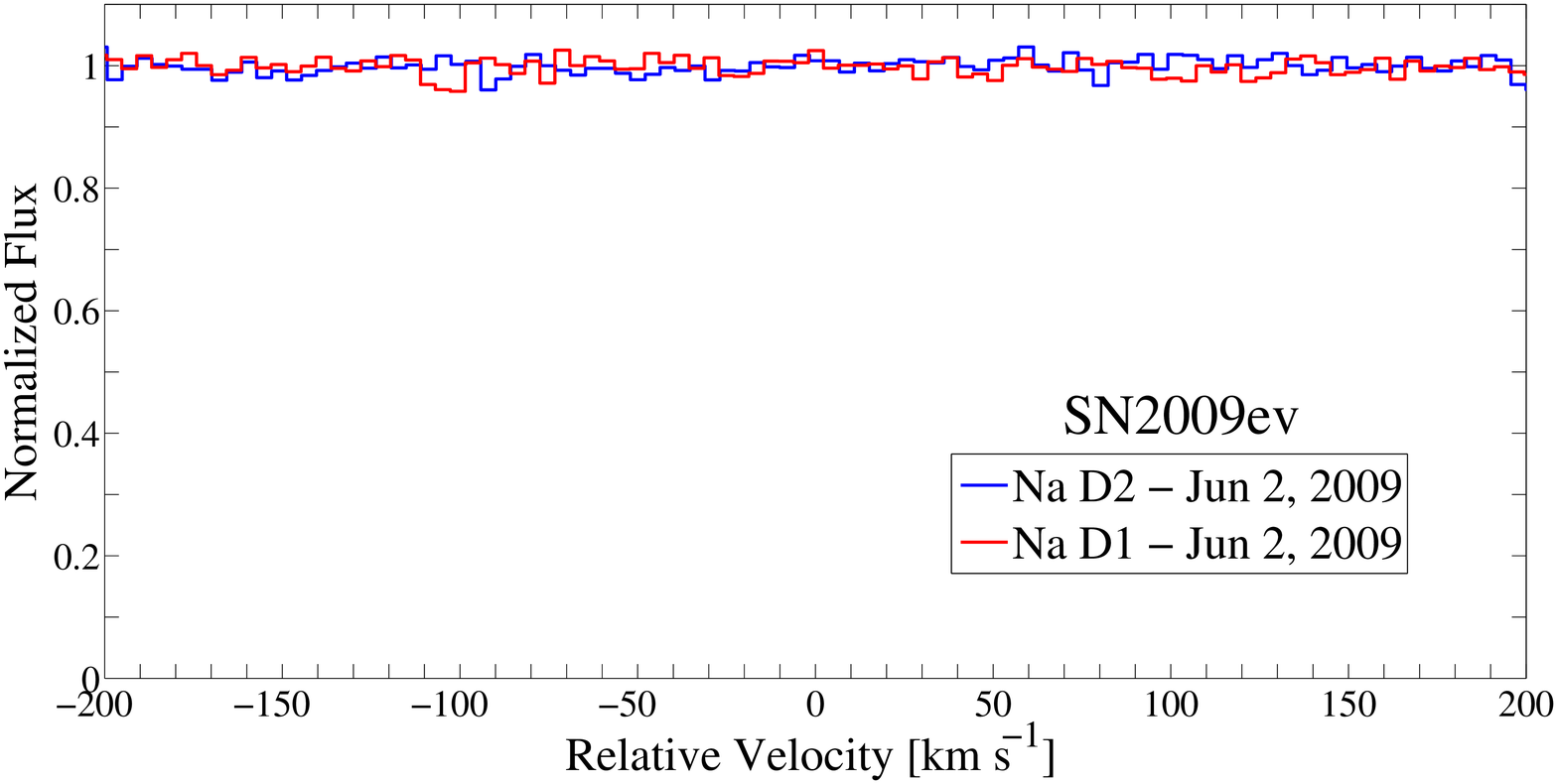} \\
\end{tabular}
\vspace{0.cm}
\begin{center}
\begin{tabular}{c}
\includegraphics[width=7.0cm]{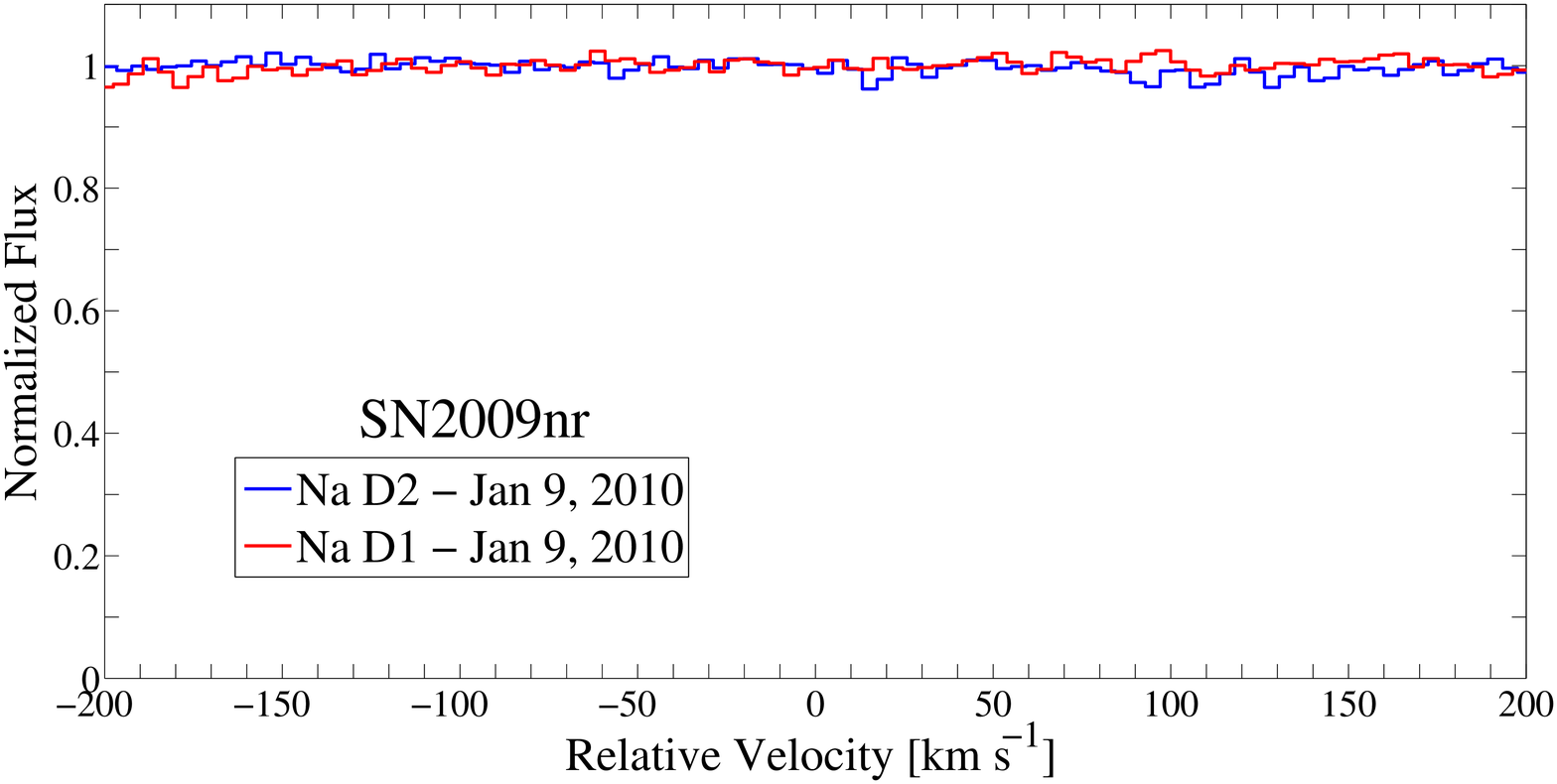} 
\end{tabular}
\end{center}
Figure S5: SN~Ia spectra that exhibit no Na~I~D absorption. The color
scheme is as described in Fig. 1.
\end{figure}

\begin{figure}
\vspace{-3.0 cm}
\begin{center}
\begin{tabular}{c c}
\includegraphics[width=6.5cm]{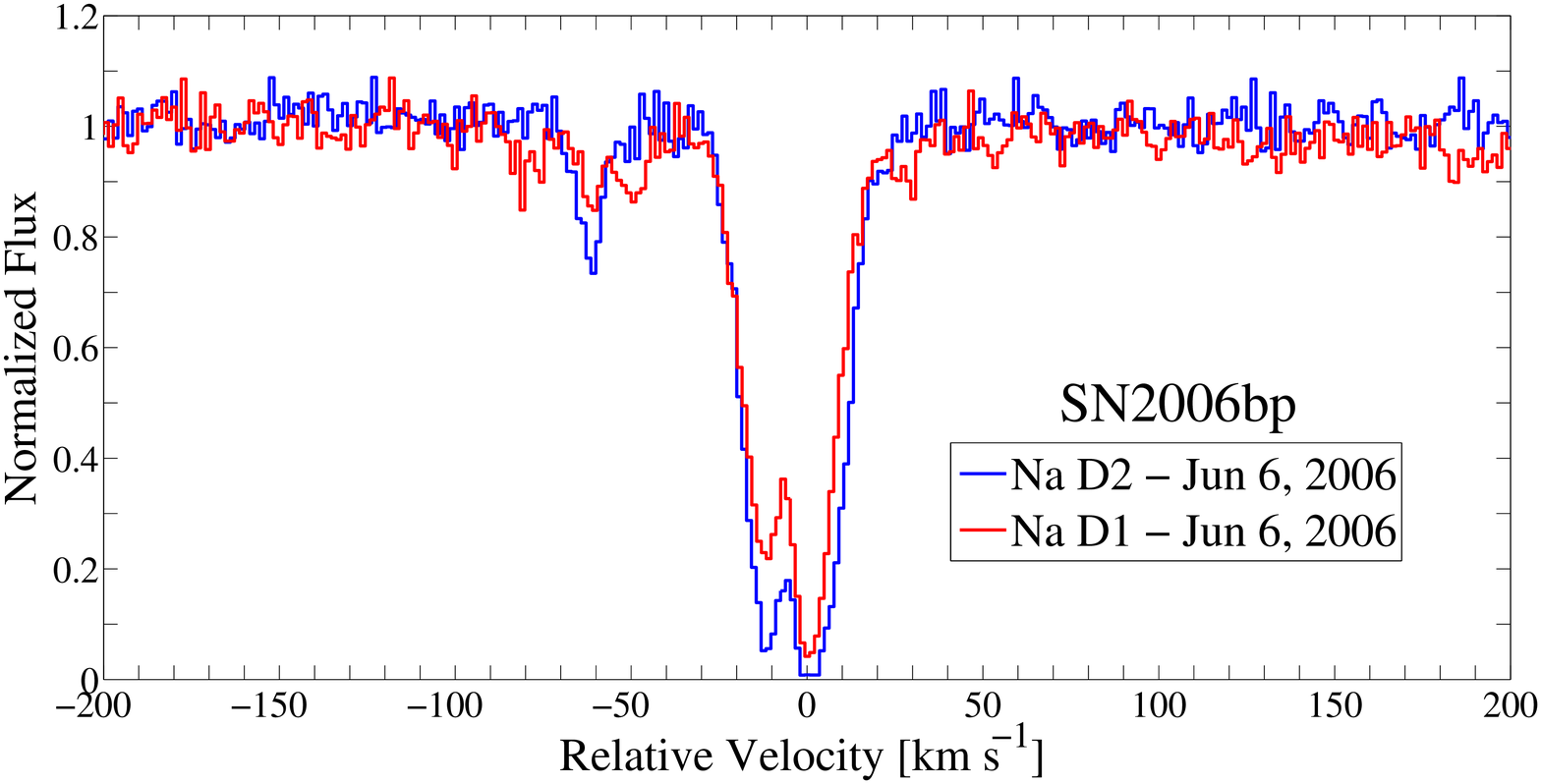} &
\includegraphics[width=6.5cm]{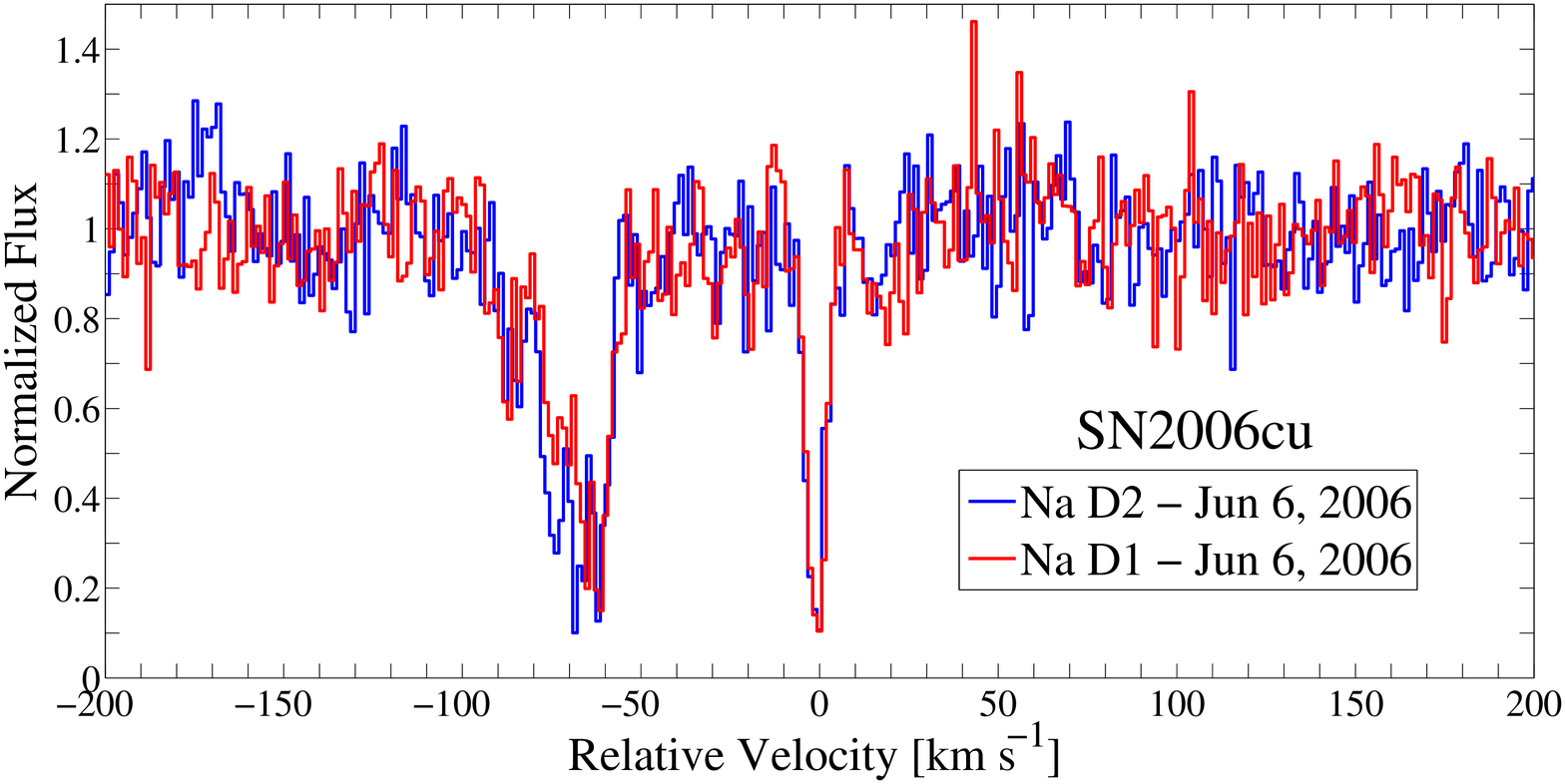}  \\
\includegraphics[width=6.5cm]{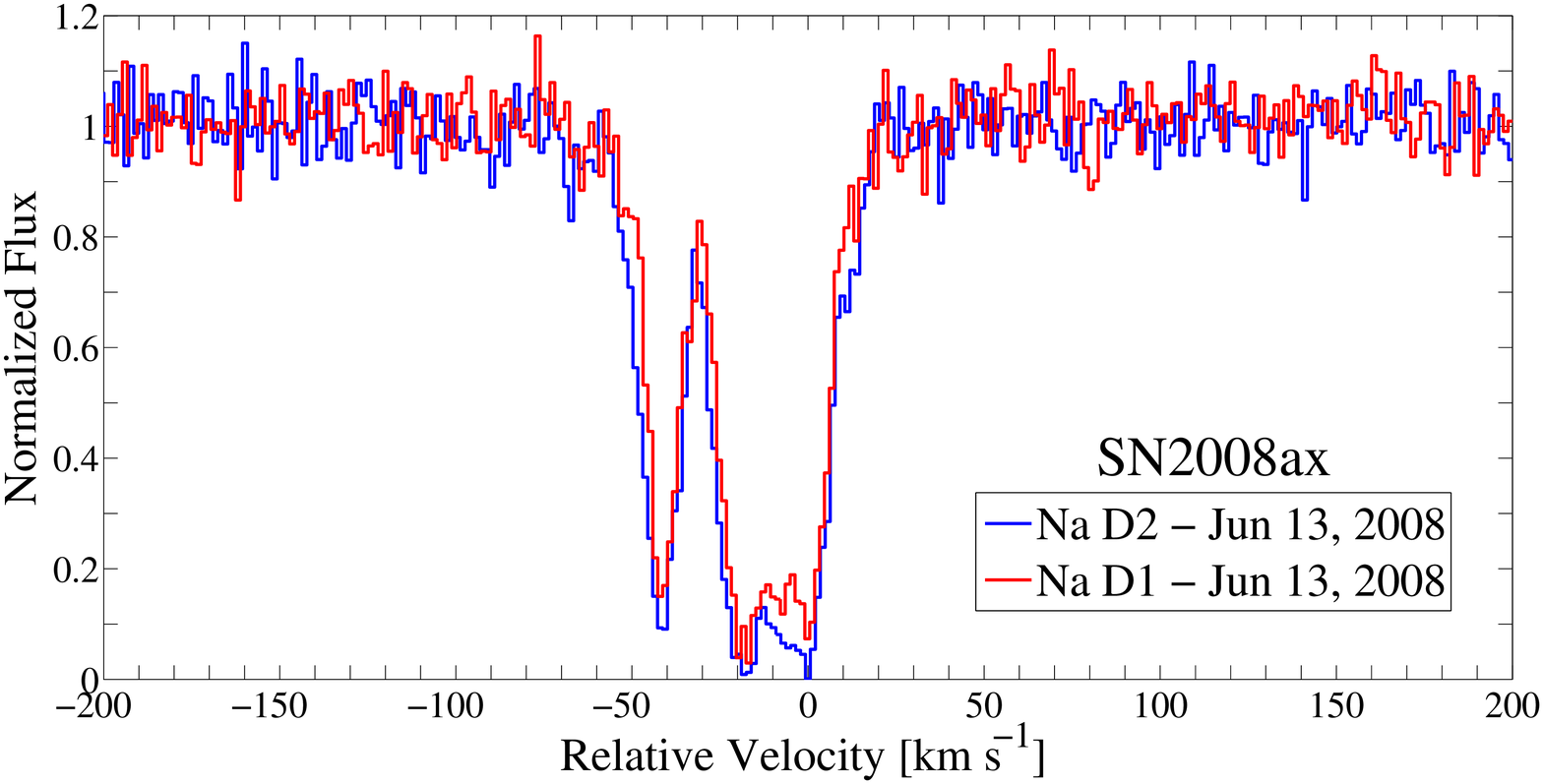}  &
\includegraphics[width=6.5cm]{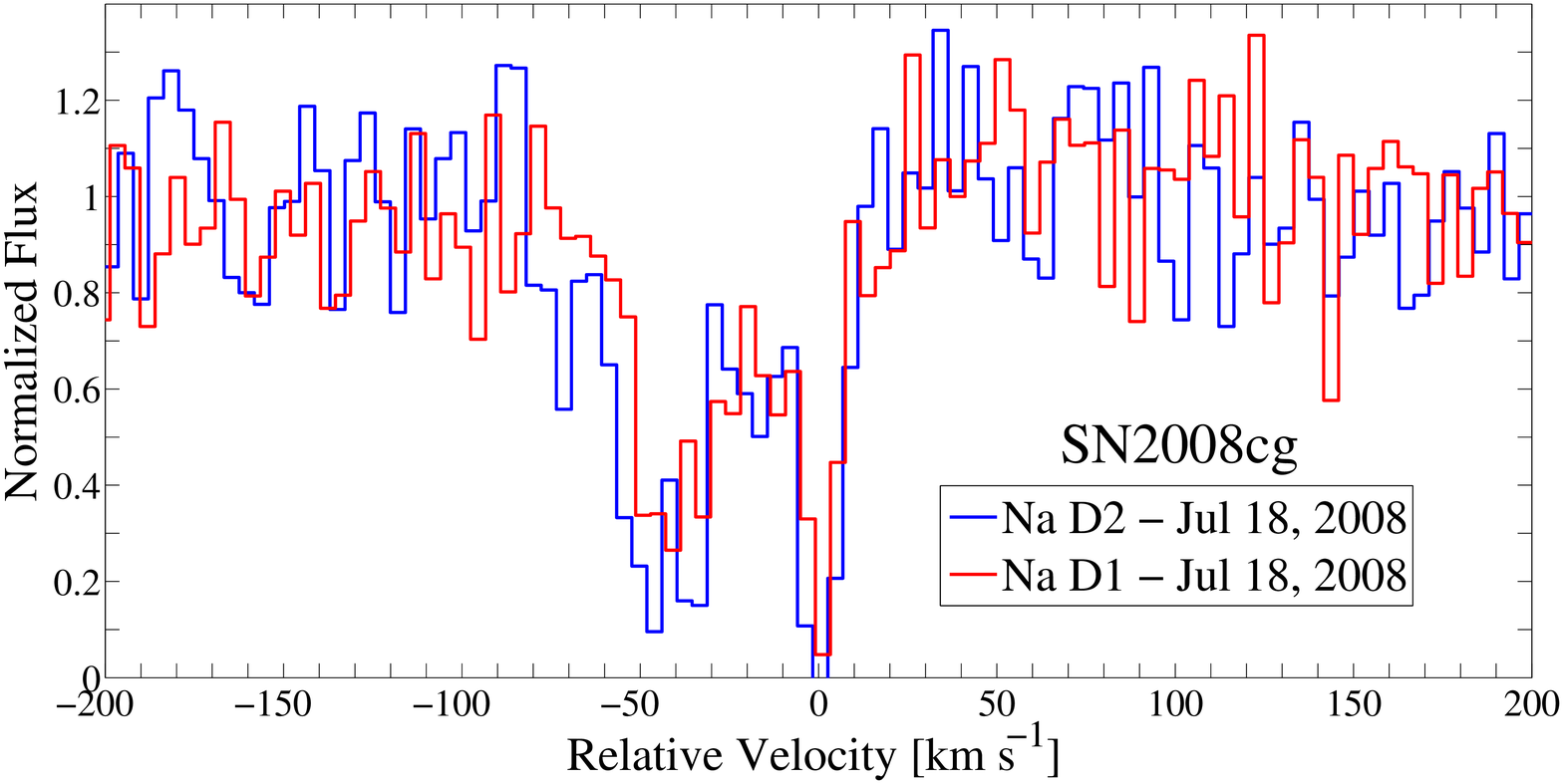} \\
\hline \\
\includegraphics[width=6.5cm]{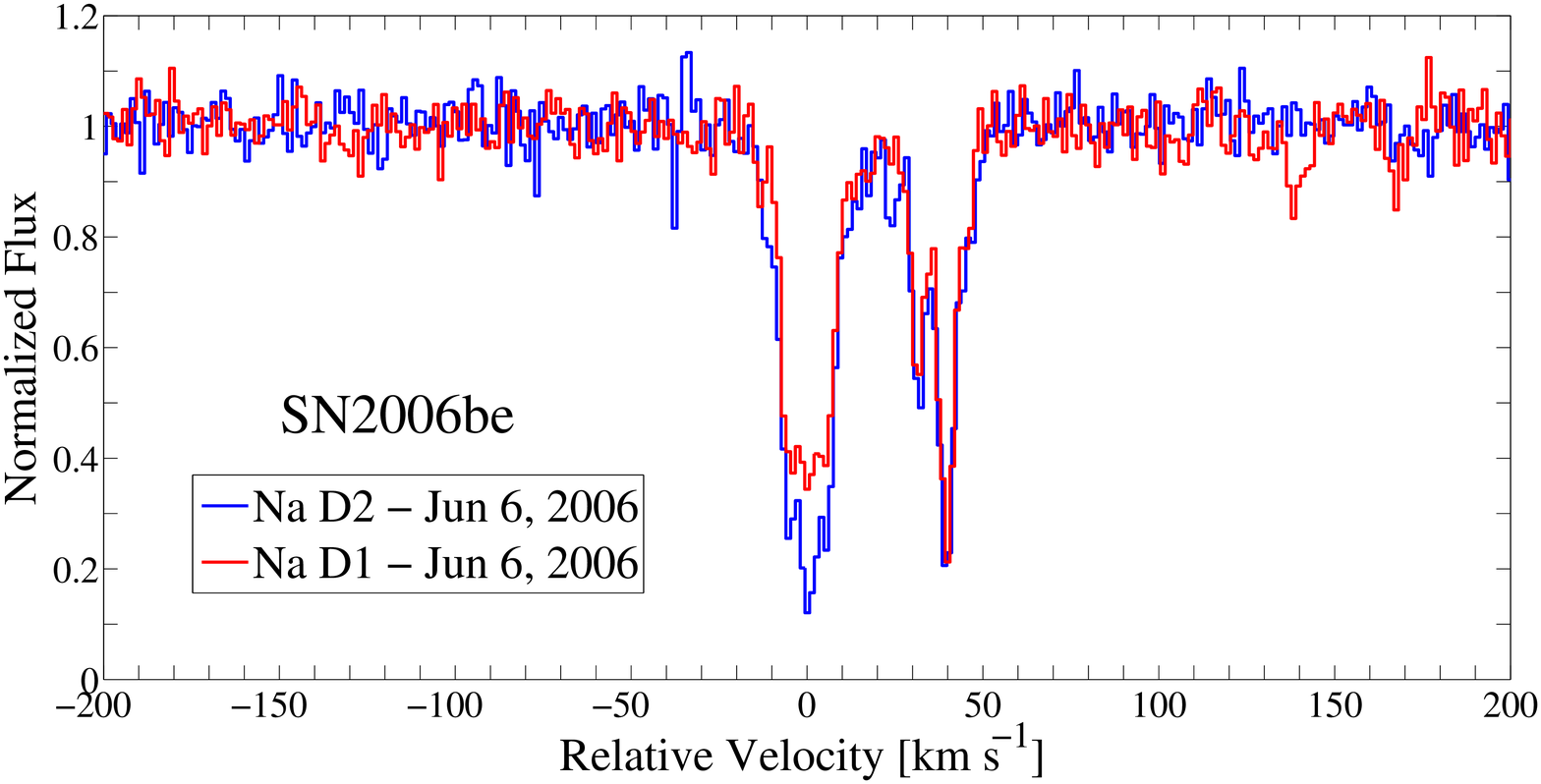} & 
\includegraphics[width=6.5cm]{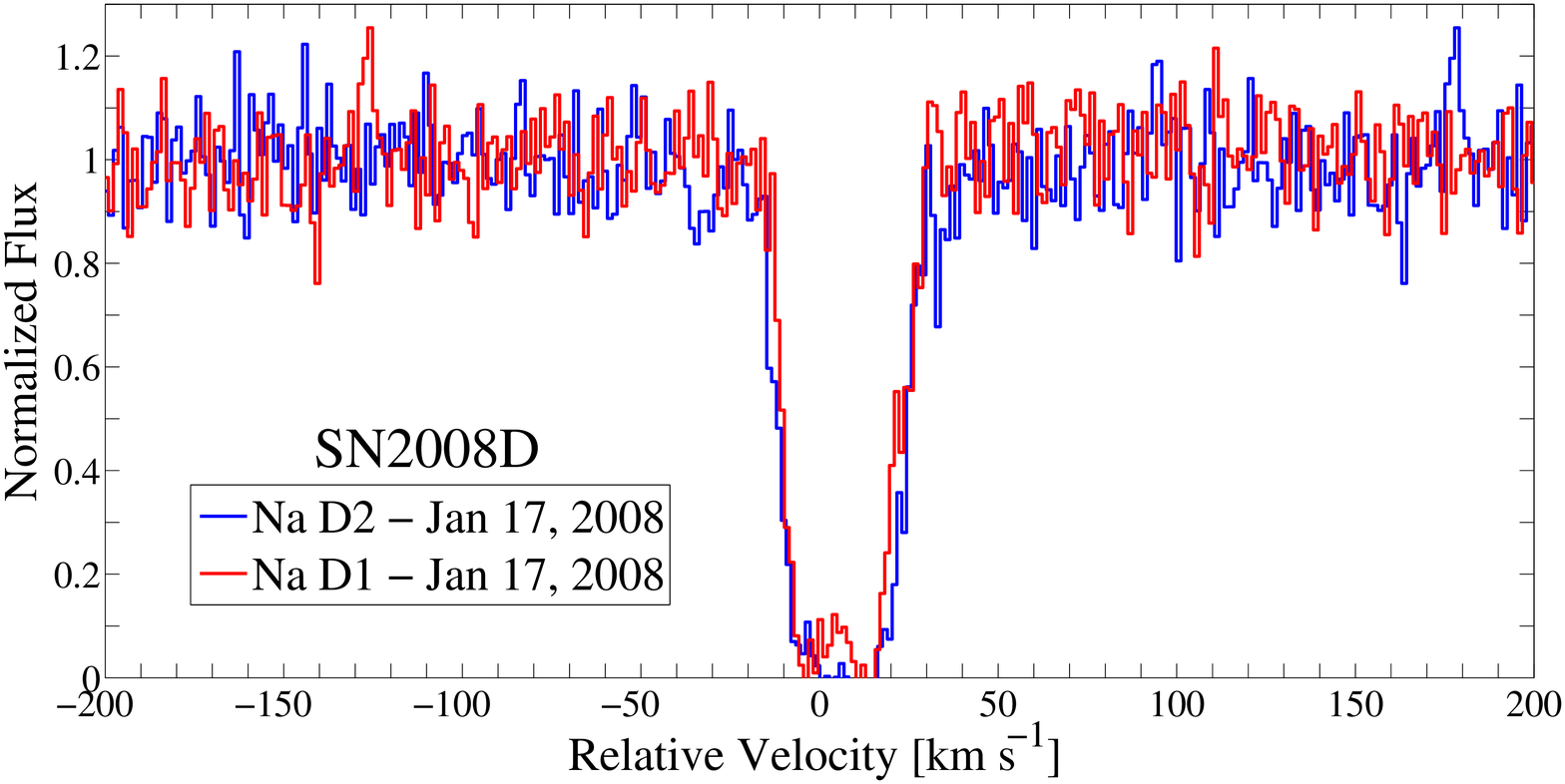}
\end{tabular}
\begin{tabular}{c}
\includegraphics[width=6.5cm]{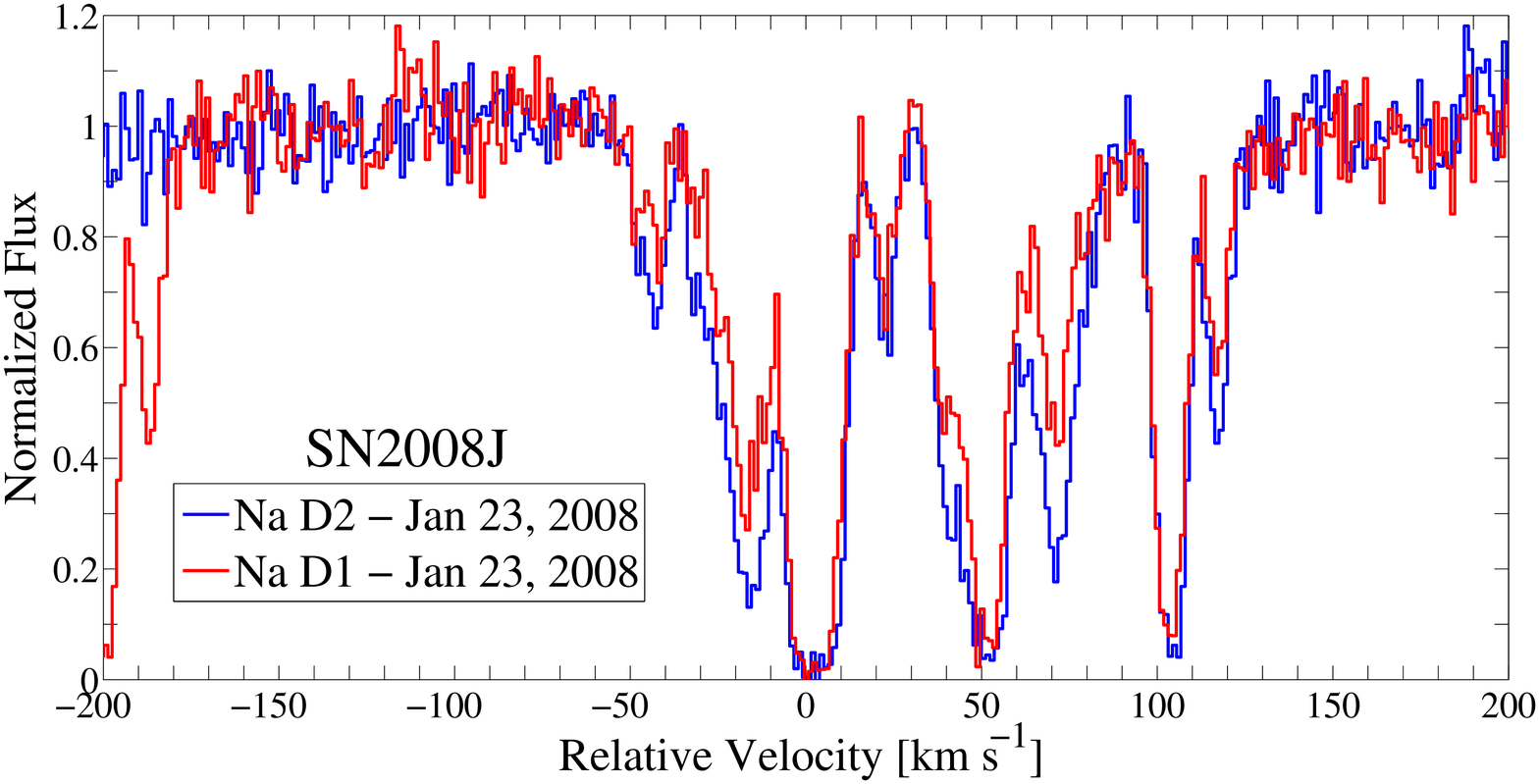}
\end{tabular}
\begin{tabular}{c c}
\hline \\
\includegraphics[width=6.5cm]{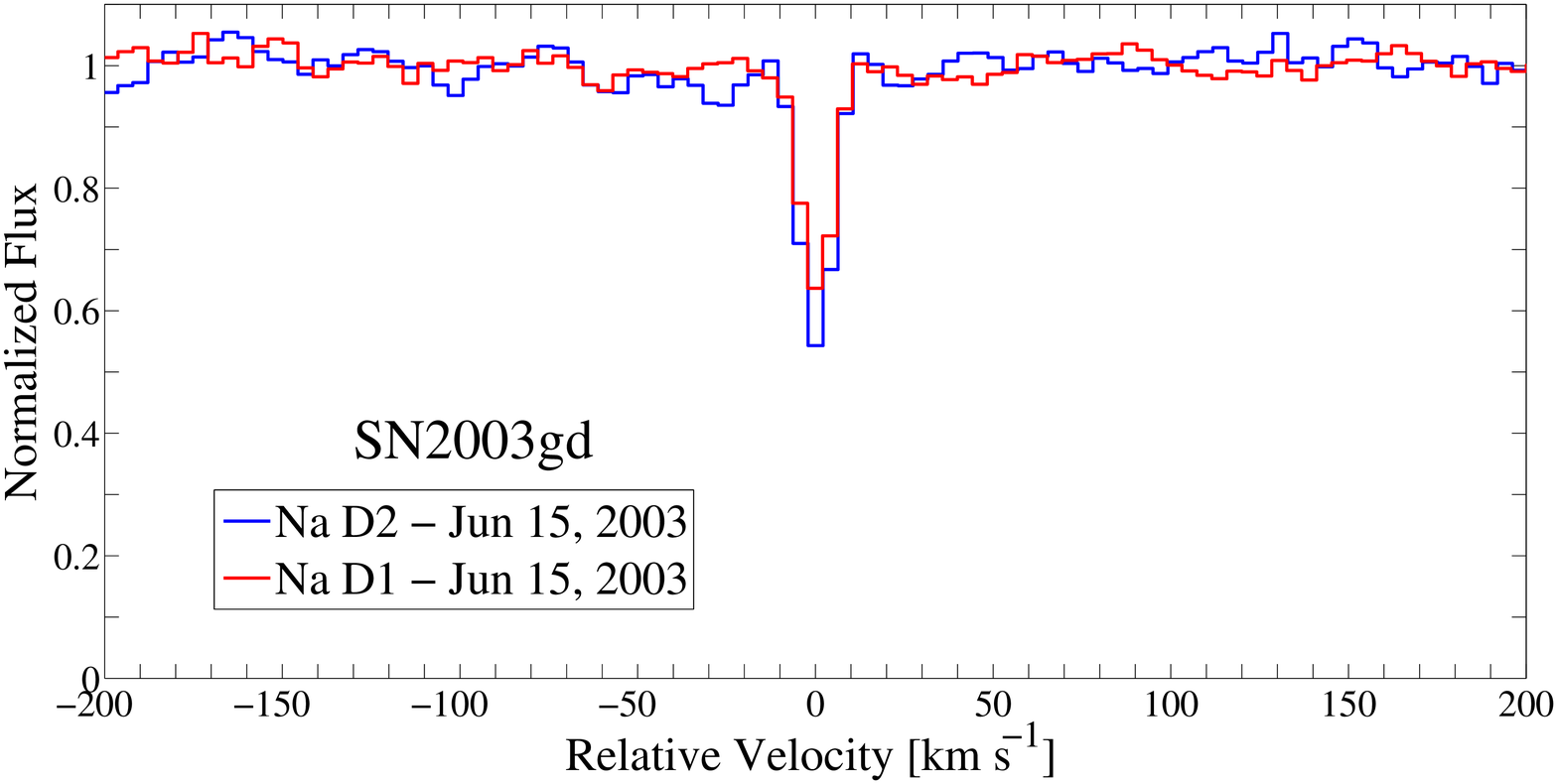} &
\includegraphics[width=6.5cm]{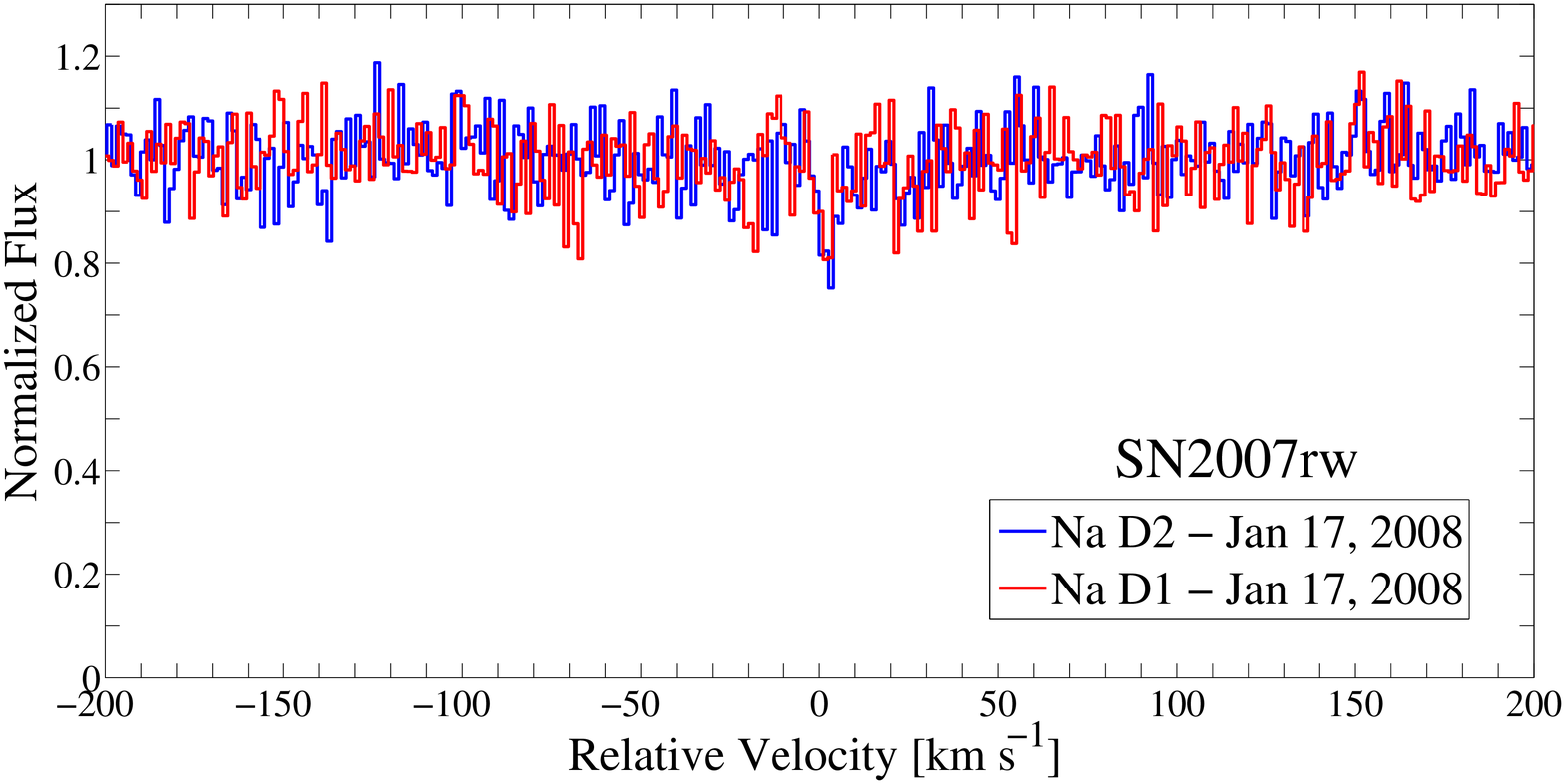} \\
\includegraphics[width=6.5cm]{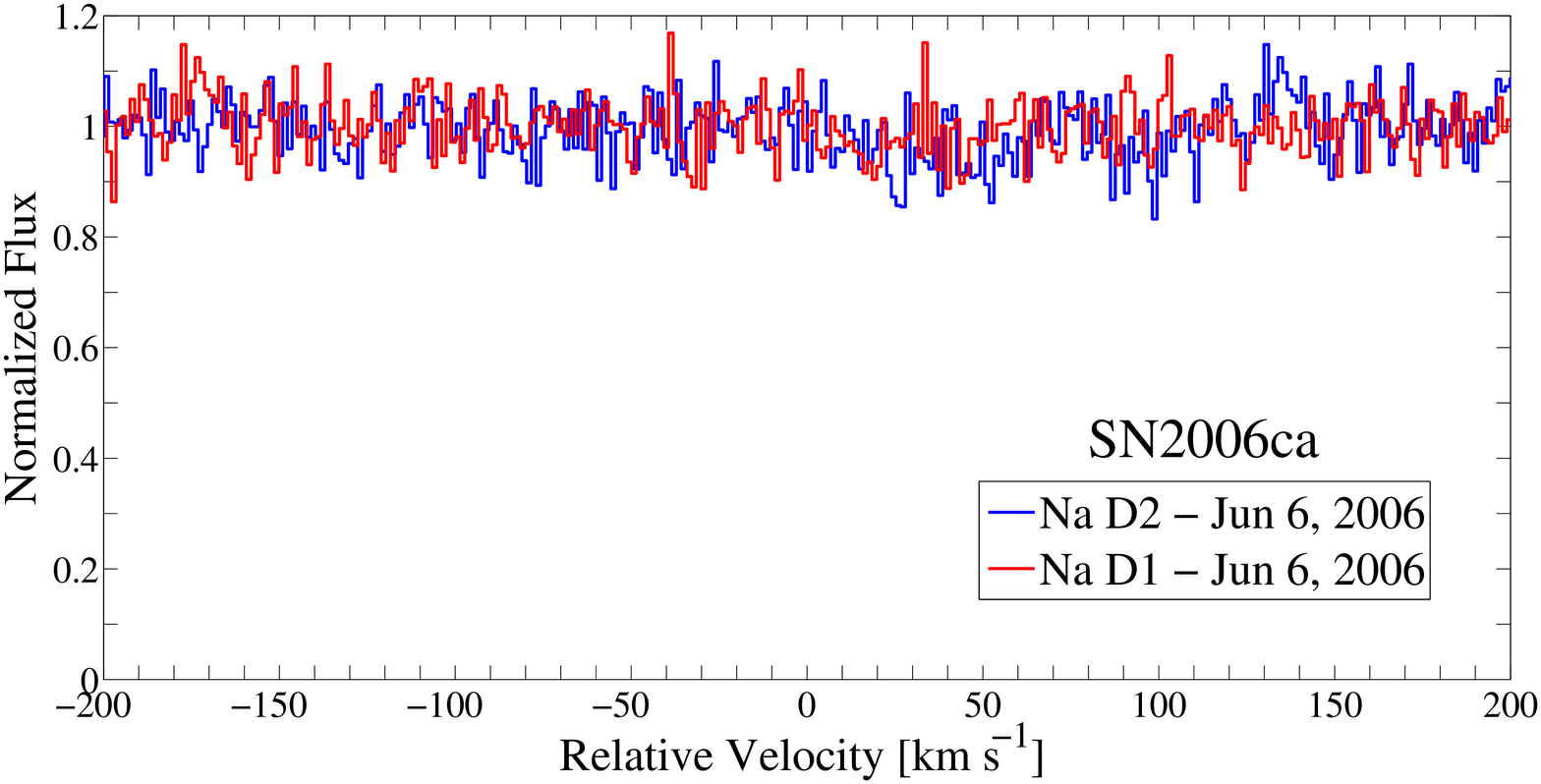}  &
\includegraphics[width=6.5cm]{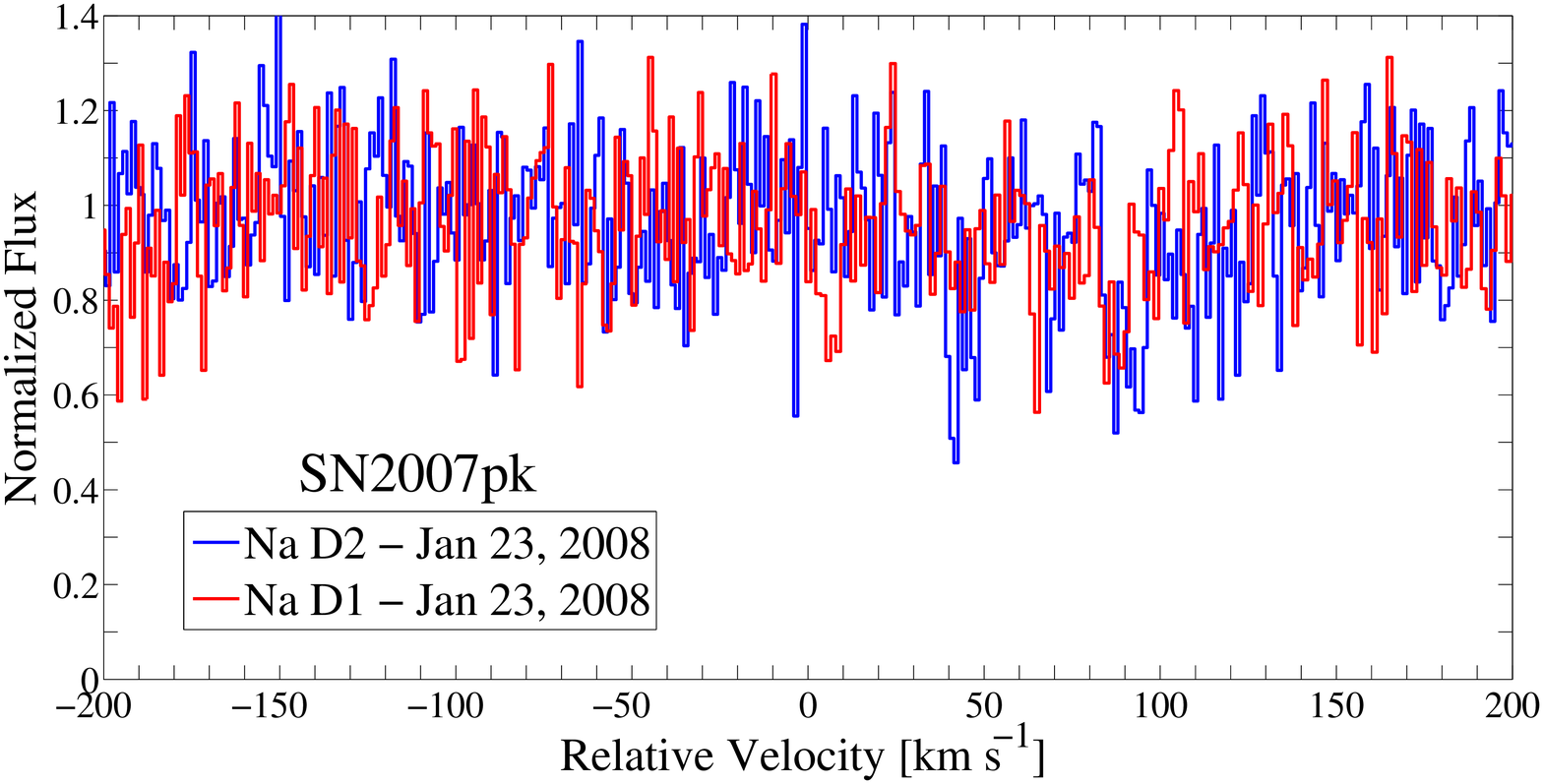} \\
\end{tabular} \\
\end{center}
\vspace{0.cm}
Figure S6: Spectra of CC~SN sample. Upper four panels are of events
exhibiting blueshifted structures. Middle three panels are of events
exhibiting redshifted structures. The bottom four panels are of events
classified as single/symmetric features (SN 2003gd and SN 2007rw) and
those that do not exhibit Na~I~D absorption features (SN 2006ca and
SN 2007pk). The color scheme is as described in Fig. 1. Analysis of
SN 2008D has been reported by Soderberg et al. (2008) ({\it 23}).
\end{figure}

\begin{figure}
%\begin{center}
\vspace{0.0cm}
\hspace{-1.25cm}
\includegraphics[width=18.0cm]{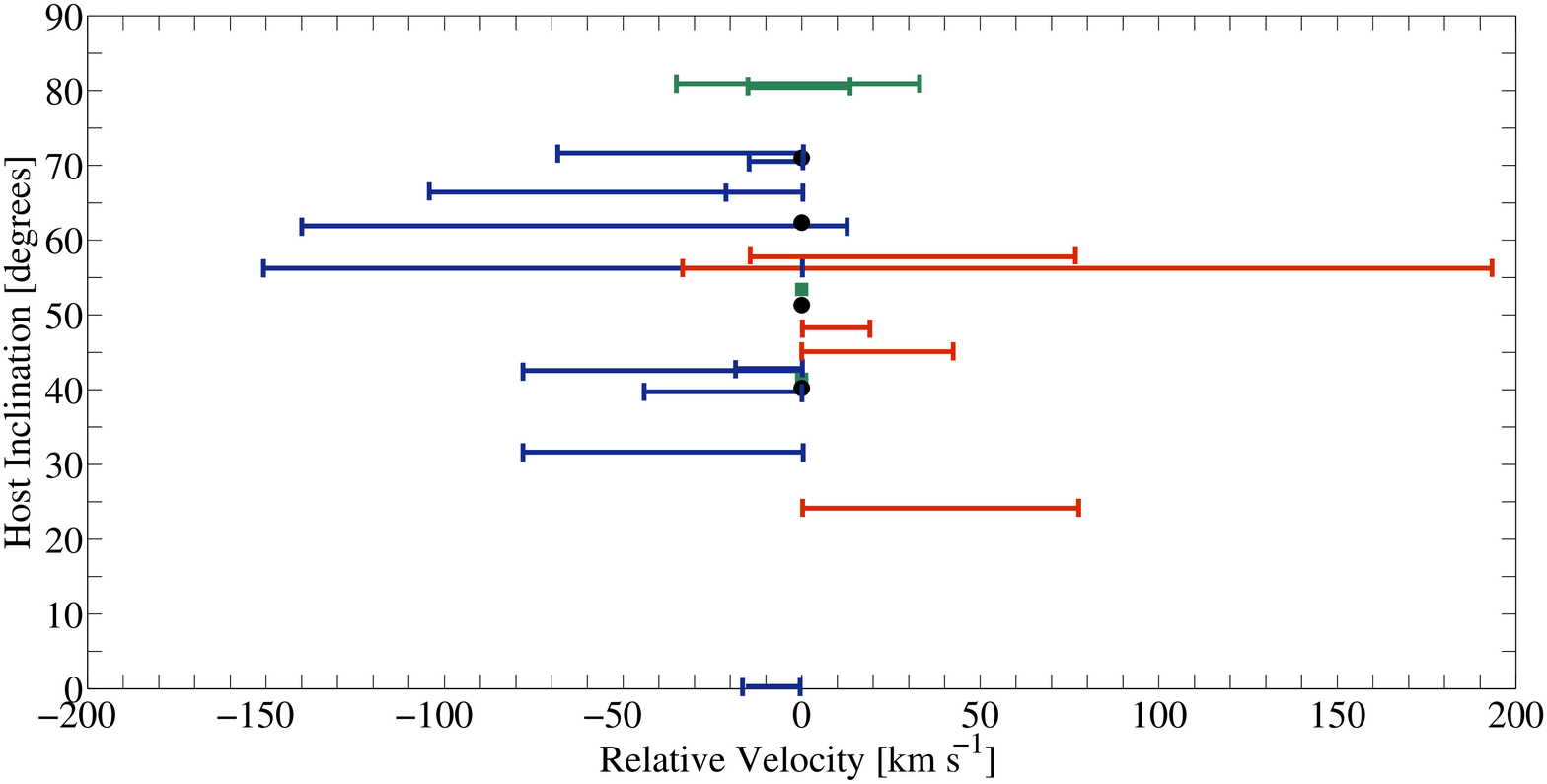} 
%\end{center}
Figure S7: The velocity span of the SN~Ia absorption features as a function of host galaxy inclination for events with spiral hosts (inclination $0^\circ$ is face on and $90^\circ$ is edge on). Inclinations are calculated using the major and minor axis given in NED ({\it 26}), $cos(i)=b/a$ (where i is the inclination, a the major axis, and b is the minor axis). Blue lines represent the events classified as blueshifted. Red lines represent redshifted events and green lines represent symmetric events. Green squares represent events with a single feature and black circles represent events with no sodium absorption detection. Galactic outflows should exhibit blueshifted absorption with a larger velocity span at lower host galaxy inclination. This is not observed in our SN~Ia sample. There seems to be no correlation between the host galaxy inclination and the velocities of the absorption features as expected from CSM. 
\end{figure}

\begin{figure}
\vspace{-2.5 cm}
\begin{tabular}{c c c}
SN 2006X, $i=32^\circ$ &
SN 2007fb, $i=71^\circ$ &
SN 2007kk, $i=40^\circ$ \\
\includegraphics[width=5.0cm]{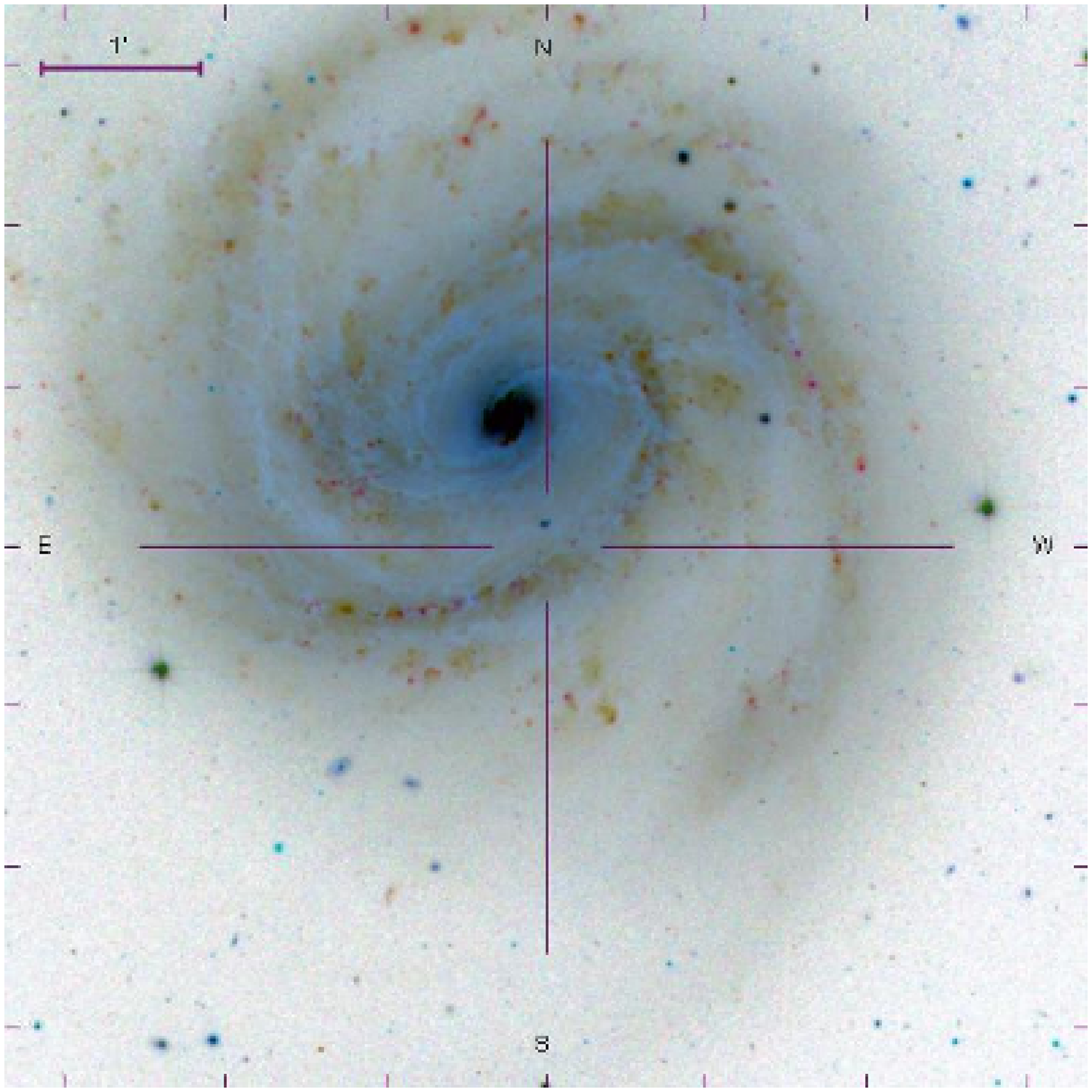} & 
\includegraphics[width=5.0cm]{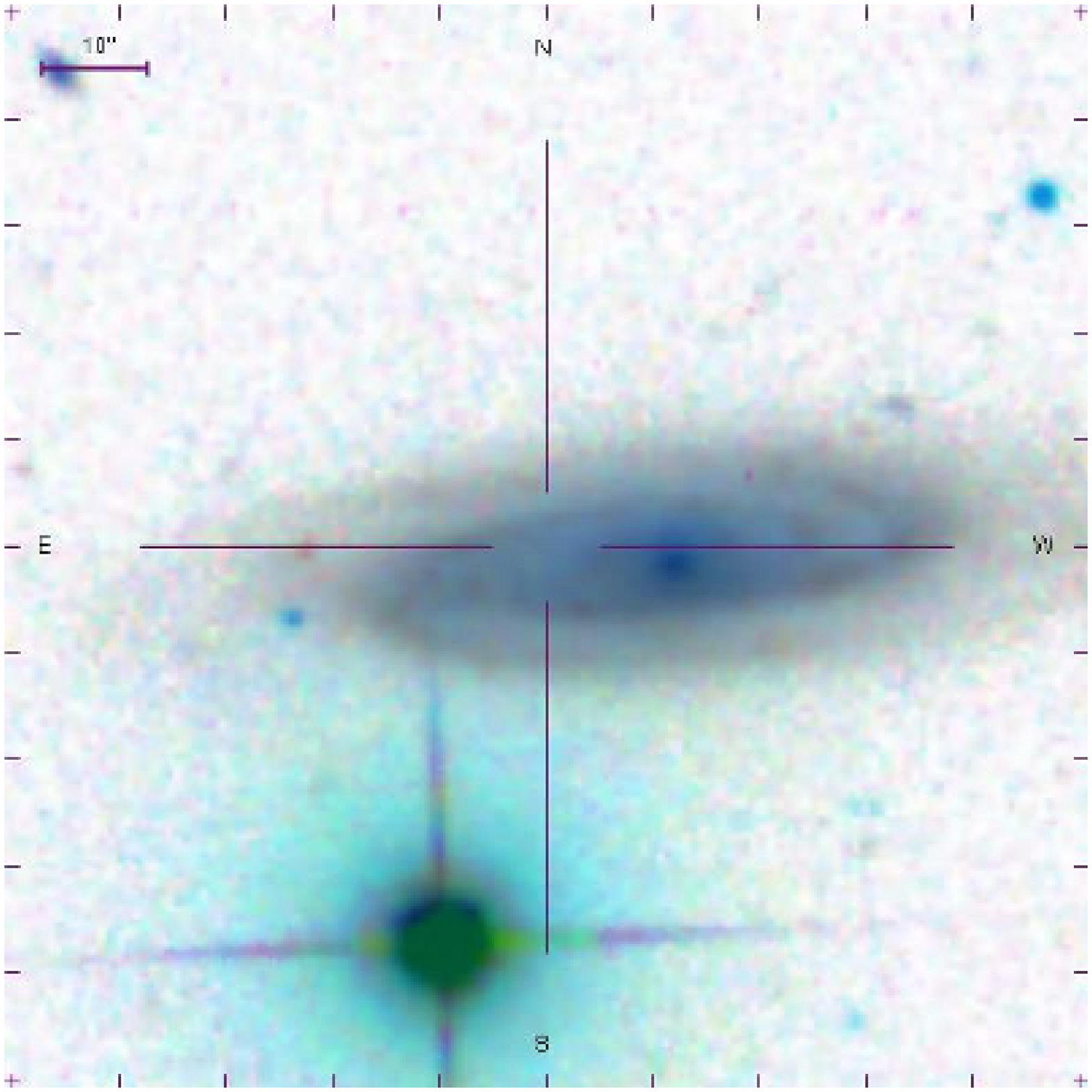} &
\includegraphics[width=5.0cm]{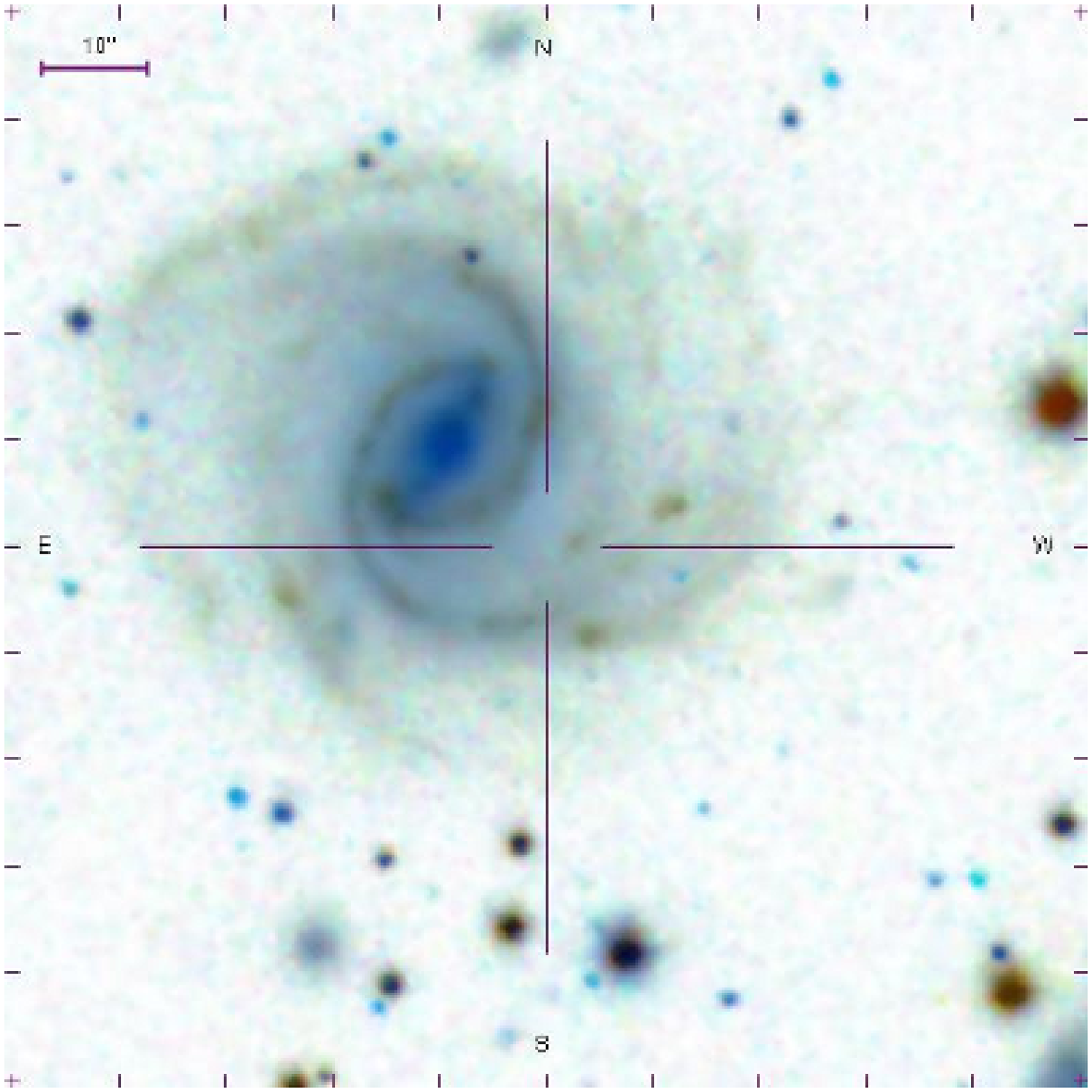} \\
SN 2007le, $i=66^\circ$ &
SN 2008C, $i=56^\circ$ &
SN 2008dt, $i=62^\circ$ \\
\includegraphics[width=5.0cm]{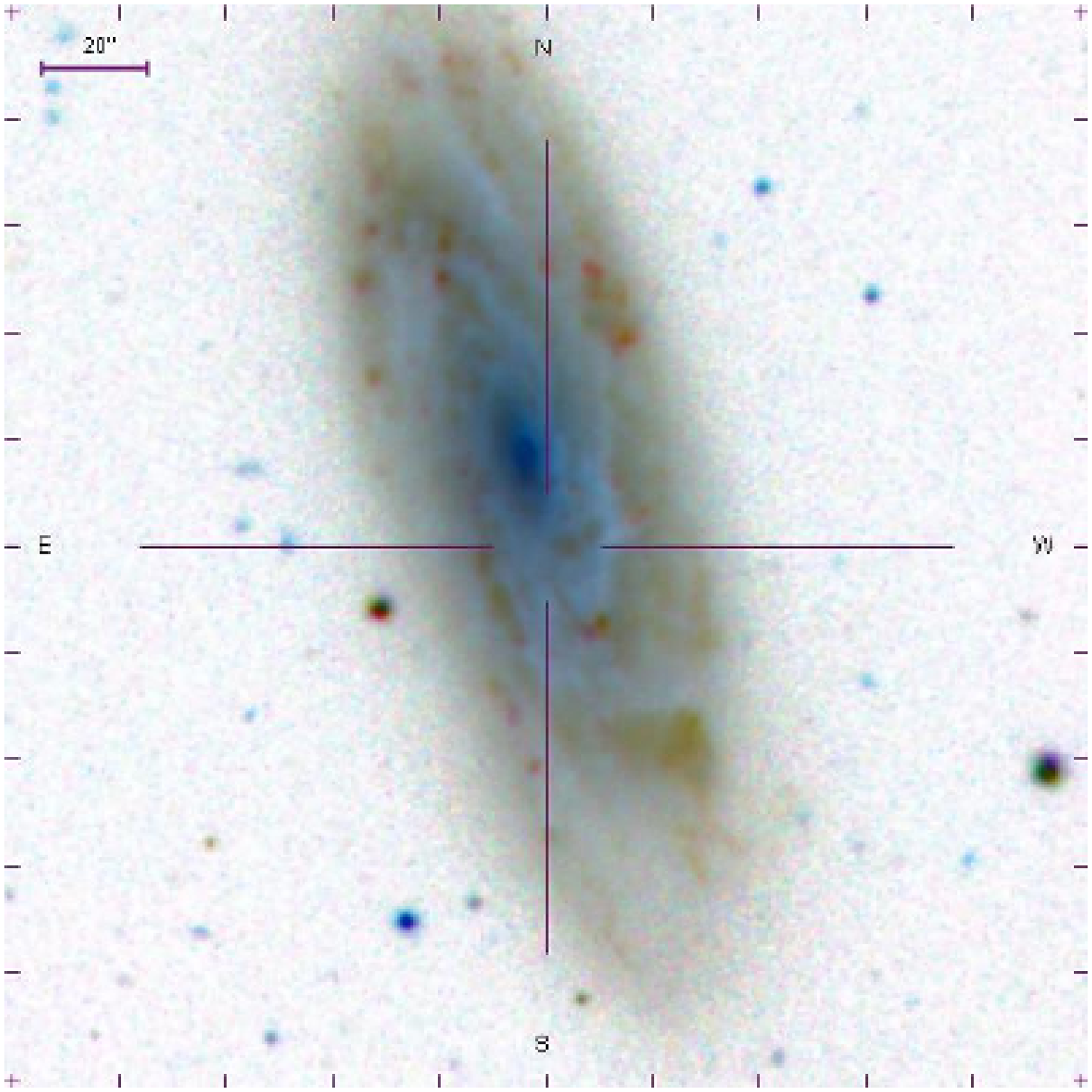} &
\includegraphics[width=5.0cm]{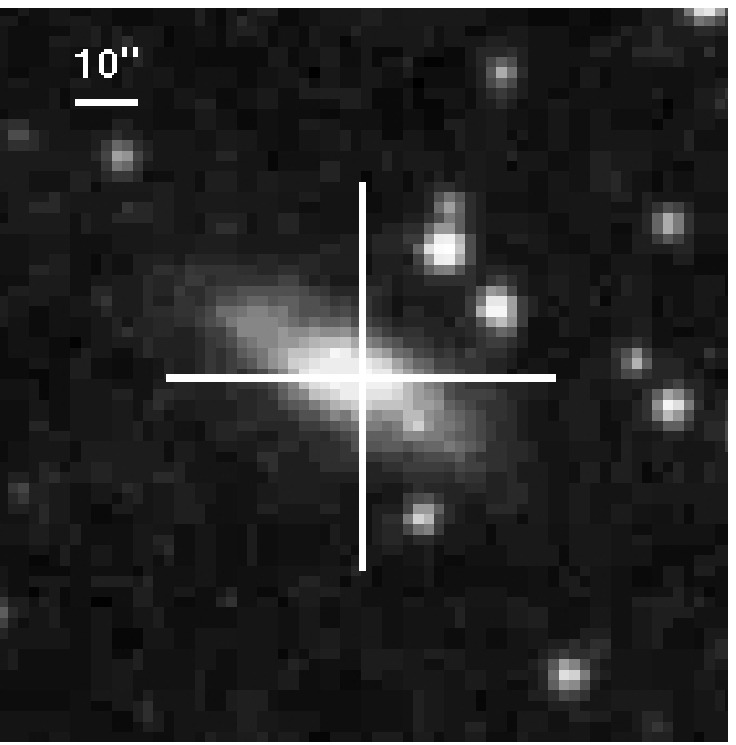} &
\includegraphics[width=5.0cm]{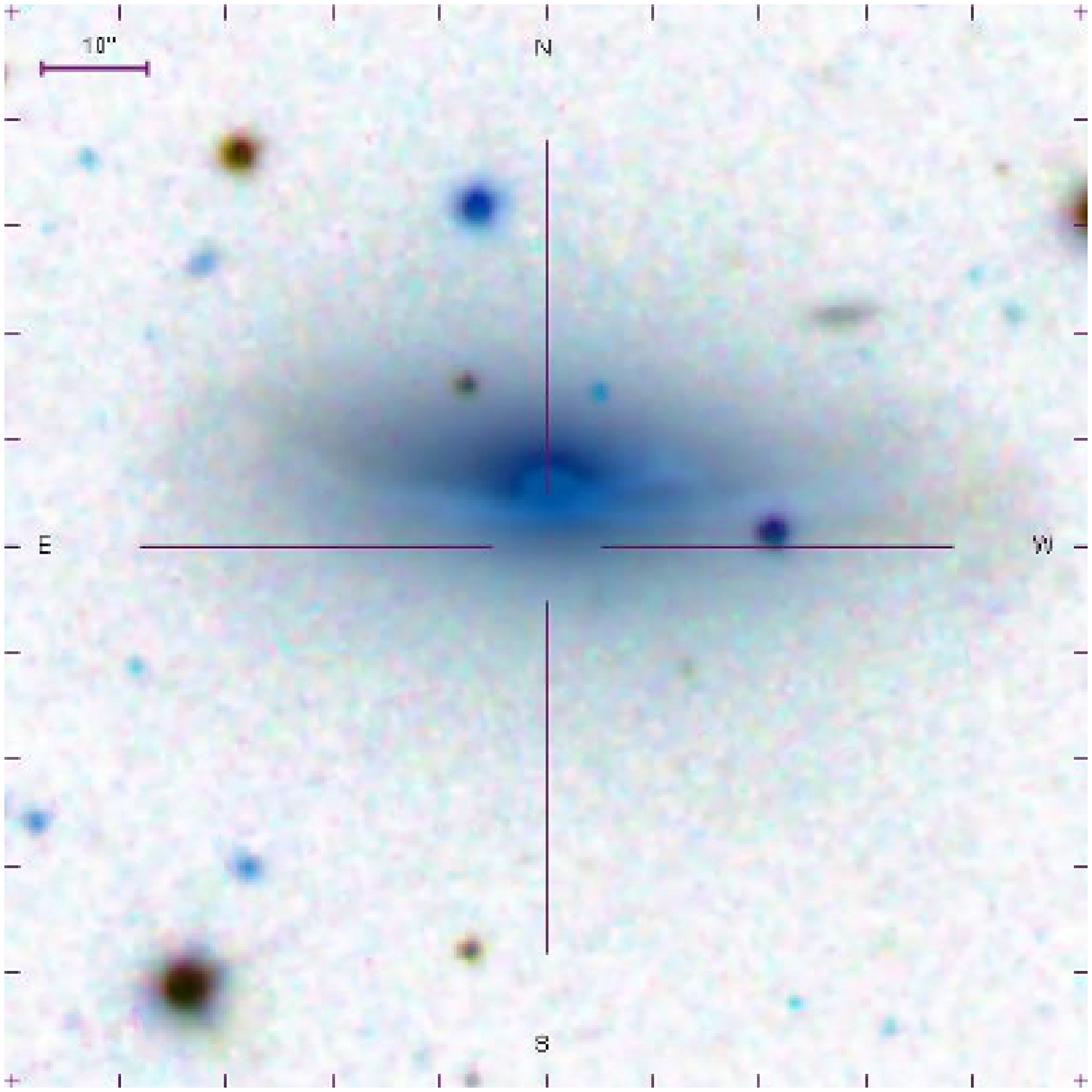}  \\
SN 2008ec, $i=43^\circ$ &
SNF20080612-003, $i=72^\circ$ &
SN 2009ds, $i=43^\circ$ \\
\includegraphics[width=5.0cm]{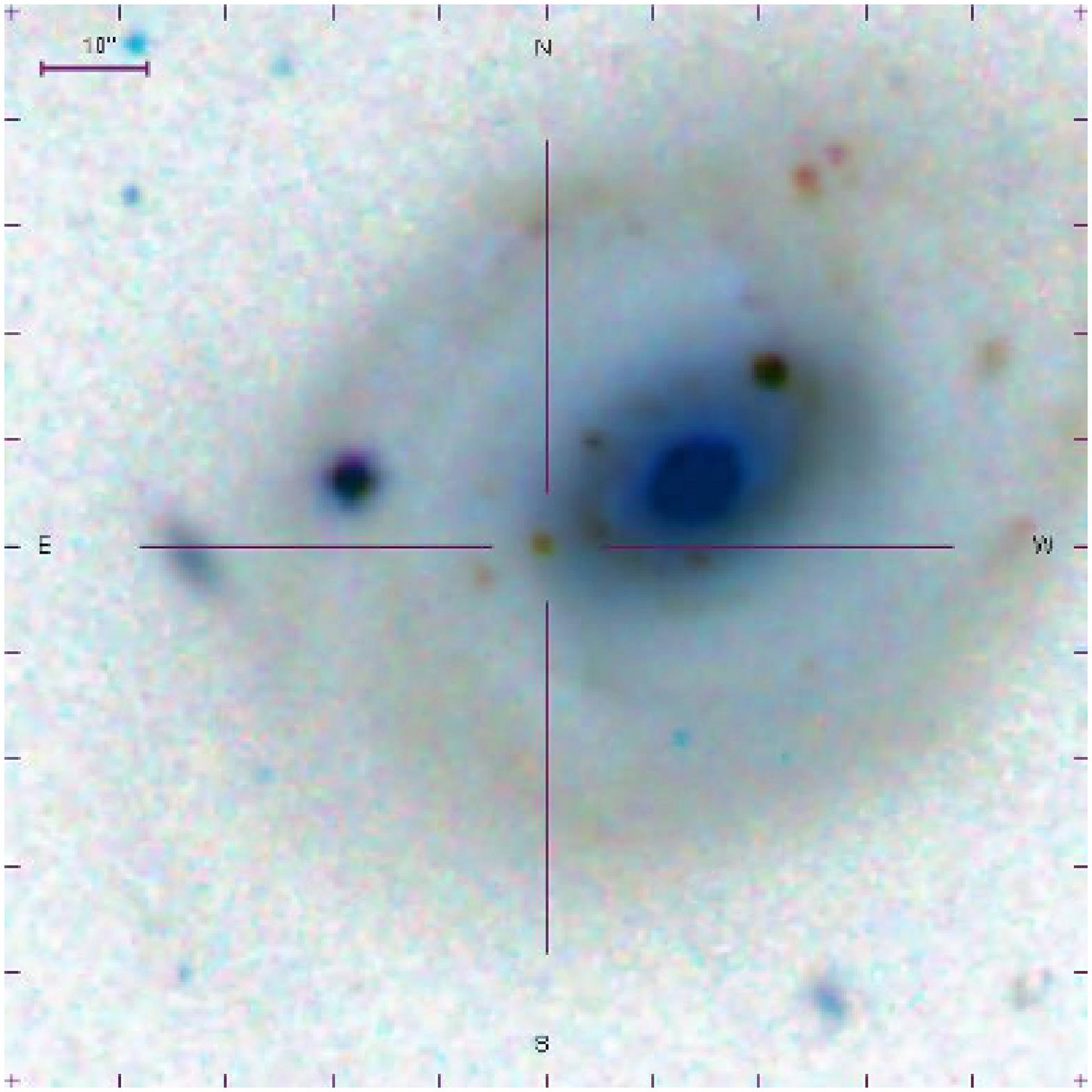} &
\includegraphics[width=5.0cm]{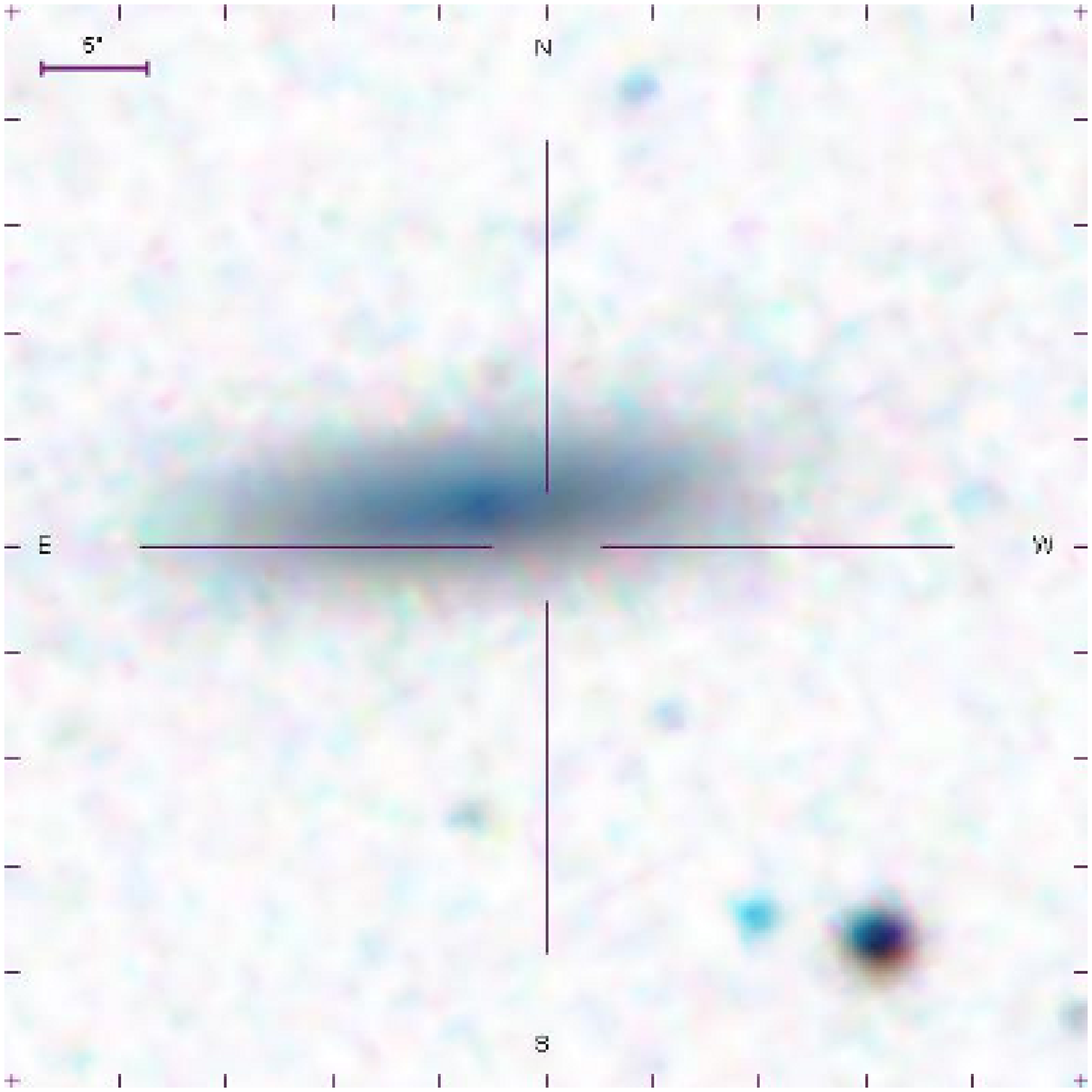} &
\includegraphics[width=5.0cm]{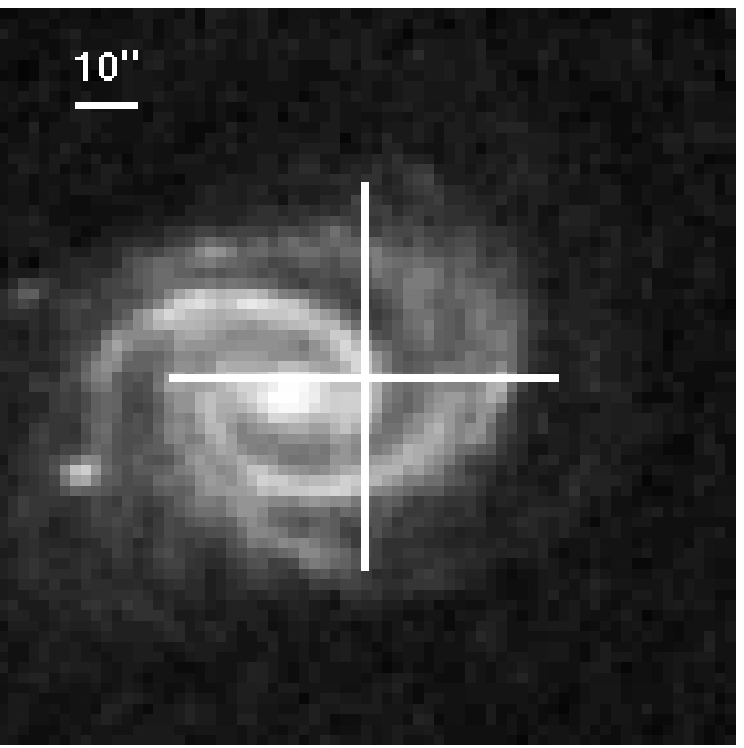} \\
\end{tabular}
\begin{tabular}{c c}
\hspace{2.5 cm}
SN 2009ig, $i=0^\circ$ &
SN 2009iw, $i=66^\circ$ \\
\hspace{2.5 cm}
\includegraphics[width=5.0cm]{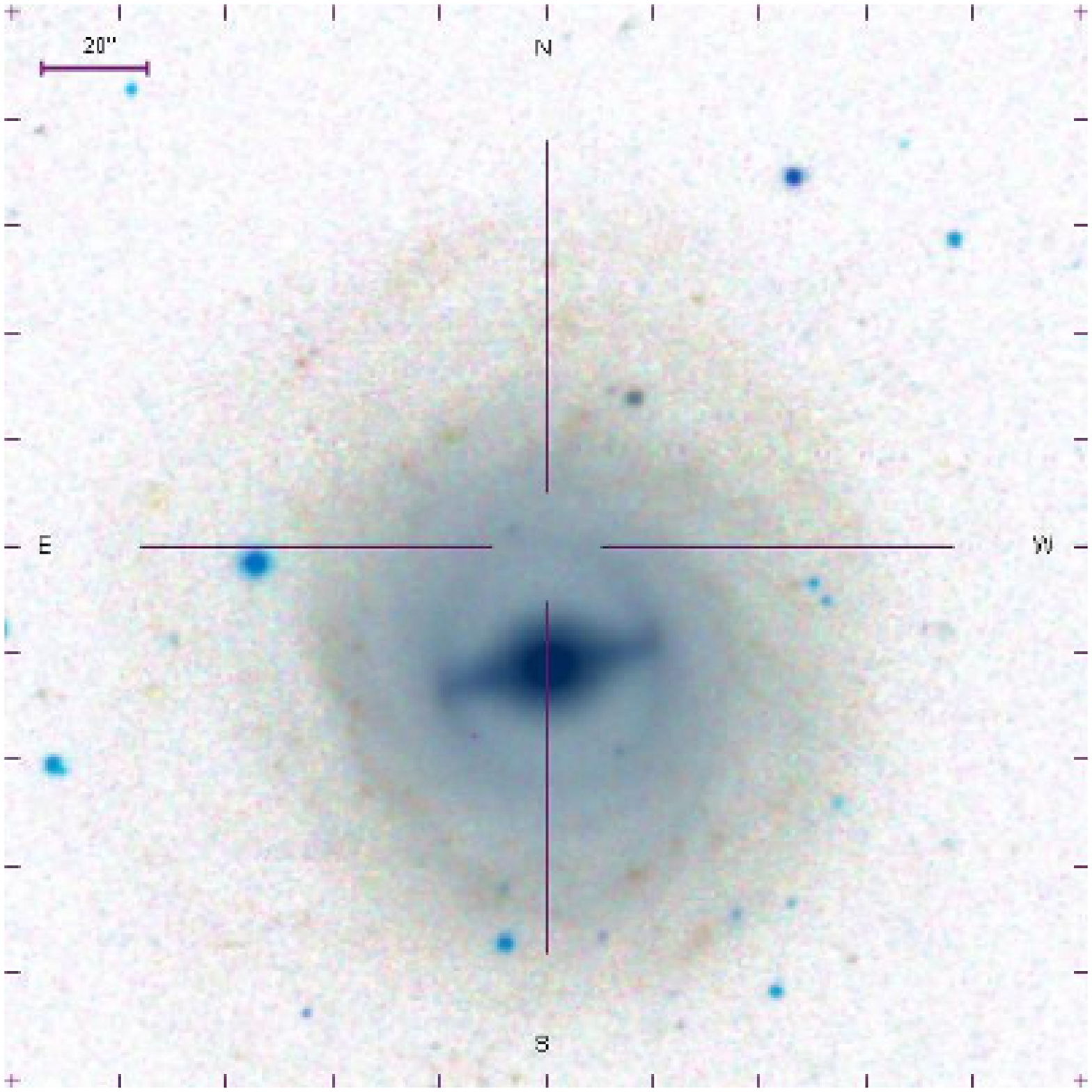} &
\includegraphics[width=5.0cm]{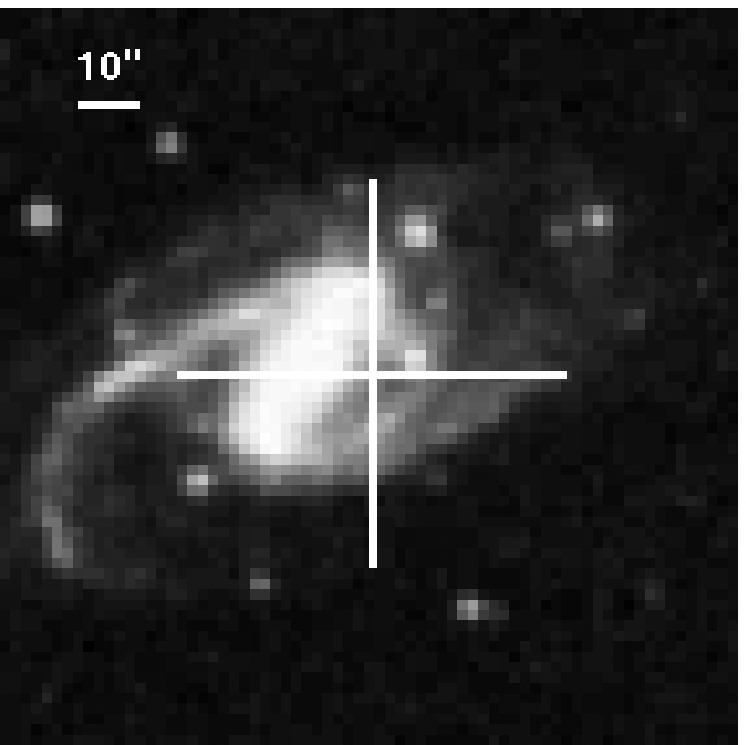}
\end{tabular} \\
Figure S8: Images of the spiral host galaxies of blueshifted SNe~Ia. The SN names and the host inclinations are given above each frame. Dark background images were taken from the ESO DSS archive (http://archive.eso.org/dss/dss). Light background images were taken using the SDSS DR8 finding chart tool (http://skyserver.sdss3.org/dr8/en/tools/chart/chart.asp). Scales are indicated at the upper left corner of each image. Cross-hairs mark the location of the SNe. 
\end{figure}

\begin{figure}
\vspace{-2.5 cm}
\begin{tabular}{c c c}
SN 2007af, $i=48^\circ$ &
SN 2007sa, $i=24^\circ$ &
SN 2009le, $i=56^\circ$ \\
\includegraphics[width=5.0cm]{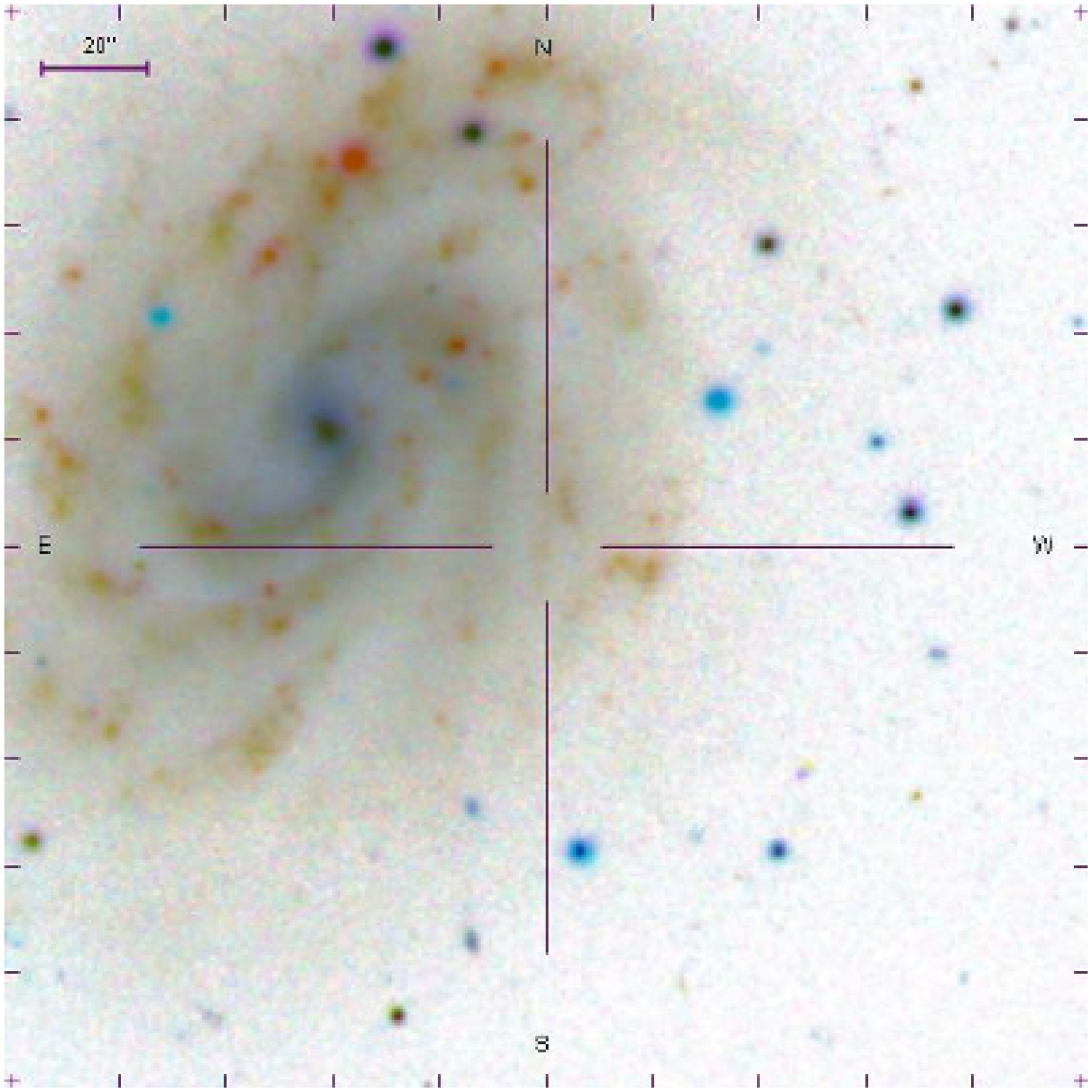} &
\includegraphics[width=5.0cm]{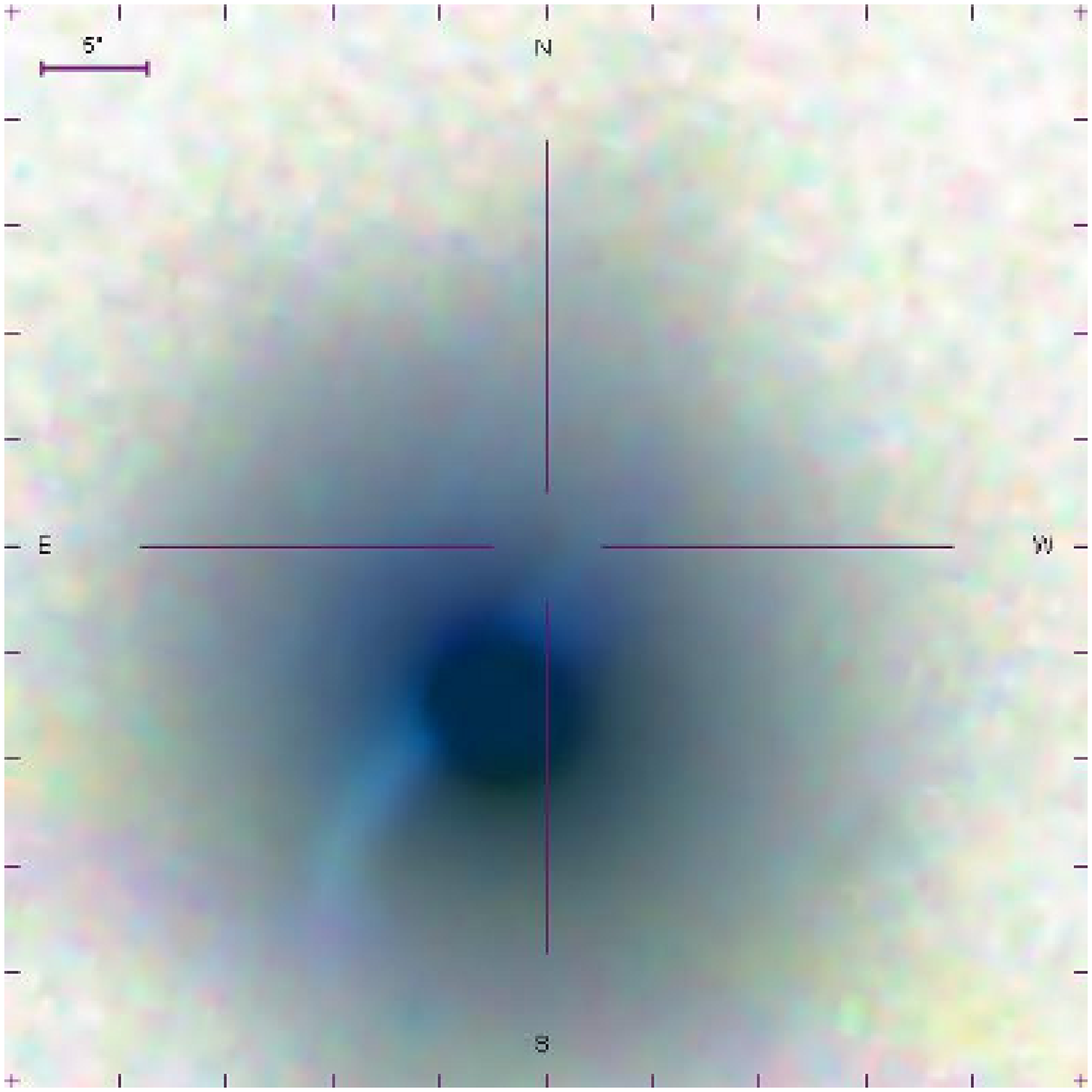} &
\includegraphics[width=5.0cm]{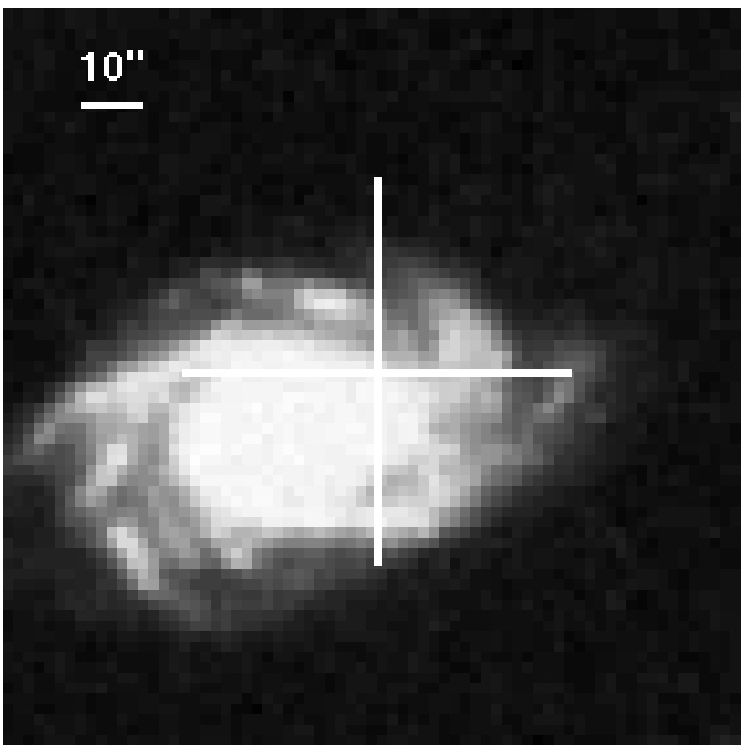} \\
\end{tabular}
\begin{tabular}{c c}
\hspace{2.5 cm}
SN 2009mz, $i=58^\circ$ &
SN 2010A, $i=45^\circ$ \\
\hspace{2.5 cm}
\includegraphics[width=5.0cm]{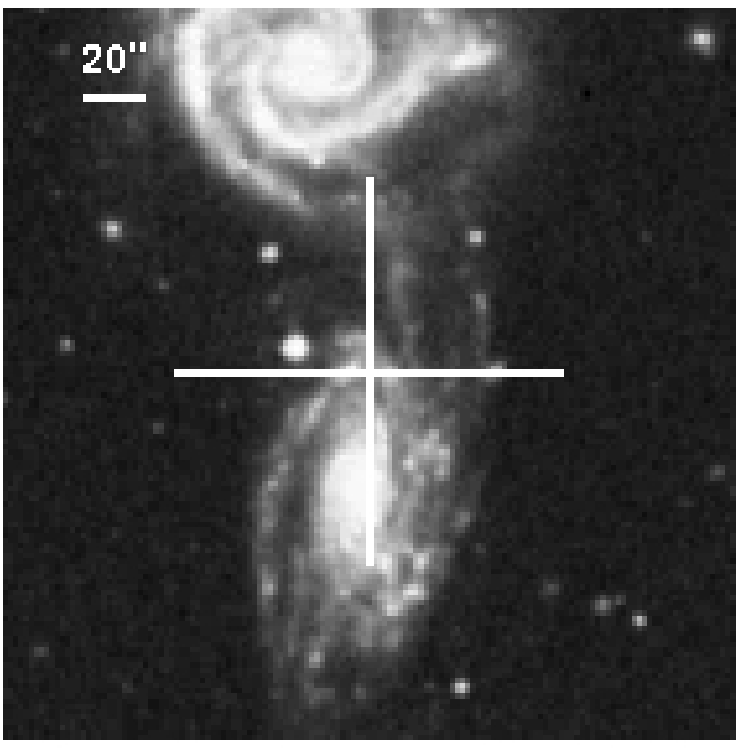} &
\includegraphics[width=5.0cm]{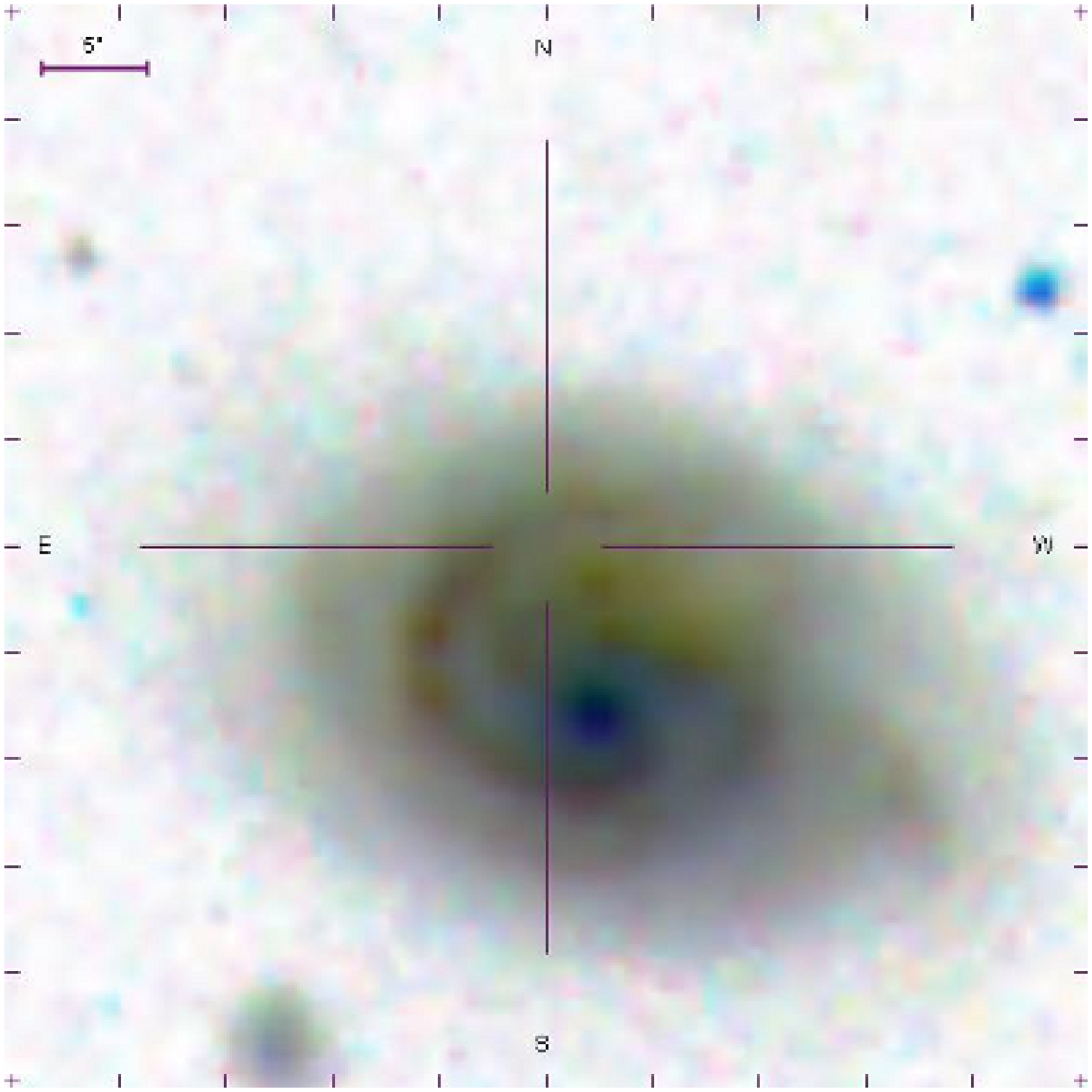}
\end{tabular} \\
Figure S9: Same as S8 for the redshifted SNe~Ia. 
\end{figure}

\begin{figure}
\vspace{0 cm}
\begin{center}
\begin{tabular}{c c}
SN 2006cm, $i=81^\circ$ &
SN 2007fs, $i=80^\circ$ \\
\includegraphics[width=5.0cm]{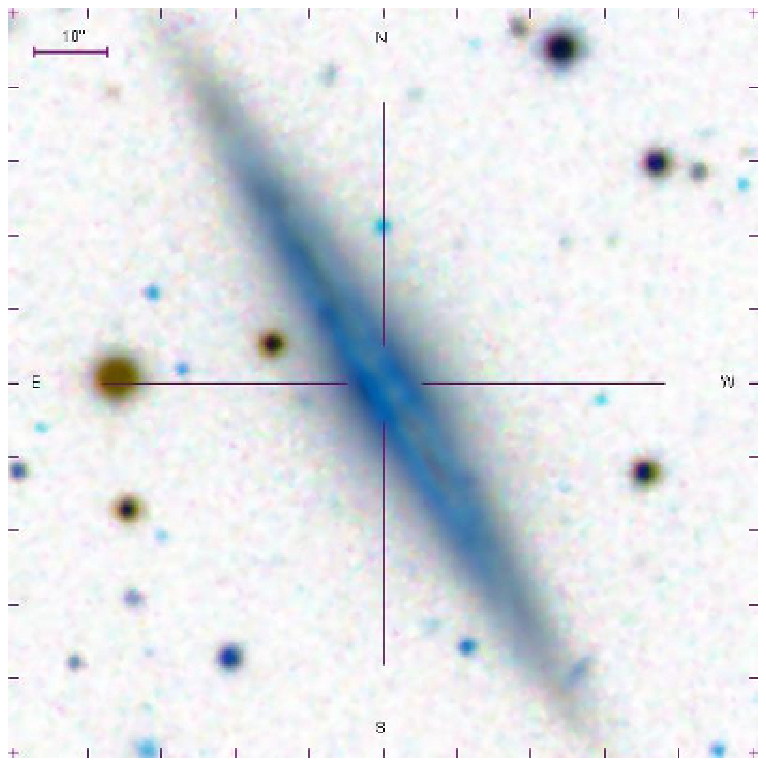} &
\includegraphics[width=5.0cm]{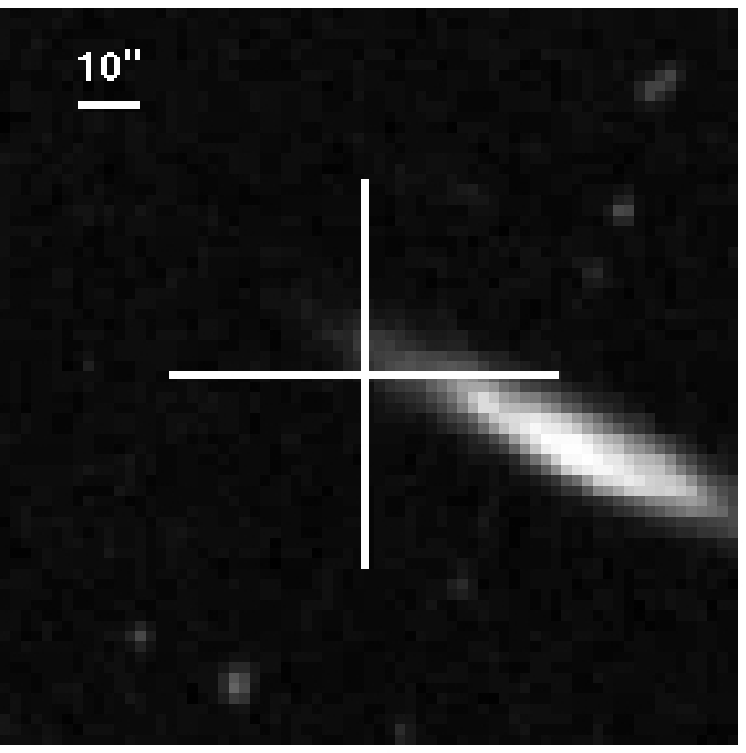} \\
SN 2007sr, $i=53^\circ$ &
SN 2010ev, $i=41^\circ$ \\
\includegraphics[width=5.0cm]{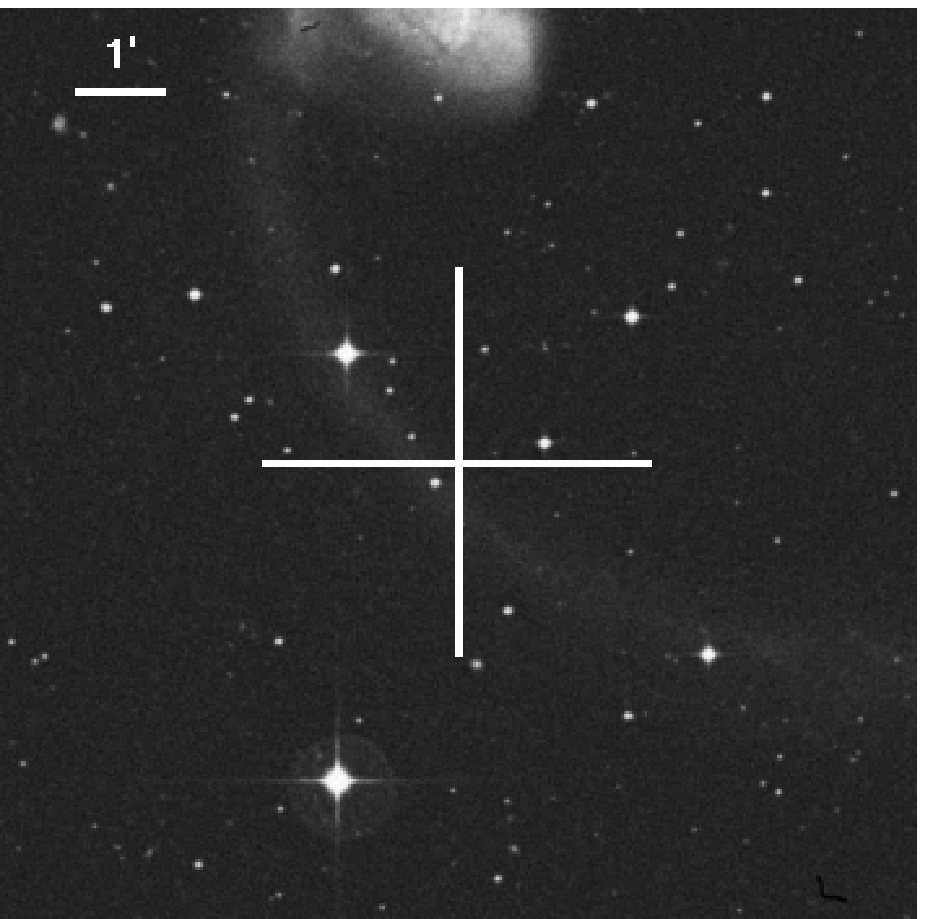} &
\includegraphics[width=5.0cm]{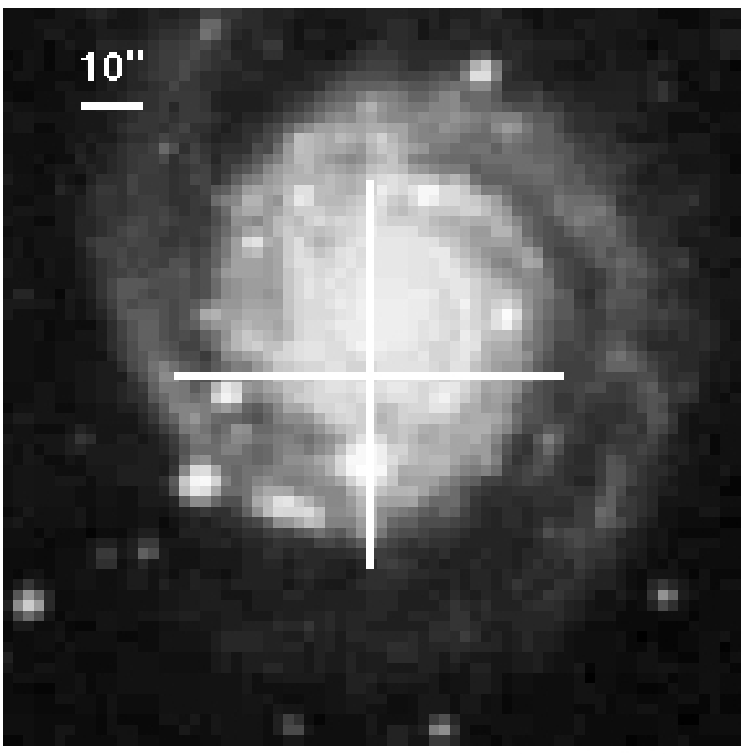}
\end{tabular} \\
\end{center}
Figure S10: Same as S8 for the single/symmetric SNe~Ia. 
\end{figure}

\begin{figure}
\vspace{0 cm}
\begin{center}
\begin{tabular}{c c}
SN 2007gj, $i=62^\circ$ &
SN 2008dh, $i=71^\circ$ \\
\includegraphics[width=5.0cm]{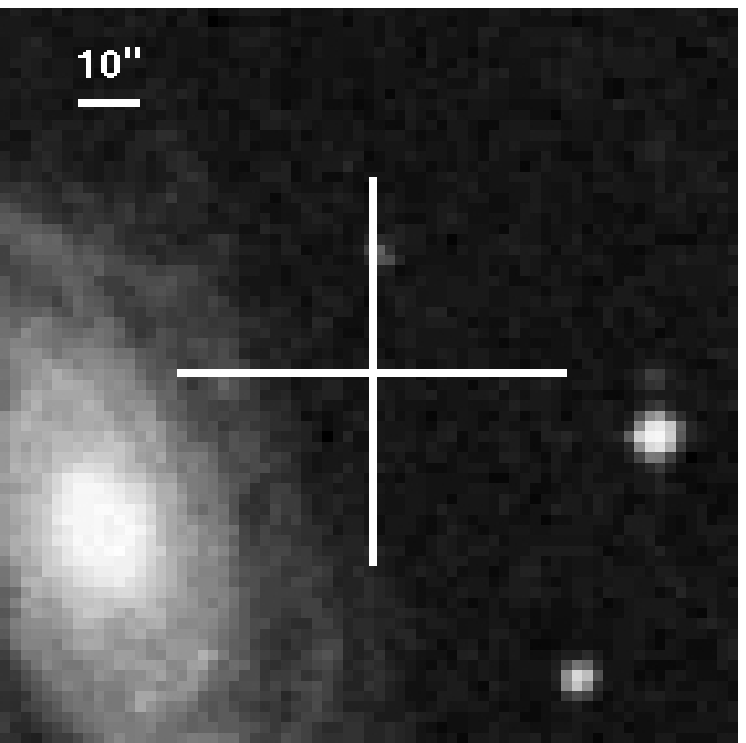} &
\includegraphics[width=5.0cm]{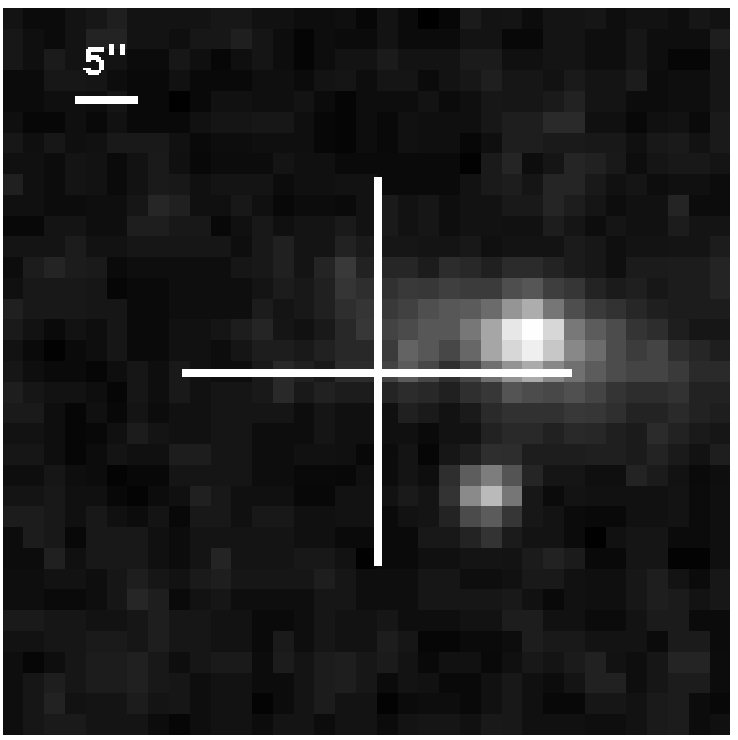} \\
SN 2009ev, $i=51^\circ$ &
SN 2009nr, $i=40^\circ$ \\
\includegraphics[width=5.0cm]{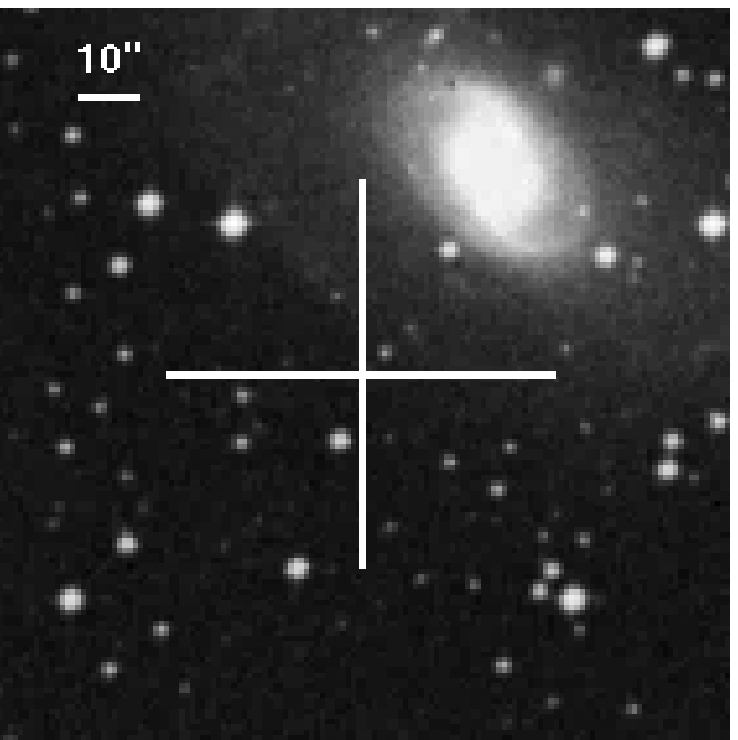} &
\includegraphics[width=5.0cm]{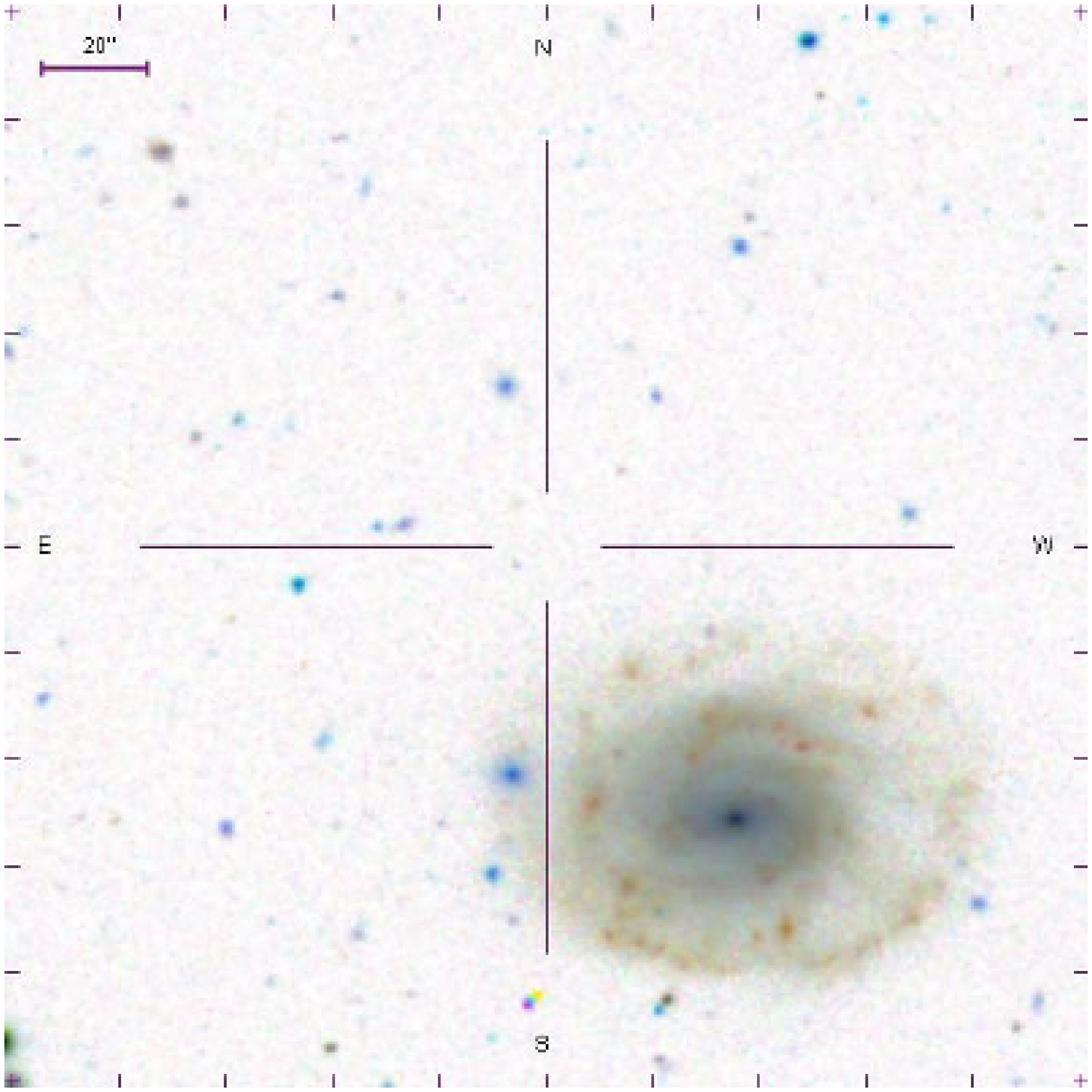} \\
\end{tabular}
\end{center}
Figure S11: Same as S8 for the SNe~Ia with no detection of sodium absorption. The inclination of the host galaxy of SN 2008dh was calculated by fitting an ellipse to the host and calculating the ratio between its minor to it's major axis.
\end{figure}

\begin{figure}
\begin{center}
\vspace{-3.0cm}
\hspace{-0.5cm}
\begin{tabular}{c}
\includegraphics[width=14.0cm]{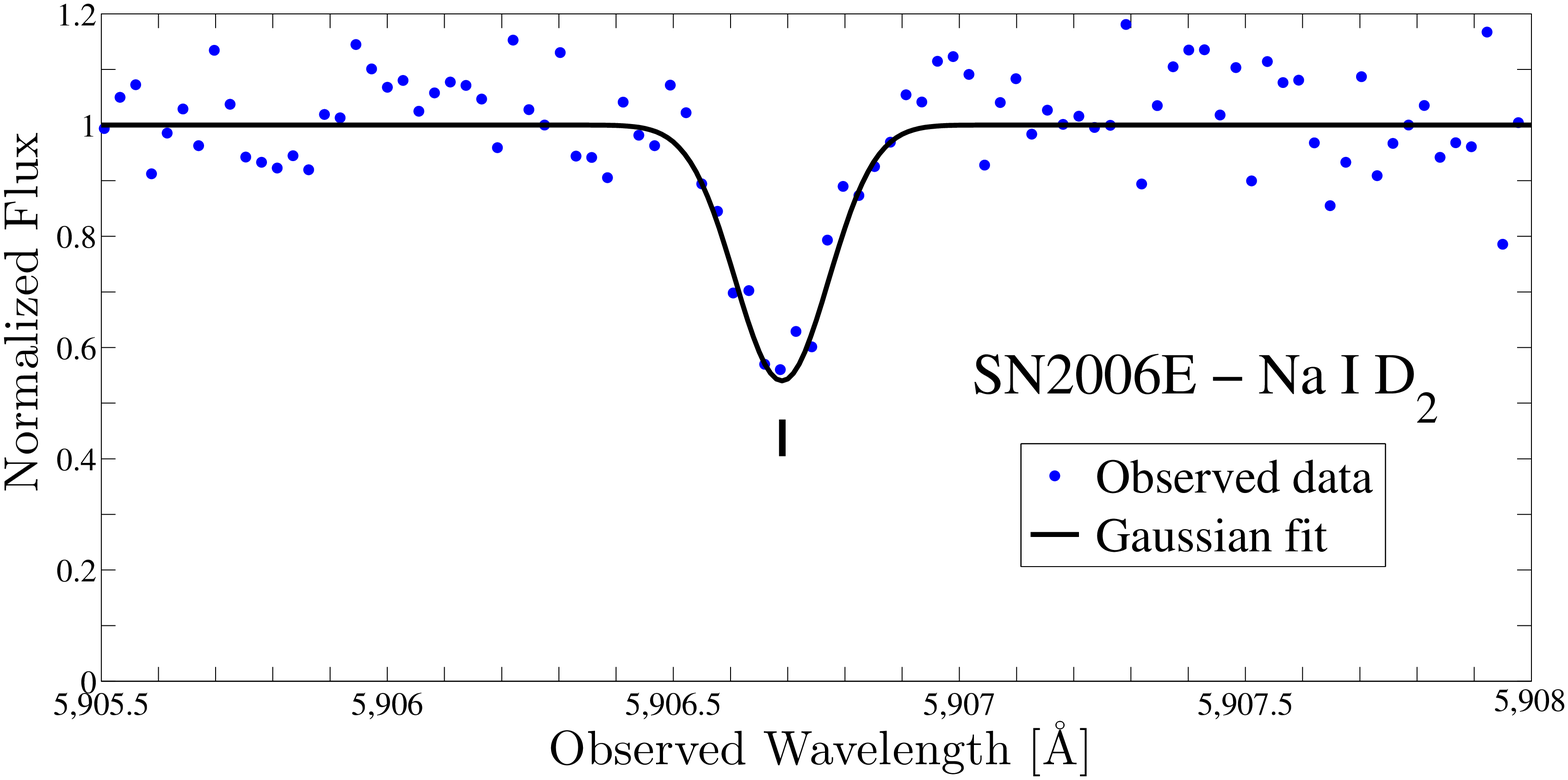} \\
\hspace{-0.35cm}
\includegraphics[width=14.0cm]{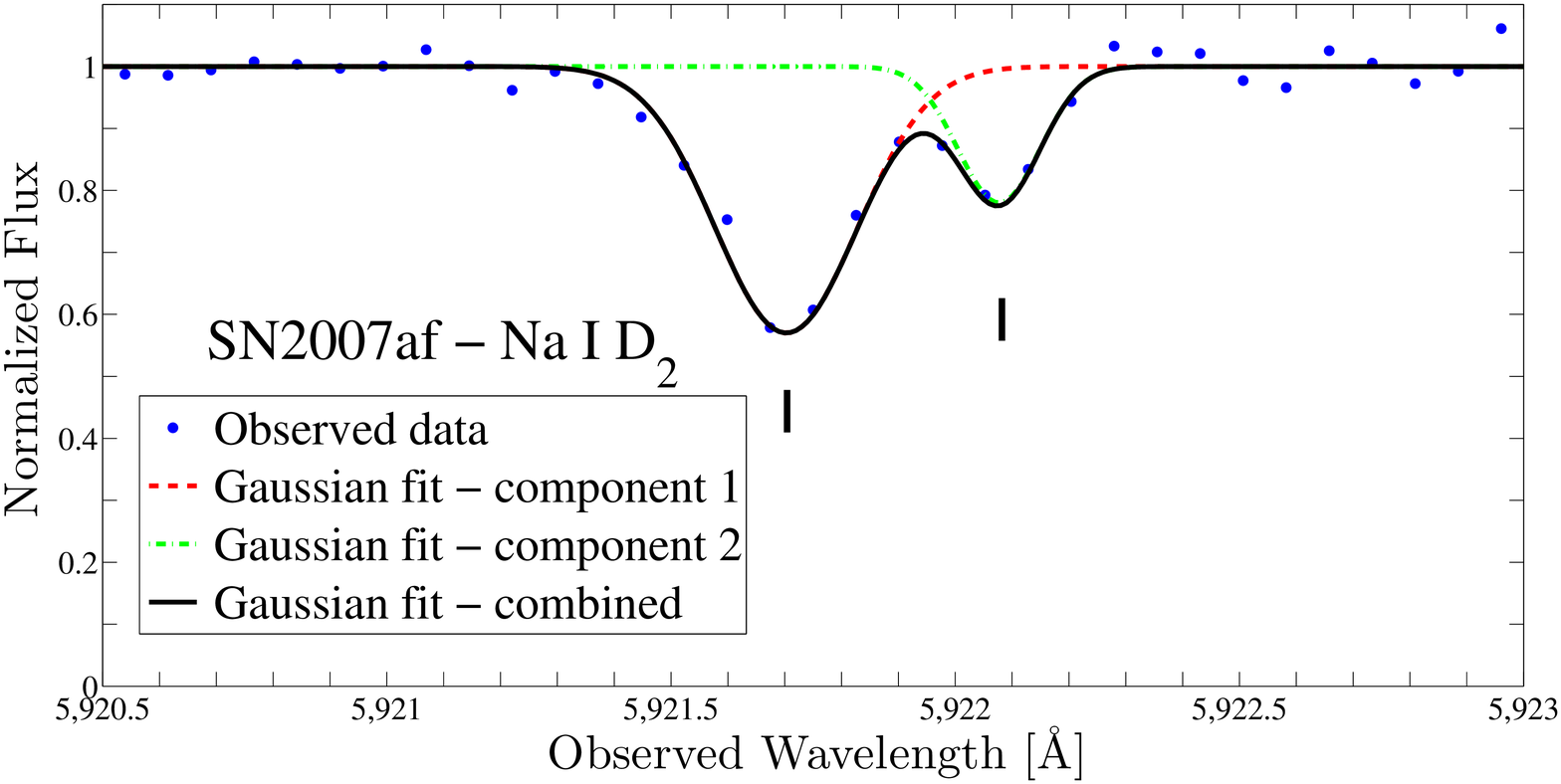} \\
\hspace{-0.35cm}
\includegraphics[width=14.0cm]{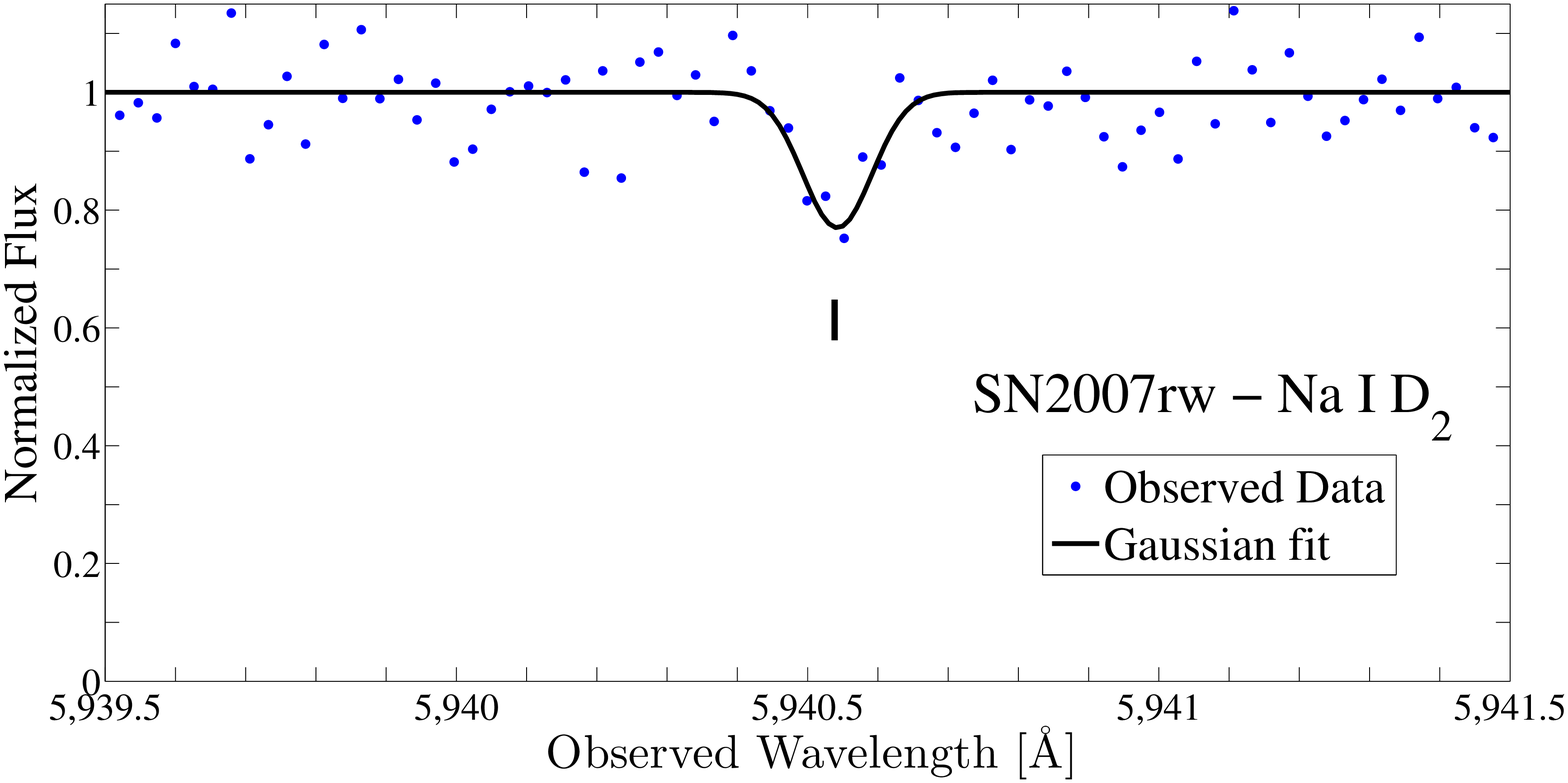}
\end{tabular} \\
\end{center}
Figure S12: Gaussian fits of the Na~I~D$_2$ absorption features exhibited in spectra of SN 2006E (upper), SN 2007af (middle), and  SN 2007rw (lower). Parameters of the fits are give in Table S4. Blue dots are the observed data. In the upper figure the black line is the Gaussian fit to the single feature. In the lower figure the red/green line is the fit to the left/right absorption feature, and the black line is the fit to both features combined. The vertical lines mark the center of the Gaussian fits.
\end{figure}

\end{document}